\pdfoutput=1
\documentclass{JINST}
\usepackage{textcomp}
\usepackage{amsmath, amsthm, amssymb}
\usepackage[ansinew]{inputenc}
\usepackage{wrapfig}
\usepackage{graphicx}
\usepackage{cite}
\usepackage{url}

\usepackage{xspace}
\newcommand{\degree}{\ensuremath{^\circ}\xspace}

\newcommand{\htp}{\ensuremath{\mathrm{H}_2^+}\xspace}
\newcommand{\nuebar}{\ensuremath{\bar{\nu}_e}\xspace}

\mathchardef\mhyphen="2D

\title{The IsoDAR High Intensity H$_{\textbf{2}}^\textbf{+}$ Transport and Injection Tests}

\author{J. Alonso$^a$, S. Axani$^a$, L. Calabretta$^b$, D. Campo$^b$, 
L. Celona$^b$, J. M. Conrad$^a$, A. Day$^c$, G. Castro$^b$, F. Labrecque$^d$, 
D. Winklehner$^a$\thanks{Corresponding author.}\\
\llap{$^a$}Department of Physics, Massachusetts Institute of Technology\\
77 Massachusetts Av., Cambridge, MA 02139, USA\\
\llap{$^b$}INFN - Laboratori Nazionali del Sud\\
Via S.Sofia 62, 95123 Catania, Italy\\
\llap{$^c$}Department of Physics, Wellesley College\\
106 Central St., Wellesley, MA 02481, USA\\
\llap{$^d$}Best Cyclotron Systems, Inc.\\
\#7-8765 Ash St., Vancouver, BC V6P 6T3, Canada\\
\\
E-mail: \email{winklehn@mit.edu}}

\abstract{This technical report reviews the tests performed at the Best Cyclotron Systems, 
Inc. facility in regards to developing a cost effective ion source, beam line 
transport system, and acceleration system capable of high \htp current output 
for the IsoDAR (Isotope Decay At Rest) experiment. We begin by outlining the requirements for the IsoDAR experiment then provide overview of the Versatile Ion Source, Low Energy Beam Transport system, spiral inflector, and cyclotron. The experimental measurements are then discussed and the results are compared with a thorough set of simulation studies. Of particular importance we note that the Versatile Ion Source (VIS) proved to be a reliable ion source capable of generating a large 
amount of \htp current. The results suggest that with further upgrades, the VIS could
potentially be a suitable candidate for IsoDAR. The conclusion outlines the key results
from our tests and introduces the forthcoming work this technical report has motivated.}

\begin{document}

\clearpage
\section{Introduction}
\subsection{The IsoDAR experiment}

\hspace{4ex} In the presently accepted 3-neutrino oscillation model, the three 
mixing angles and mass squared splitting values are relatively well known
\cite{nakamura:particle_physics}. There are, however, intriguing anomalies that 
do not fit within the three-flavor paradigm and suggest new physics beyond the 
standard model. These anomalies can be resolved with the hypothesis of a (3~+~N) 
sterile neutrino model, in which there are three light mostly-active neutrino mass states and 
N massive mostly-sterile neutrino mass states. Among the anomalous experiments, 
LSND observed a 3.8~$\sigma$ excess in \nuebar 
events \cite{aguilar:lsnd}, MiniBooNE observed a low energy $\nu_e$ and \nuebar excess of 
3.4~$\sigma$ and 2.8~$\sigma$ in a muon-neutrino flavored beam respectively ~\cite{aguilar:miniboone}, and the short baseline reactor experiments observed a \nuebar deficit from unity of 0.943$\pm$0.023~\cite{mention:reactor_anomaly}.

The IsoDAR experimental program is being developed as the first stage of the 
DAE$\delta$ALUS experiment to primarily investigate the aforementioned anomalies 
by addressing the sterile neutrino hypothesis. IsoDAR will accurately map the 
\nuebar disappearance oscillation neutrino wave within a 1 
kiloton-class liquid scintillating (LS) detector to determine the number of 
extra sterile neutrino flavors. The IsoDAR neutrino source relies on a 
low-energy, high-current (5 mA) H$_2^+$ beam impinging on a $^9$Be target
producing neutrons which flood a sleeve containing highly enriched $^7$Li, 
thereby creating $^8$Li (neutron capture) which decays via a well-understood  
$\beta$-decay \cite{bungau:daedalus}. The usage of \htp can mitigate the 
current limitations constraining the previous generation of cyclotrons and will 
enable IsoDAR to produce a scientifically useful and interesting neutrino flux. 
IsoDAR, paired with an observatory like KamLAND \cite{abe:kamland}, would 
observe 8.2 $\times$ 10$^5$ reconstructed inverse beta-decay events in five 
years. With this data set, IsoDAR could decisively test the N = 1 and N = 2 
sterile neutrino oscillation models within the confidence intervals of the 
anomalous experiments, allow precision measurement of $\nuebar-\mathrm{e}^-$ 
scattering\cite{conrad2014precision}, and search for production and decay of exotic particles\cite{aberle2013whitepaper}.

\begin{figure}
\begin{center}
\includegraphics[width=1\columnwidth]{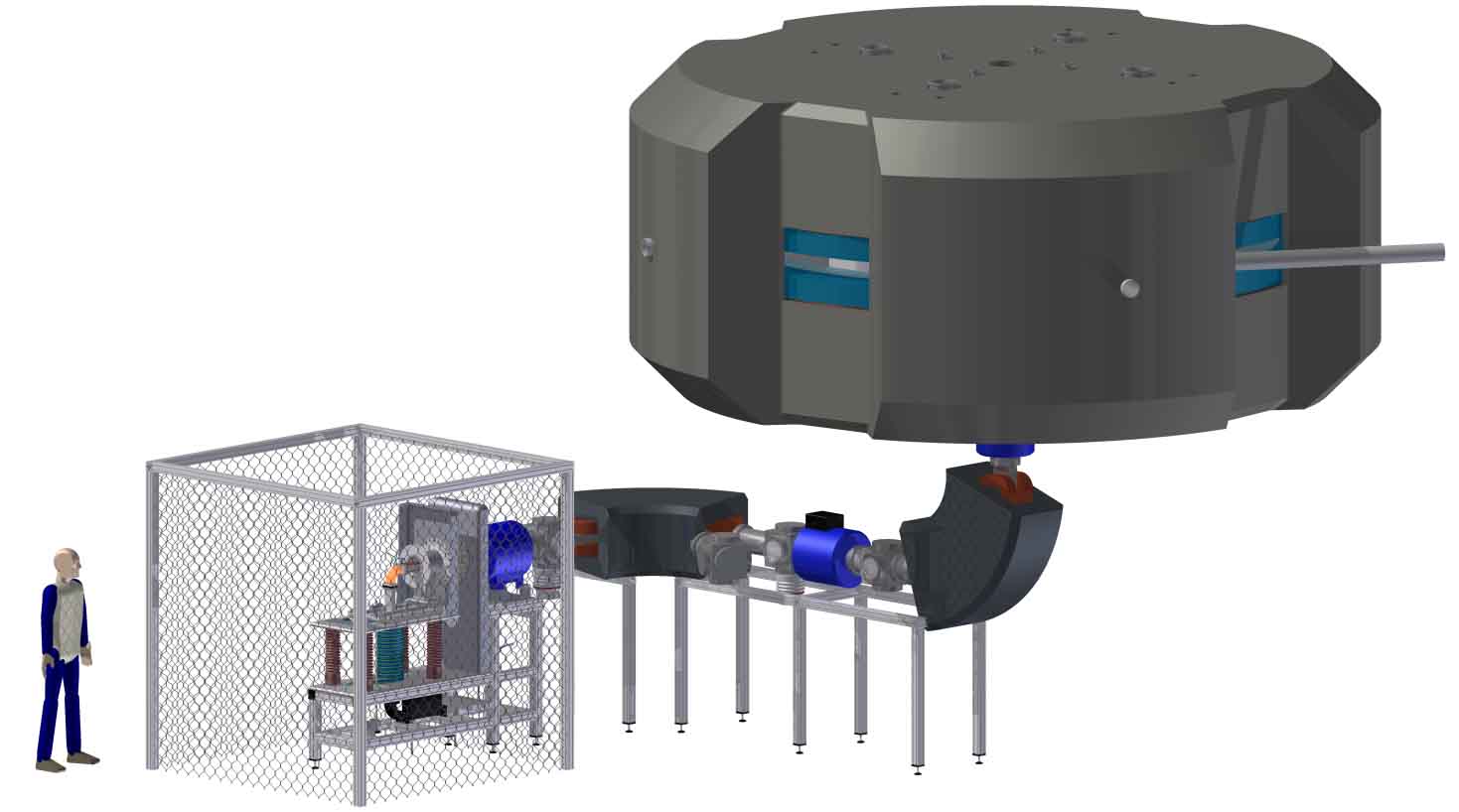}
\caption{Relative size of the IsoDAR injection system and the compact cyclotron. 
The accelerated \htp beam exits the image on the right and is transported to the 
antineutrino-production target approximately 50~m downstream.}
  \label{fig:cyclotron_cad}
 \end{center}
\end{figure}

IsoDAR is split into three main components: the IsoDAR injection system, the compact 
cyclotron (see Figure~\ref{fig:cyclotron_cad}), and the antineutrino-production target. 
The latest test stand built at the Best Cyclotron Systems, Inc. (BCS) facility in Vancouver, 
Canada, was adapted to test the concept of the current IsoDAR injection system and is the subject of this technical 
report. The outcome of these initial tests, and this report, represent the early stages of exploration into the development of the IsoDAR injection system.

\subsection{IsoDAR injector requirements}
\label{marker1}
\hspace{4ex}To accomplish the physics goals of the IsoDAR experiment, outlined 
in \cite{adelmann:isodar}, the IsoDAR injector is required to address three 
challenges: space-charge effects, high intensity \htp from an ion source, and efficient 
injection into the cyclotron.

\begin{enumerate}

\item Space-charge effects: A limiting factor constraining the maximum current in 
modern day cyclotrons is space-charge. Self-generated electric fields due to a high 
charge density cause the beam emittance to increase, making it more difficult to 
efficiently inject, accelerate, and extract ions from a cyclotron. 
The generalized perveance, $K$, a measures of the strength of space-charge, is given by:

\begin{equation} 
\mathrm{K} = \frac{q \mathrm{I}\cdot (1-\gamma^2\mathrm{f}_\mathrm{e})}{2 \pi \epsilon_0 m \gamma^3 \beta^3},
\label{eqn:space-charge}
\end{equation}

\cite{reiser:beams} with q, I, m, $\gamma$, and $\beta$ the 
charge, current, rest mass, and relativistic parameters of the particle beam, 
respectively and $\mathrm{f}_\mathrm{e}$ the space charge compensation factor. 
The higher the value of K, the stronger the space-charge effects. 
IsoDAR has specifically chosen \htp to overcome space charge limitations. 
Each \htp ion provides 2 protons after stripping thus reducing the necessary
beam current by a factor of 2. 
So, accounting for the changes in mass and velocity, this leads to a reduction of 
K by a factor of 1.4 compared to protons.
Space charge effects can be further reduced by maximizing the beam energy 
($\gamma^3\beta^3$ term), controlling space-charge compensation 
through beam line pressure ($\mathrm{f}_\mathrm{e}$ term), and avoiding 
electrostatic potentials in the beam line (e.g. Einzel lenses) whenever 
possible.

\item Intensity of \htp ion source: To capture 5 mA of \htp in the cyclotron, the 
ion source may need to provide up to 50 mA of \htp, depending on the combined 
efficiency of the buncher, beam transport, inflection, and RF capture. While many proton ion sources readily achieve 50 mA, this is at the 
edge of what \htp ion sources have produced so far.

\item Injection into the cyclotron: The axial injection into the cyclotron for 
the more magnetically rigid \htp ions (compared to protons -- H$^+$), requires an 
unusually large spiral inflector to redirect the beam into the center plane of the cyclotron. The large spiral inflector is constrained by the physical space in the 
central region, allowable beam emittance, and maximum voltage across the inflector plates before electrostatic breakdown.
\end{enumerate}

The adaptations for \htp ions to the BCS test stand facility were aimed at 
addressing the viability of overcoming these challenges.

\section{Experimental setup}

\subsection{Injector test stand layout \label{sec:bcs_layout}}
\begin{figure}[!t]
    \centering
    \includegraphics[width=1.0\columnwidth]{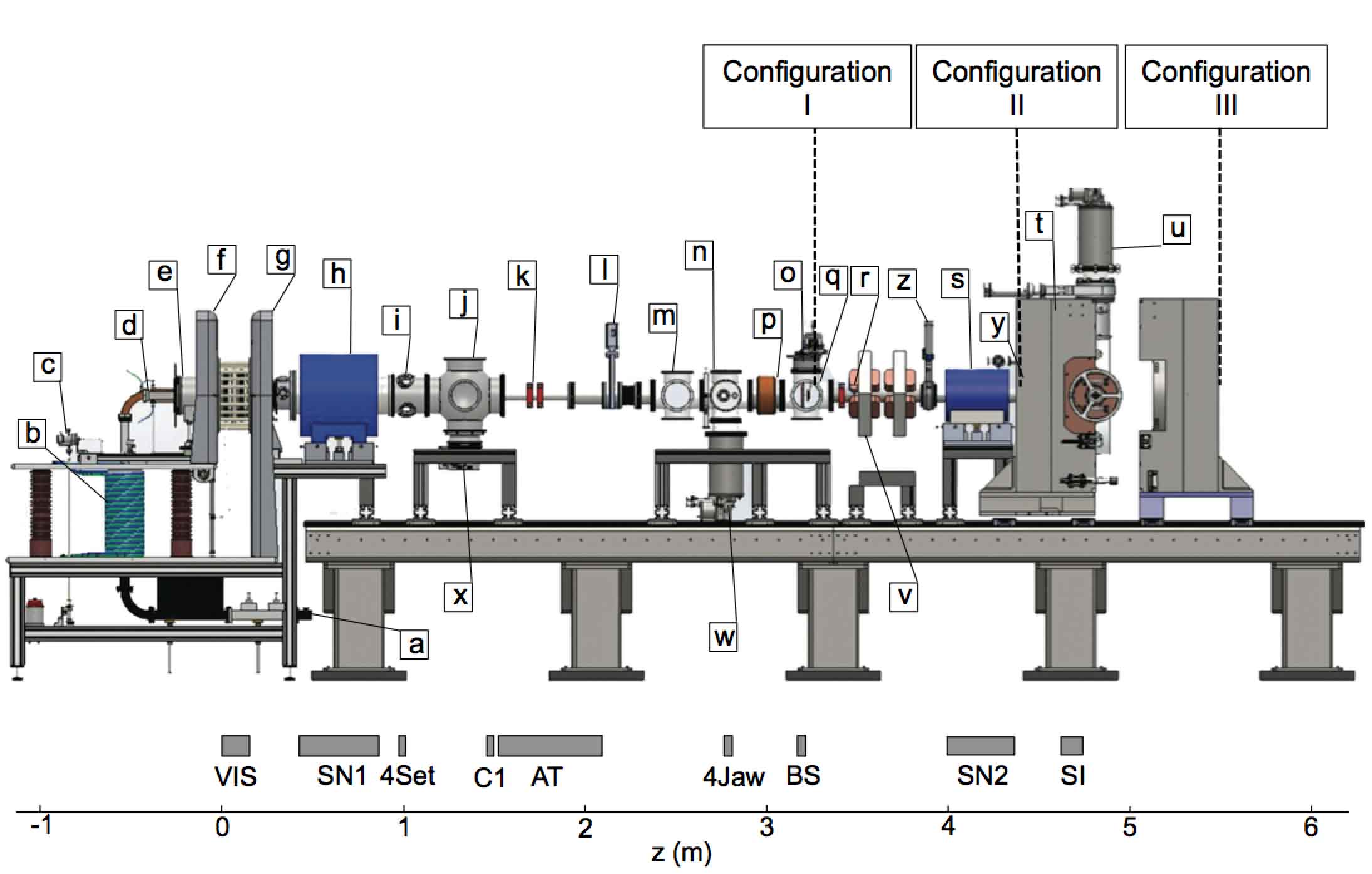}
    \caption{A complete schematic of the BCS test stand. The enumerated 
             components are: 
             [a] RF magnetron, 
             [b] graded, insulated waveguide, 
             [c] hydrogen mass flow controller, 
             [d] quartz window, 
             [e] plasma chamber, 
             [f] high voltage side, 
             [g] groud side and upstream ion gauge, 
             [h] Solenoid 1 (SN1), 
             [i] 4-Sector collimator (4Set), 
             [j] water-cooled collimator (C1), 
             [k] first set of steering magnets on a small adapter tube (AT),
             [l] first gate valve,  
             [m] 4-Jaw collimator (4Jaw), 
             [n] Retarding Field Analyzer (RFA) and downstream ion gauge, 
             [o] interchangeable Faraday cup (FC) and pneumatic beamstop (BS), 
             [p] DC Current Transformer (DCCT), 
             [q] Allison-type Emittance Scanner (ES), 
             [r] second set of steering magnets, 
             [s] Solenoid 2 (SN2), 
             [t] 1 MeV test cyclotron, 
             [u] cyclotron cryopump, 
             [v] quadrupole focusing magnets, 
             [w] beam line cryopump, 
             [x] 1500 l/s turbo pump, 
             [y] spiral inflector (SI), 
             [z] second gate valve. 
             A list of commonly used abbreviations can be seen at the bottom of
             the figure as well as the approximate length of the beam line. The
             gray boxes represent the approximate locations and sizes of the
             important component used in the simulations (cf. Section
             \protect\ref{sec:simulations}). The end points for Configurations 
             I-III as described in the text are also indicated.}
    \label{beamline_schematic}
\end{figure} 
\hspace{4ex} The ion injector is divided into four individual sections: the 
Versatile Ion Source (VIS), the low-energy beam line transport (LEBT) system, the 
spiral inflector (SI), and the cyclotron. The experimental layout of the components 
can be seen in Figure~\ref{beamline_schematic} along with a list of commonly referred to abbreviations and their location along the beam line.
During the runs in the summers of 2013 and 2014, three different
configurations were used to study the beam dynamics in the LEBT and 
the spiral inflector performance:

\begin{itemize}
	\item Configuration I: The first tests were performed with the beam line
		  extending 3.5 m beyond the extraction aperture. As seen in Figure
		  \ref{beamline_schematic} marked by
		  grey boxes, the main beam shaping elements were a solenoid magnet
		  (SN1) close to the source, and several sets of collimators. An
		  emittance scanner (ES) and a Faraday cup (FC) were installed in two 6-way 
		  crosses towards the end.
	\item Configuration II: The beam line was later extended to 5.5 m, with the
	      addition of another solenoid (SN2) and a quadrupole doublet. Most 
	      tests using the quadrupole doublet did not yield any improvements for 
	      cyclotron injection and thus it was mostly unused. For the rest of
	      the paper, it will be considered a simple beam pipe. The emittance scanner
	      and the Faraday cup were moved to the new end of the beam line.
	\item Configuration III: Finally, the cyclotron was attached to the beam 
	      line end, replacing the emittance scanner and Faraday cup. 
	      The rest of the LEBT is identical to Configuration II.
\end{itemize}

Configurations I-III are labeled at the top of Figure \ref{beamline_schematic} 
and indicate the end points of the three beam line configurations.

\subsubsection{Versatile Ion Source (VIS) \label{sec:vis}}
\hspace{4ex} The VIS is a 2.45 GHz, non-resonant microwave ECR ion source that 
was constructed at the Istituto Nazionale di Fisica Nucleare, Laboratori Nazionali 
del Sud (INFN-LNS). It was developed as an evolution of the TRIPS source
\cite{celona:trips} in order to produce a robust, high current, continuous wave 
(CW) proton beam. In the VIS, ions are produced by collisions with electrons 
resonantly heated by 2.45 GHz microwaves which are supplied through an electrically insulated waveguide. The magnetic field necessary for electron cyclotron resonance 
is provided by two permanent magnet solenoids.
This is a convenient configuration as virtually no power has to be transferred to 
the high voltage platform.
Due to its efficiency and high-current characteristics, it was 
felt that it could potentially be a good source for \htp ions, and
so was shipped to Vancouver and incorporated into the BCS test stand for the experiments conducted by the collaborative effort of the MIT IsoDAR group, BCS staff, and the team from INFN-LNS \cite{castro:vis2}. 

The VIS consists of elements [a]-[g] in Figure
\ref{beamline_schematic}. Microwave power is provided by a Sairem 
2 kW 2.45 GHz magnetron, with flexible operation in Continuous Wave (CW) or pulsed 
mode. Diatomic hydrogen gas is injected into the plasma chamber through an El-Flow
F-200CV mass flow controller. 
A 100 kV, 100 mA FuG Elektronik GmbH power supply provides the
high voltage to the platform and ion source body relative to
the grounded beam line, thus defining the ion beam energy.
Typically, due to the high relative humidity in Vancouver, the VIS was operated at $<66$ kV source potential, while a suppressor
voltage of $<-3.5$ kV was used to avoid electrons streaming from
the LEBT into the plasma chamber with the aim to prevent 
electrostatic breakdown and discharges.

\begin{figure}
    \centering
    \includegraphics[width=0.7\columnwidth]{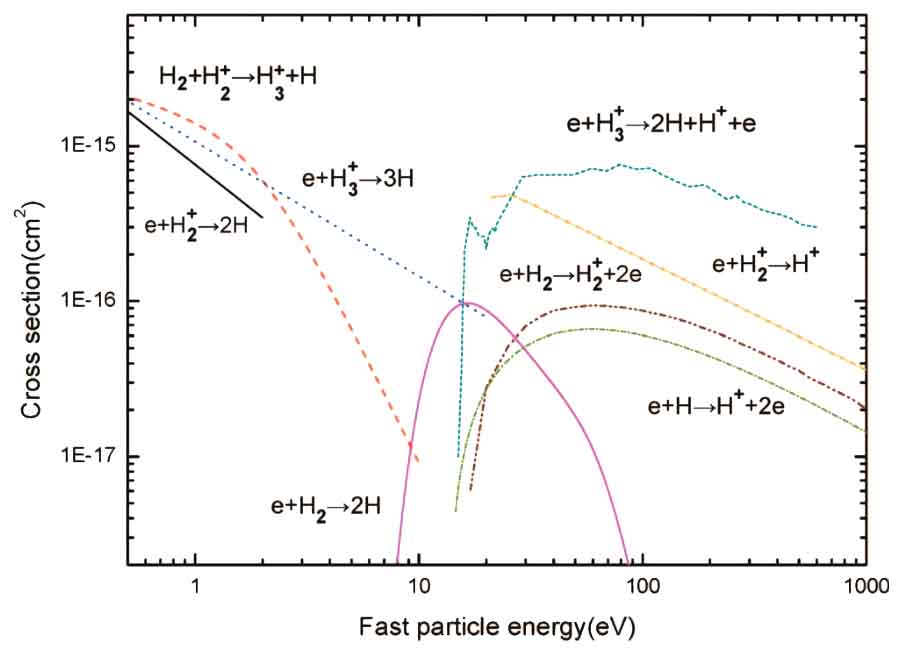}
    \caption{Relevant cross sections for the main physical processes in a
             hydrogen ion source. From \protect\cite{xu:current}.}
    \label{cross}
\end{figure}

The dominant processes leading to production and loss of \htp in a plasma 
ion source are \cite{xu:current}:
\begin{equation} \label{production}
     \begin{aligned}
        \mathrm{H}_{2} + \mathrm{e}^{-} &\longrightarrow \htp + 2e^{-} \\
        \htp + e^{-} &\longrightarrow \mathrm{H}^{+} + \mathrm{H} + e^{-} \\
        \htp + \mathrm{H}_{2} &\longrightarrow \mathrm{H}_{3}^{+} + \mathrm{H} \\
	    \htp + e^{-} &\longrightarrow 2\mathrm{H}^{+} + 2e^{-}. \\
     \end{aligned}
\end{equation}
and the relevant cross sections are plotted in Figure \ref{cross}.

Atomic physics arguments show that the ratio of \htp to H$^{+}$ ratio is strongly affected by electron density n$_e$ and ion lifetime
in the plasma $\tau_i$ \cite{zhang:ion}. In particular the increase of
$\tau_i$ favors proton generation since it increases the probability
that \htp collide with electrons leading to their break-up 
into protons. $\tau_i$ can be modified in many ways, for example by
modifying the magnetic field or the dimension of the plasma chamber.
Since the VIS magnetic field is generated by permanent magnets
\cite{celona:trips}, modifying the magnetic field required shifting 
the magnetic assembly along the plasma chamber. These are further
investigated in Section \ref{sec:vis_experiment}.

The extraction system of the VIS is a typical accel-decel tetrode system with the following electrode voltages used during these tests:
The source (plasma aperture) at $+55$ kV to $+65$ kV, the Puller at $-2$ kV to $-4$ kV, and the remaining two electrodes grounded. The plasma 
aperture had a diameter of 8 mm, while the apertures of the remaining
electrodes were 12 mm.

\subsubsection{Low Energy Beam Transport (LEBT)}
\hspace{4ex} The LEBT (elements [h]-[z]) shown in Figure \ref{beamline_schematic}, is a 
transport system that applies dynamic modifications and diagnostic services to 
the beam. In particular, it must focus, align, and separate the ion species 
prior to injection into the spiral inflector. The LEBT was originally designed
using the ray-tracing code TRACK \cite{aseev:track}, not including space charge effects. The measured beam parameters were later compared with WARP
\cite{grote:warp1} simulations (cf. Section \ref{sec:simulations}).

Focusing of the beam is accomplished using both solenoids and a
quadrupole-doublet. 
The asymmetric particle rotation combined with the radial fringe 
fields of the two solenoids, SN1 and SN2 (element [h] and [s]), 
provide a second order focusing to the beam. The focusing length
associated with each solenoid is proportional to the mass 
squared times the velocity squared, $m^2 v^2$, and inversely proportional to the 
charge of the ion squared times the magnetic field squared, $q^2 B_{eff}^2$. The 
difference in $m/q$ for protons and \htp allows the solenoids to selectively 
focus either ion species while dispersing the other. This was the primary method of ion species separation within the beam line. The dispersed ion species 
collides with the water-cooled collimators (elements [j] and [m]), or the vacuum 
tube walls. 
The rotatable quadrupoles (element [v]), can serve to better match the
beam to the spiral inflector before entering the cyclotron (element
[w]). However, in practice they were not used as they appeared to have no noticeable effect on beam transmission.

The LEBT was aligned using a theodolite located downstream of the cyclotron, sighting with fiducials specifically made for various parts of the beam line (e.g. axial opening of the cyclotron).
Each component's axis was aligned to the axis of the VIS extraction hole to within $\pm0.25$ mm. 
Beam skew arising from the remaining misalignment was compensated by using dipole steering magnets (elements [k] and [r]).
Two gate valves (elements [l] and [z]), interlocked with the pressure
and relevant power supply sensors separate the vacuum system into three parts: the VIS, the LEBT, and the cyclotron. 
Vacuum in the upstream side of the LEBT was sustained by a 1500 l/s turbo pump (element [x]),
and in the downstream LEBT and cyclotron by cryopumps (elements [w] and [u]). 
Hot filament ion gauges were located at the ion source exit (element [g]), in the 
middle of the beam line (element [n]), and in the cyclotron (element [t]).
Experimental results pertaining to beam transport can be found in Sections
\ref{sec:transport} and \ref{sec:separation}.

\subsubsection{The spiral inflector \label{sec:spiral_inflector}}

\begin{figure}[!t]
    \centering
    \includegraphics[width=0.5\columnwidth]{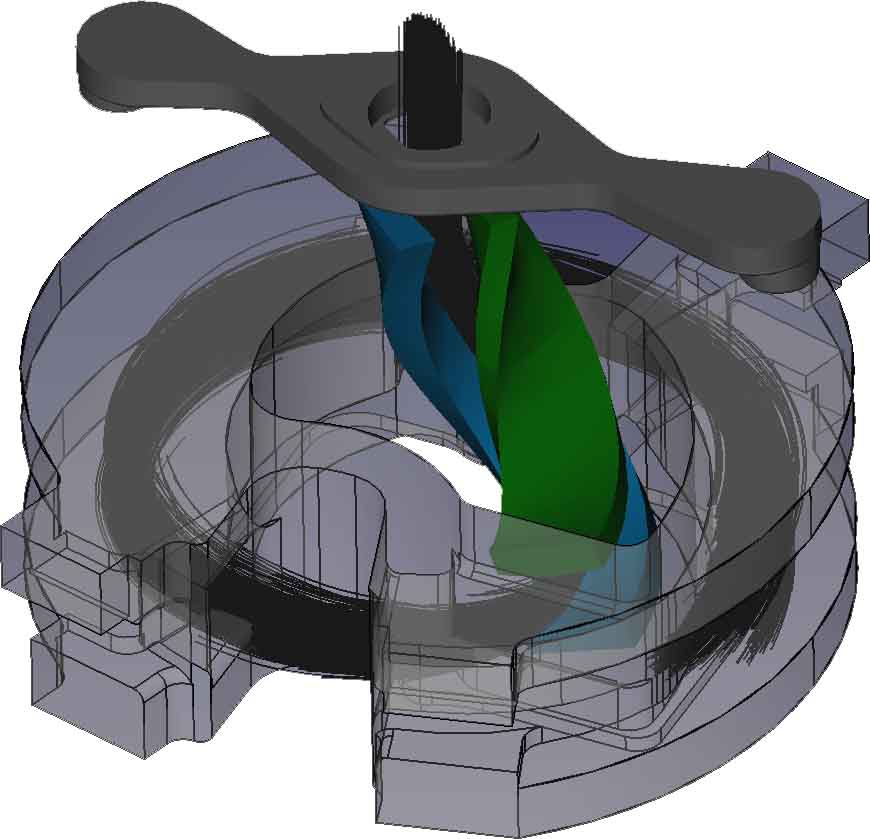}
    \caption{CAD rendering of the spiral inflector and particle trajectories. 
             The positive and negative electrostatic deflectors are shown in 
             blue and green respectively. The cyclotron magnetic field in this 
             image is directed vertically upwards. Particles enter the spiral 
             inflector via the rectangular grounded collimator (solid gray) and 
             are guided into the cyclotron mid-plane by means of the cyclotron 
             magnetic field and the electrostatic potential between the 
             electrodes. The copper housing (transparent gray) isolates the 
             spiral inflector from the RF fields driving the
             cyclotron. 
             The recesses on both sides of the housing (one of them visible on 
             the lower left) provide space for the tips of the dees.}
    \label{fig:spiral_inflector}
\end{figure} 

\hspace{4ex} The axial injection of an ionized beam into a cyclotron is usually 
realized using an electrostatic device called a spiral inflector, which consists
of two curved electrode deflectors (see Figure~\ref{fig:spiral_inflector}). The 
electrostatic potential between these electrodes is able to bend the beam 90\degree
from the axial line to the median plane of the cyclotron. The helical trajectory of the beam is determined both by the shape of the electrodes (electric field) and the
magnetic field of the cyclotron as the beam is bent into the median plane. The applied voltage depends on the velocity of the beam as well as the rigidity of the ions. 

The spiral inflector tested at BCS was designed to mimic the conditions required 
for the IsoDAR inflector, its primary defining characteristic being the 15 mm gap
between the electrodes. Compared to other similar designs \cite{jongen:cyclotron1},
this is rather large because it has to take into account the larger beam size due 
to space-charge effects. 
The spiral inflector has a copper housing that surrounds the electrodes to minimize the interaction between the electric fields generated by the dees and the
electrostatic fields between the electrodes. 
To minimize beam striking the uncooled electrodes, a water-cooled, grounded, rectangular collimator shields the spiral inflector entrance. 

The preliminary design of the spiral inflector called for a nominal voltage of $\pm11$ kV and a tilt angle of 16\degree for optimal beam injection at 60 keV. The total height of the device was approximately 80 mm.  These parameters were found by using a MATLAB code based on an analytic theory for spiral inflectors
\cite{bellomo:axial_injection, toprek:spiral_inflector}. The electrode shape was
calculated using VectorFields OPERA \cite{opera:online}.
This design was later modified to account for some effects due to fringe fields on 
the electrodes. One of these modifications was to reduce the overall length of each
electrode by 5 mm.

First turn acceleration is achieved by the dee tips extending into recesses in the
spiral inflector housing. The shapes of the dee tips and the spiral inflector 
housing have been designed to guide the particles from the spiral inflector exit 
to the acceleration region while providing the necessary energy gain and beam
focusing.

The final simulation (neglecting space-charge effects) of the spiral inflector was able to transport a 60 keV beam with a normalized emittance of 0.62 $\pi$-mm-mrad at a transmission efficiency of 100$\%$.

\subsubsection{The test cyclotron}

\begin{figure}
\begin{center}
\includegraphics[width=0.7\columnwidth]{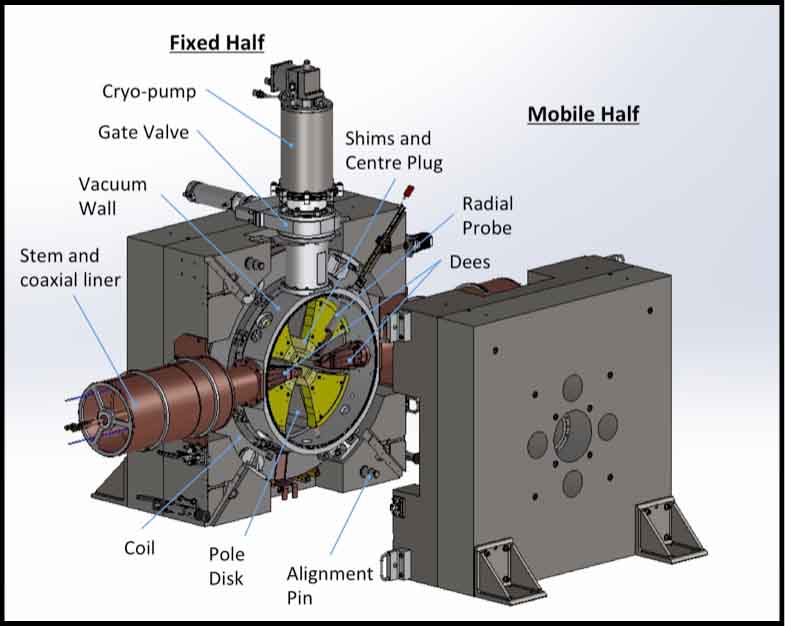}
\caption{Cyclotron component layout (see also Figure 
         \protect\ref{fig:cyclo_inside}).}
  \label{fig:cyclotron}
\end{center}
\end{figure} 

\hspace{4ex} The test cyclotron components include a magnetic structure, a vacuum system and an RF system (see Figure
\ref{fig:cyclotron}). The magnetic structure consists of an
electromagnet split in two halves, each enclosed in an identical steel
structure. The combined steel assembly has an outer dimension of 1.1 m W $\times$ 1.1 m H $\times$ 0.9 m L and opens up at the cyclotron median plane.
Both halves of the cyclotron are mounted on rails which allow for quick and easy access to the inside of the cyclotron for adjustment and changes to the inflector and central region. 
There are five ports at which radial probes can be inserted to measure beam 
currents and profiles in the cyclotron. Four of them are located in the four hill regions and one in the bottom valley, this is most easily seen in Figure \ref{fig:cyclo_inside}.
During the experiments, three radial probes were available, to be inserted at
any three of these locations (only one is shown in Figure \ref{fig:cyclotron}). 
The inner structure is composed of permanent pole disks on which rest machinable 
components (shims and center plug) designed to replicate the IsoDAR cyclotron 
magnetic field. In order to generate the average magnetic field of 1.1 Tesla, 
the two large coils are powered by two highly stable 125~A, 80~V power supplies 
connected in parallel. The energy gain by the particles is limited to 1 MeV to 
avoid material activation.

The main tank is brought down to an operating pressure of 
$1.33\cdot10^{-6}$
mbar by a combination of a 605~l/min rotary vane pump and a 1500~l/s cryopump. 
The pressure is measured by both a thermocouple and a hot filament ion gauges at 
their respective operational range.

Like the magnetic structure, the RF system is composed of permanent and 
interchangeable parts to achieve different cyclotron configurations. The 
accelerating gaps and resonator geometry are determined by a set of dees, 
center post, and center liners that are specific for each configuration. Coarse 
tuning is done manually by adjusting shorting plates between the stems and the 
coaxial liners and fine tuning is carried out by motor-driven tuners and couplers.
The front-end stage of the 20~kW RF amplifier is a wideband solid-state driver 
directly connected to the final stage tuned for the operational frequency of 
49.2~MHz. The amplifier has been tuned to a maximum CW output power of 13~kW 
(17~kW equivalent in pulsed mode). The resonators have been designed to reach up 
to 70~kV of accelerating dee voltage. Despite our best effort, arcing and cavity 
losses only allowed a dee voltage of $<63$ kV at the peak power of 17~kW.
The accelerating voltage was estimated using a capacitive pick-up probe.
The probe was calibrated based on measurements of x-rays
(using the endpoints of the bremsstrahlungs spectra) with
a detector interfaced with a vacuum aperture looking into the cyclotron chamber.
Large fluctuations limited the value of these pick-up probe measurements and
it appears that the actual voltage in the central-region accelerating
gaps was even lower than the estimates from the x-ray measurements. Results of
the injection and acceleration studies will be reported in Section 
\ref{sec:cyclotron_injection}.

\subsection{Beam line diagnostics}
\hspace{4ex} This section describes the diagnostic instrumentation used to make measurements regarding beam emittances and beam currents. The data gathered using a retarding field analyzer, which pertains to the space-charge compensation of the beam line, will be discussed in subsequent papers, so is not included in this technical report. A description of the Allison-type emittance scanner, Faraday cup, and beamstop follows. 

\subsubsection{Allison-type emittance scanner}
\begin{figure}[!t]
    \centering
    \begin{minipage}{.38\linewidth}
        \includegraphics[width=\linewidth]{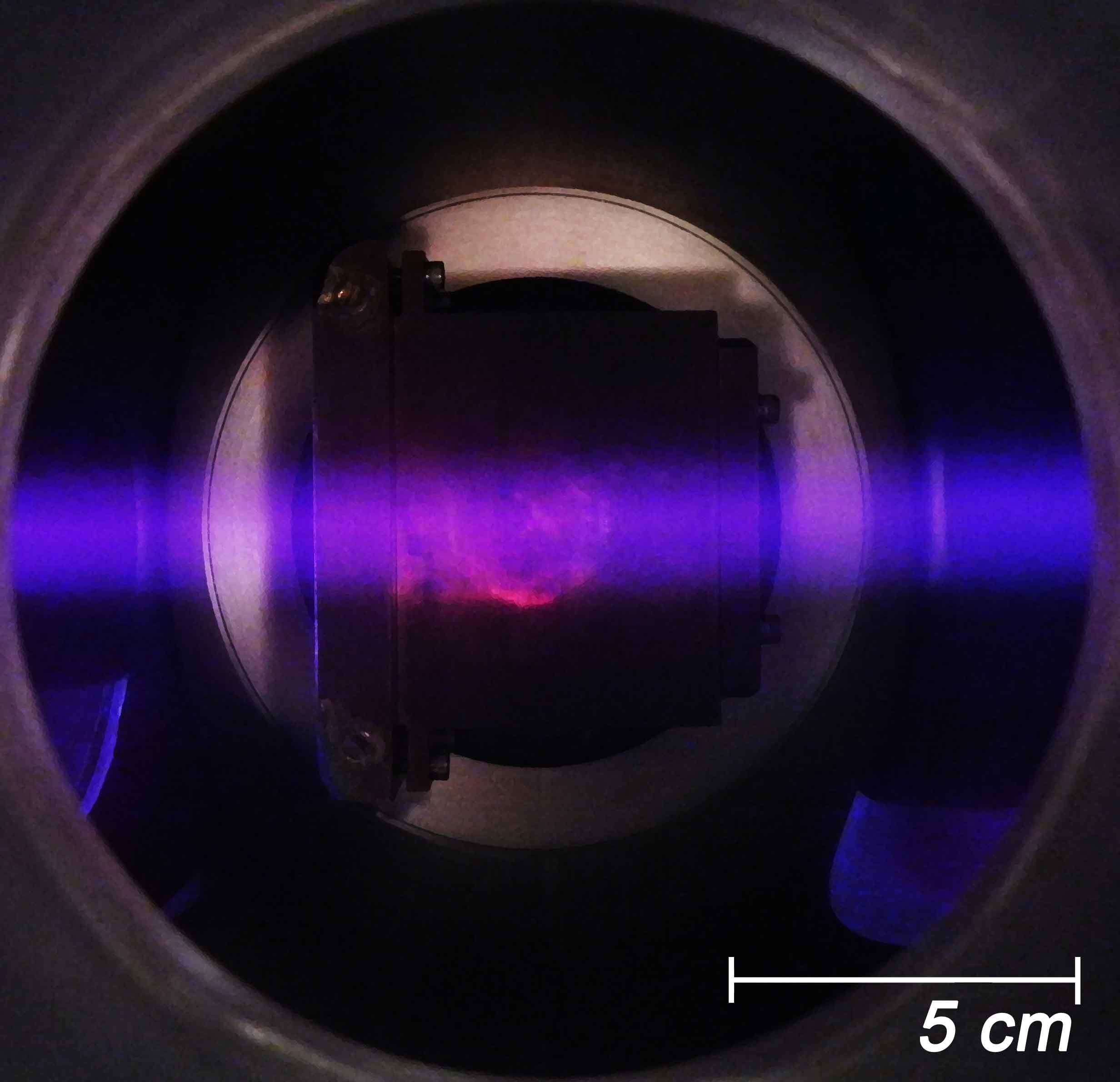}
    \end{minipage}
    \hspace{.05\linewidth}
    \begin{minipage}{.45\linewidth}
        \includegraphics[width=\linewidth]{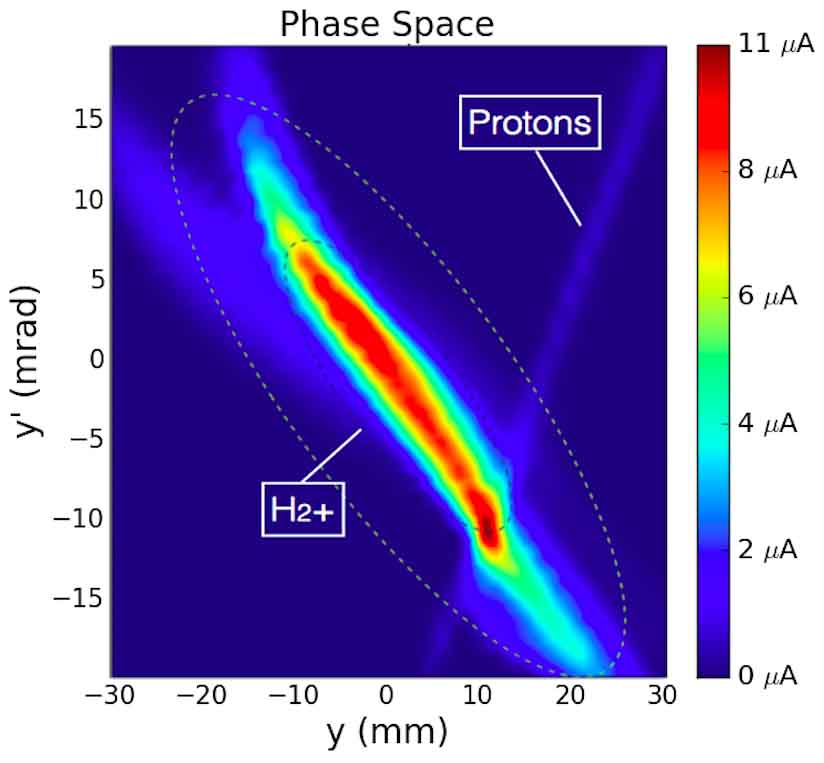}
    \end{minipage}
    \caption{Left: An increased beam line pressure allowed for the beam to be 
             easily seen. A photograph of the beam was taken through a quartz 
             window situated directly below the emittance scanner at Configuration I
             (indicated in Figure 2) with 340 A supplied to SN1. The base of the
             emittance scanner can be seen above the beam. 
             Right: An example normalized vertical emittance plot of the beam is 
             shown. The blue dotted ellipse represents the 1-rms emittance and 
             the green dotted ellipse represents the 4-rms emittance. The color bar
             represents the relative beam current and should not be interpreted 
             as a direct measurement of total current. The protons and \htp 
             are indicated. Here, the \htp is coming into focus while the protons 
             are highly divergent. 
             What appears to be a distortion in the \htp beam in the lower right is
             caused by the superposition of \htp and protons.}
    \label{sample}
\end{figure}
\hspace{4ex} The kinematics of the beam are described by the particle ensemble 
representation in 6-dimensional phase space: 3-dimensions for its spatial 
components x, y, and z, and 3-dimensions for its momentum components x', y', and 
z'. The Allison-type emittance scanner (element [q]) of Figure
\ref{beamline_schematic}, is a device to measure the emittance of the beam.
If $\langle x^2 \rangle$ is the second moment of the beam's horizontal phase space
distribution $f(x, x')$; defined as:
\begin{equation}
\langle x^2 \rangle = \frac{\iint x^2 f(x, x') dxdx'}{\iint f(x, x') dxdx'}
\end{equation}
and analogous for $\langle x'^2 \rangle$ and $\langle xx' \rangle$, then the 
horizontal rms emittance can be defined as:
\begin{equation}
\epsilon_{rms} = \sqrt{\langle x^2 \rangle\langle x'^2 \rangle - 
\langle xx' \rangle^2}
\hspace{15pt} \left[mm\mhyphen mrad\right]
\end{equation}
Similarly the vertical emittance can be calculated by replacing x with y and x' with y'.
The emittance scanner samples either the x-x' or the y-y' phase-space by means
of two slits, electrostatic deflection plates, and a small Faraday cup. 
The spatial information comes from the position of the emittance scanner,
actuated by a stepper motor, and the angle information from the voltage applied
to the deflection plates located between the two slits. The current recorded
in the Faraday cup for each pair of x and x' gives the distribution $f(x, x')$.
A sample emittance plot is shown in Figure~\ref{sample} (right).

\subsubsection{Faraday cup and beam stop} 
\hspace{4ex} A Faraday cup is a device typically made out of copper that measures the net current incident on the interior face of a cup-shaped (sometimes a cone or a semi-sphere) conductor. This cup is connected to an ammeter or current amplifier, while an electron suppression electrode is mounted upstream to provide an negative electrostatic
potential to turn back electrons emitted from the cup interior surface upon the impact of beam ions. 
Both cup and electron suppressor are shielded by a grounded aperture or housing. 
The tests at BCS employed a Faraday cup to accurately measure the total transported beam current at the end points of Configuration I and II. When the cyclotron was connected to the beam line, the Faraday cup was replaced by a beam stop ([o] in Figure \ref{beamline_schematic}), that could be pneumatically moved out of the way to not obstruct the beam. This beam stop (and its upstream collimator) were insulated, so could also read current. Electron suppression of the beam stop was 
achieved by placing a strong permanent magnet close-by. This did not lead to full suppression and readings from the beam stop have a higher uncertainty (cf. Section
\ref{sec:errors}).

\paragraph{Electron Suppression} \label{electron_sup} \mbox{}\\
\begin{figure}
    \centering
    \includegraphics[width=0.55\columnwidth]{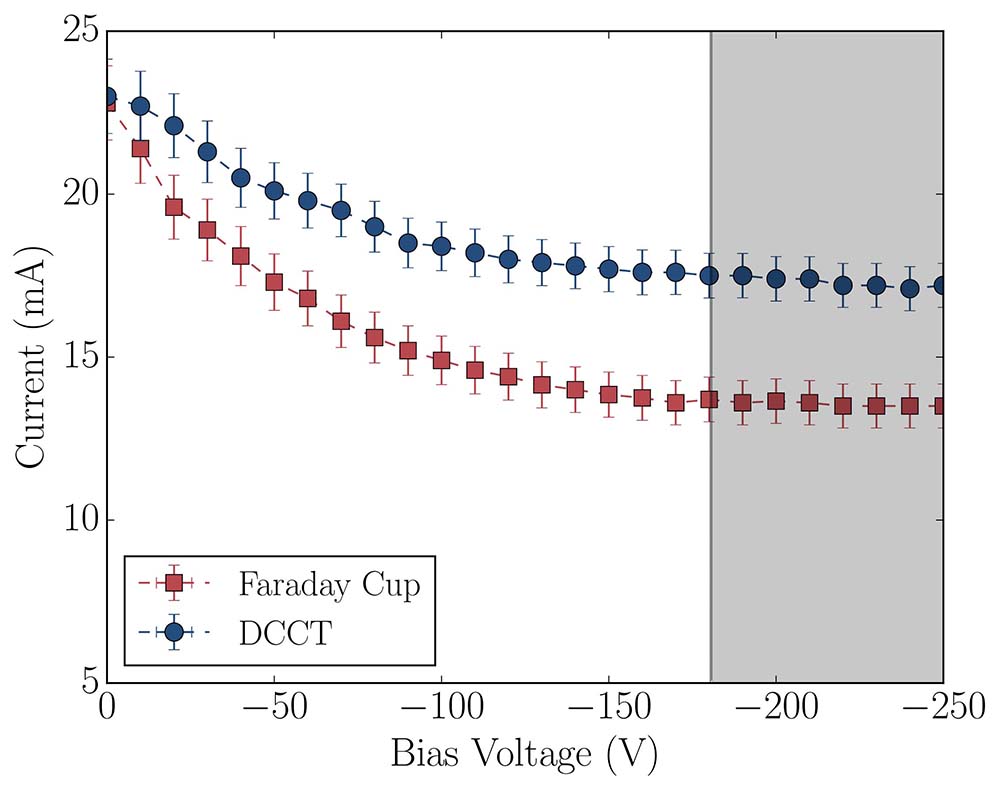}
    \caption{The effect of measured beam current as a function of electron 
             suppression voltage on the Faraday cup and DCCT. The current 
             stabilizes when a negative suppression voltage of -180 V is 
             applied.}
    \label{electronsup}
\end{figure} 
\hspace{4ex} When a beam strikes the surface of a metal, electrons can be ejected from the surface. If the electrons are not suppressed (returned to the surface of the metal), this appears as a positive current. Electron suppression of the Faraday cup was accomplished by application of a negative bias to a dedicated electrode between the cup and the grounded entrance aperture. In order to determine the effect of electron suppression, a set of measurements was taken with the Faraday cup located at the endpoint of Configuration I and a Bergoz DC Current Transformer (DCCT) located one position further upstream. As shown in Figure \ref{electronsup}, an electron suppression voltage of approximately -180~V was sufficient to prevent electrons from escaping the Faraday cup. Whenever we refer to a Faraday cup measurement, adequate suppression can
be assumed.
It should be noted that an electrostatic electron suppression voltage was not needed 
for measurements on the radial probes within the cyclotron because the parallel magnetic
field effectively suppressed the escaping electrons.
Due to the inability to suppress all the back-streaming electrons artificially increasing the measured current in the DCCT, 
we removed the DCCT and unless otherwise noted, all subsequent current measurements were done using only the electron suppressed Faraday cup.

\subsection{Data acquisition}\label{sec:daq}
\hspace{4ex} Beam currents were measured using custom-built bipolar current-sampling 
data-acquisition modules, connected through shielded 50 Ohm cables.
The sampling rate of the input stage was 5.5 kHz, with data averaged and stored in a 
buffer at a 66 Hz rate. Readout from the PLC buffer to the personal computer 
was done through Ethernet at a 5 Hz polling rate. 

\begin{figure}[!t]
    \centering
    \includegraphics[width=0.9\columnwidth]{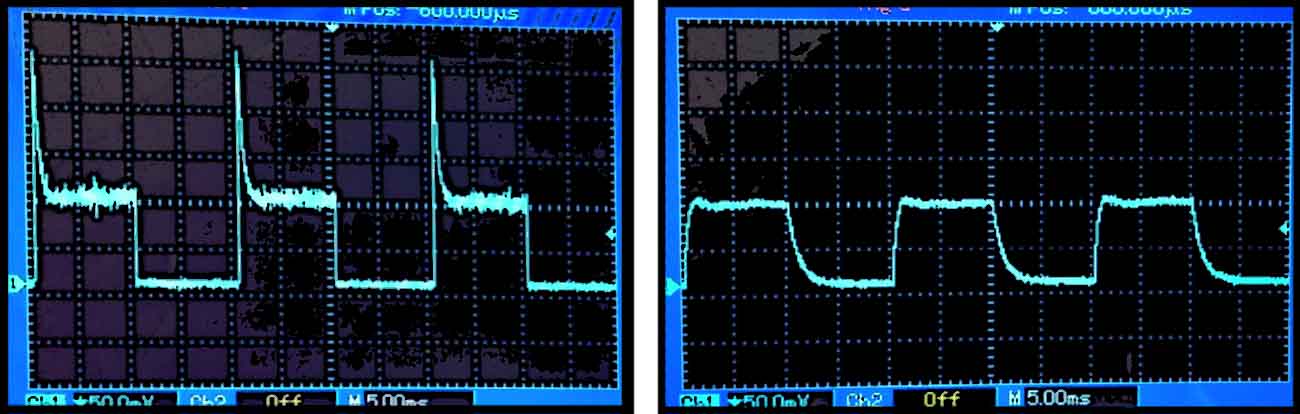}
    \caption{Two oscilloscope traces of the beam pulse. Left: Unmodified pulse 
             shape input into DAQ. Right: Capacitor inserted into DAQ board 
             acts to average the pulse and limits the preglow pulse height. 
             Giving a more accurate current measurement.}
    \label{cap}
\end{figure}

Initially, upon switching to pulse-mode (required because of power limitations 
in the cyclotron RF system), it was noticed that the current 
amplifiers were overestimating the current in each pulse. This was found to be 
due to the pulse detection algorithm in the current amplifiers and the pulse 
shape. An oscilloscope trace of several \htp pulses is shown in Figure~\ref{cap} 
(left). The quick rise in the pulse reflects the high initial rate of H$_2^+$
production and may be referred to as "pre-glow" \cite{xu:current}.
This can be explained by the 
production processes shown in \ref{production}. The production of protons is a 
two-step process involving energetic collisions with free electrons in the 
plasma chamber, whereas the \htp is readily produced after a single process 
involving the H$_2$ gas. Since the protons production will therefore have a longer 
rise time, an initially high H$_2^+$ pulse followed by an exponential decay to 
the equilibrium state is expected. 
(The trace in Figure \ref{cap} represents the Faraday cup signal when beam focusing is set to optimize for \htp transmission to the cup.  When the beam line is set to focus protons into the cup, the pre-glow peak is missing, replaced by a slow rise-time pulse with a time constant of about 1 ms.)
The pulse detection algorithm in the current 
amplifiers would average over the initial pre-glow and plateau currents, and 
overestimate the actual amount of current in each pulse. A low-pass filter
was inserted between the BNC input and data acquisition module to provide a more stable and even input signal (see Figure~\ref{cap}, right).

\subsection{Uncertainty estimates \label{sec:errors}}
\hspace{4ex} In keeping with the goal of this paper as an examination of the concept rather than a demonstration, we have conservatively estimated the uncertainties for the different types of measurements and only provide upper limits for the maximum errors. The following subsections describe these estimates, both statistical as well as systematic in nature, for the various types of measurements.

\subsubsection{CW beam measurement uncertainties}
The majority of the tests performed before injecting beam through the spiral inflector were conducted with a continuous (DC) beam. Associated with each measurement we identify an uncertainty associated with the measuring device as large as the least significant digit of measurement, as well as include a noise estimate if there was variation associated with the measurement. 

\paragraph{Faraday Cup}\mbox{}\\
\hspace{4ex} We assume no systematic errors on the fully electron suppressed Faraday cup
as described in Section \ref{electron_sup}. The statistical variations were
typically on the order of $\pm 2\%$. We conservatively report all beam current 
measurements in the Faraday cup with an uncertainty of $\pm 5\%$.

\paragraph{Beam stop}\mbox{}\\
\hspace{4ex} The electron suppression on the beam stop was achieved by placing permanent
magnets on the front and back of the beam stop plate, thereby creating a 
more or less parallel magnetic field that guides secondary electrons back 
to the plate. This cannot be considered full electron suppression and thus
a systematic error of -10\% is assumed in addition to the statistical error
of $\pm 5\%$.

\paragraph{Beam Stop Front Plate and Spiral Inflector Aperture}\mbox{}\\
\hspace{4ex} Both of these plates did not have any significant electron suppression through
either a negative electrode or a magnetic field. They are both considered 
completely unsuppressed. Taking the curve in Figure \ref{electronsup} as a 
basis, we see that the suppressed current is $\sim 60\%$ of the 
unsuppressed one. This was measured for a copper cup with impinging \htp
ions which is similar to the other unsuppressed surfaces. However, the Faraday
cup also has a certain depth which may recover some of the escaping electrons. Taking this into account, we conservatively estimate the systematic error as -50\% with an additional uncertainty of $\pm 20\%$ and thus report the 
current measured on an unsuppressed copper surface as 
$\mathrm{I}_\mathrm{actual} = 0.5 \cdot \mathrm{I}_\mathrm{meas.} 
\pm 0.2 \cdot \mathrm{I}_\mathrm{meas.}$.

\subsubsection{Pulsed beam measurement uncertainties}
\hspace{4ex} As discussed in Section \ref{sec:daq}, the pulsed beam exhibited a 
pre-glow effect. A capacitor was added, effectively working as a low-pass
filter mostly evening out the peak at the beginning of the pulse.
Comparison of measured values with oscilloscope traces lead to estimating the
systematic error for any pulsed beam as $\pm5\%$ from this effect, in addition
to the previously discussed uncertainties for the various devices.

\paragraph{Paddle and Radial Probes}\mbox{}\\
\hspace{4ex} Inside the cyclotron a strong magnetic field on the order of 1 T was present
which leads to full electron suppression during the current measurements.
We thus report the measured currents with a base statistical error of 
$\pm 5\%$.
Additional systematic errors from probes picking up RF signals
from the dees during acceleration tests are taken into account by 
subtracting the background obtained by inserting
the beam-stop and measuring the signal without ion beam in the cyclotron.

\subsubsection{Emittance measurement uncertainties}
\hspace{4ex} The uncertainties in the reported measurements come from the quadrature combination of uncertainties in the measuring devices and observed electronic noise. Further, we add an uncertainty of $\pm5\%$ due to the electron suppression described in Section \ref{electron_sup}. 
There is an additional large systematic error from emittance scans not 
covering the full phase space area of the beam. This systematic error can be determined by simulations and is included in the values reported in the text. However, the graphs 
in Figures \ref{fig:phase_spaces_con_I} and  \ref{fig:emittance_diameter} do not 
include the systematic modification. Instead, the simulations were restricted to the same limits, showing good agreement.

\subsubsection{Dee voltage uncertainties}
\hspace{4ex} Due to the highly unreliable pick-up probe for measurements of the dee voltage
and large fluctuations in the dee voltage itself, we do not estimate an
uncertainty for the dee voltage. Instead we report the values as they were read, 
but suggest that the actual values are more closely related to the best fit obtained from the OPERA comparisons to measured transmission efficiencies (cf. Section 
\ref{sec:opera}).

\section{Experimental measurements}

\subsection{Overview}\label{sec:overview}

\hspace{4ex} In March of 2013 the VIS source, high-voltage platform and supply, 
microwave generator, a large-bore solenoid plus controls, power supplies and 
miscellaneous supporting equipment were shipped from Catania, Italy to the BCS 
facility. The shipment was met by a team from the INFN-LNS, who unpacked the 
equipment and interfaced it with the heavy-duty rails supporting the beam line 
equipment and the cyclotron. 
During the summer of 2013 initial tests were performed by our MIT group, staff of BCS and the Catania group. The first results were reported in \cite{alonso:vis1}.
Based on these results, the months between the summers of 2013 and 2014 were used to perform improvements to the data-acquisition system, beam diagnostic devices, cyclotron RF system, and (as described below) to the VIS.
The results summarized in this Section were mainly obtained in the 2014 run.\\

To begin addressing the IsoDAR injector requirements enumerated in Section 
\ref{marker1}, the tests primarily sought to:

\begin{enumerate}
  \item characterize and optimize the VIS for \htp operation (Section
        \ref{sec:vis_experiment}),
  \item separate the ion species in the transport line (Section \ref{sec:separation}),
  \item determine transport efficiencies of the beam line and the spiral inflector 
       (Section \ref{sec:transport}),
  \item capture beam into the cyclotron (Section \ref{sec:cyclotron_injection}),
  \item verify accuracy of beam line simulations (Section \ref{marker2}).
\end{enumerate}

\subsection{Characterizing the VIS \label{sec:vis_experiment}}
\begin{figure}[!t]
    \centering
    \begin{minipage}{.38\linewidth}
        \includegraphics[width=\linewidth]{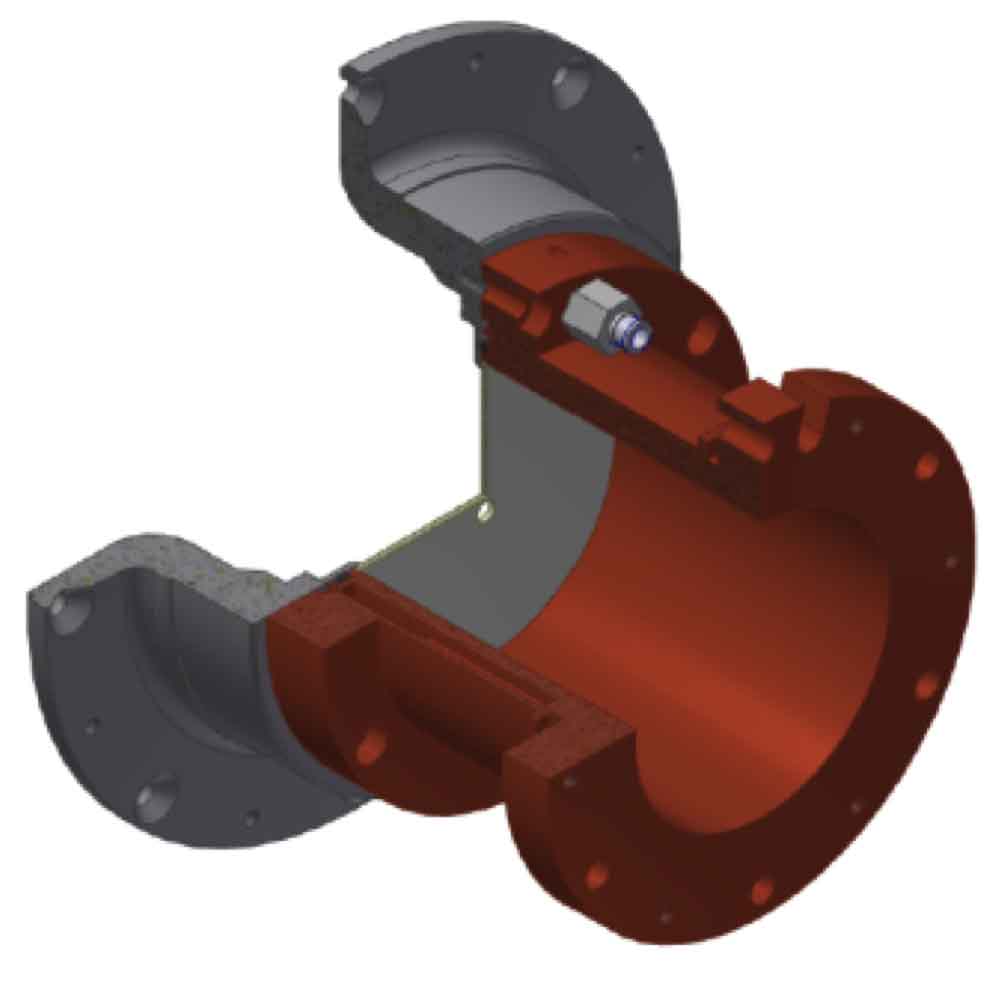}
    \end{minipage}
    \hspace{.05\linewidth}
    \begin{minipage}{.35\linewidth}
        \includegraphics[width=\linewidth]{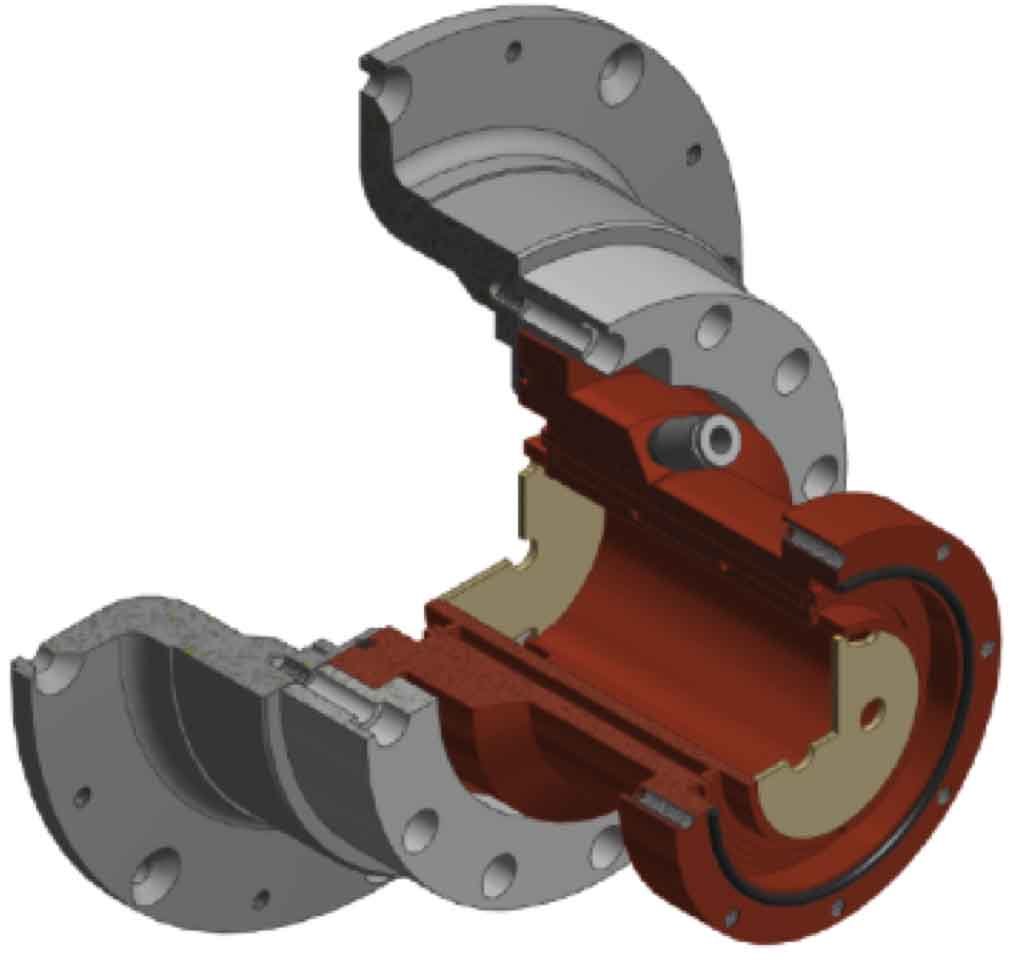}
    \end{minipage}
    \caption{The plasma chamber designs. Left: The standard chamber: 10 x 5 cm 
             (length x radius); Right: The small chamber1: 10 x 2.5 cm.}
    \label{fig:new_chamber}
\end{figure}
\hspace{4ex} The results obtained in 2013 showed that the VIS
could provide good \htp currents. However, as the source design
was optimized for proton currents, as expected 
\htp suffered in comparison to proton currents. As one raised the microwave power, a stable discharge first occurred at about 400 watts. At this point the total extracted beam current was about 30 mA, of which 8 to 10 mA was \htp, the remainder being protons.  As the microwave power was raised, the proton current increased, up to a maximum of about 40 mA at 1400 W, however the \htp current remained constant at $8-9$ mA.  The higher power increased the ion production, but also increased the breakup of \htp into protons.  

Due to the fact that the VIS is using a permanent magnet structure which was optimized for proton production, the increase of 
\htp fraction was achieved solely by reducing the ion confinement time through reduction of the plasma chamber diameter. While the original cylindrical plasma chamber had a diameter of 10 cm and a 
length of 10 cm, the new one has a diameter of 5 cm still with a
length of 10 cm. This new chamber, and a revised microwave 
coupling section, were brought to Vancouver midway through the
summer of 2014. In the companion paper \cite{castro:vis2} 
these modifications are described, and results obtained on the BCS test stand are reported. To summarize: As expected, the \htp fraction increased to almost 50\% at low microwave powers (about 
150 watts now), and the best \htp current achieved now improved to
about 12.2 mA. Small adjustments in the location of the permanent
magnet ring were performed to optimize the \htp current value. Additional adjustments to the plasma chamber shape might improve
\htp currents further.

All beam line measurements reported in this paper, as well as all the cyclotron injection studies were performed with the original, standard plasma chamber.

\subsection{Ion species separation}\label{sec:separation}
\begin{figure}[!t]
    \centering
    \includegraphics[width=0.8\columnwidth]{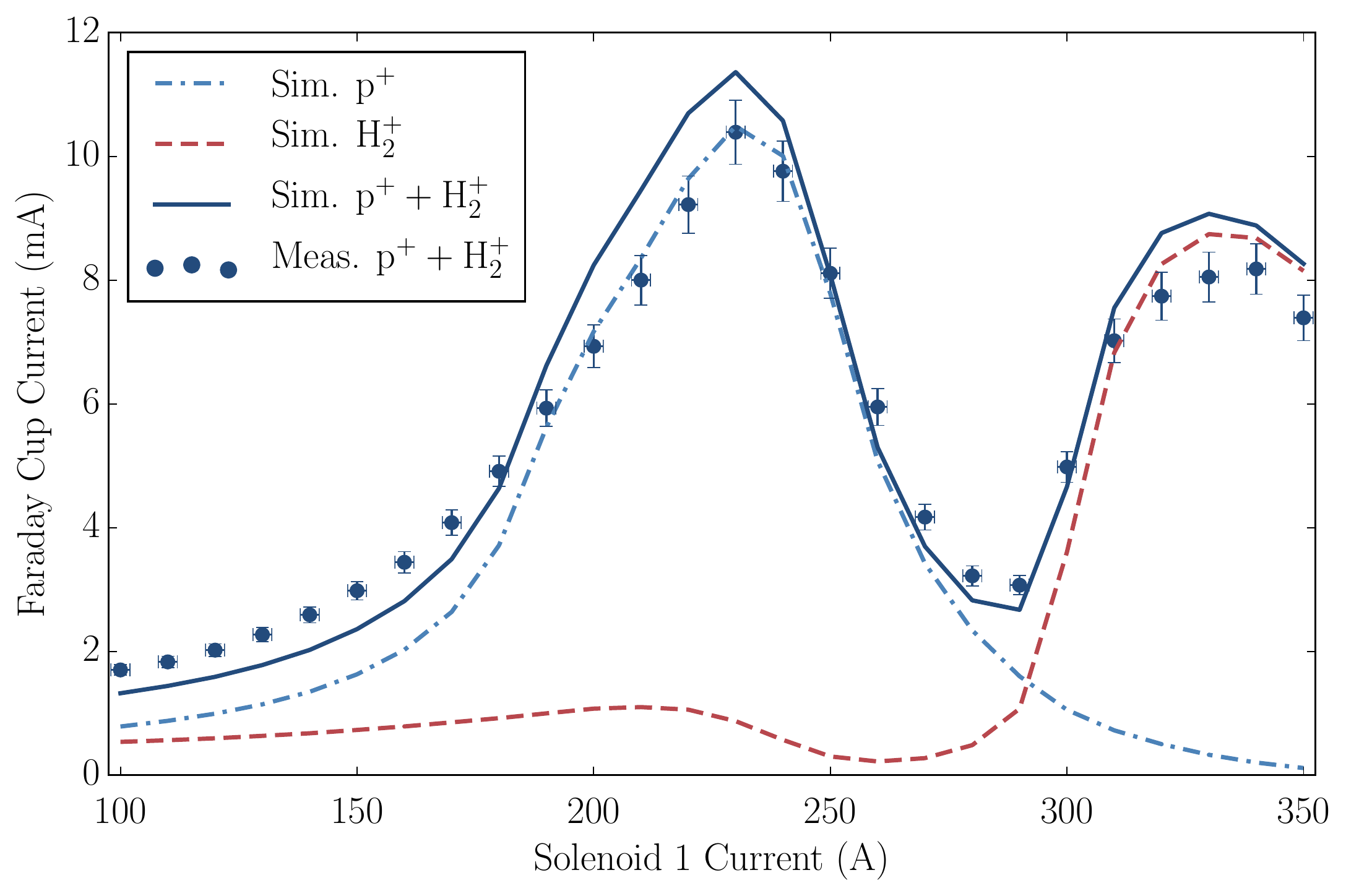}
    \caption{Configuration I, measured and simulated beam currents as a function of SN1
    			current. The beam current is measured in an 
             electron suppressed Faraday cup located behind a 56 mm aperture 
             at the end point of Configuration I. Extraction voltage 
             was maintained at 60/-3.2 kV, RF power at 400~W, and H$_2$ gas mass 
             flow controller at 46$\%$.}
    \label{fig:fc0_vs_sol1}
\end{figure}

As the BCS test stand was initially designed to transport H$^-$ ions (for the development of 
a LEBT lines for isotope-producing cyclotrons), there were no provisions for 
magnetically analyzing the beam produced by the VIS. In particular, to measure the current of an ion species, a Faraday cup and emittance scanner were used. Since the VIS was mounted 
straight along the axis of the rails, the protons, H$_2^+$, and perhaps a 
small fraction of H$_3^+$ all propagate along the same axis from the source 
towards the cyclotron. Typically, in order to separate the ion species, a dipole magnet would be used but due to space constraints an alternative method was required. Simulations indicated that a solenoid with a longitudinal magnetic field strength of roughly 2.7 kG, placed 50~cm downstream from the source extraction aperture, could effectively be used to separate the various ions. As seen in Figure \ref{fig:fc0_vs_sol1}, the measured current on the Faraday cup at the Configuration I end point agreed well with the beam line simulation; the relative current at both the proton focus (SN1: 240 A) and at the H$_2^+$ focus (SN1: 340 A), was in good agreement. 

\subsubsection{Beam shape studies, formation of a hollow beam}\label{sec:hollow}
\begin{figure}[!t]
    \centering
    \includegraphics[width=0.63\columnwidth]{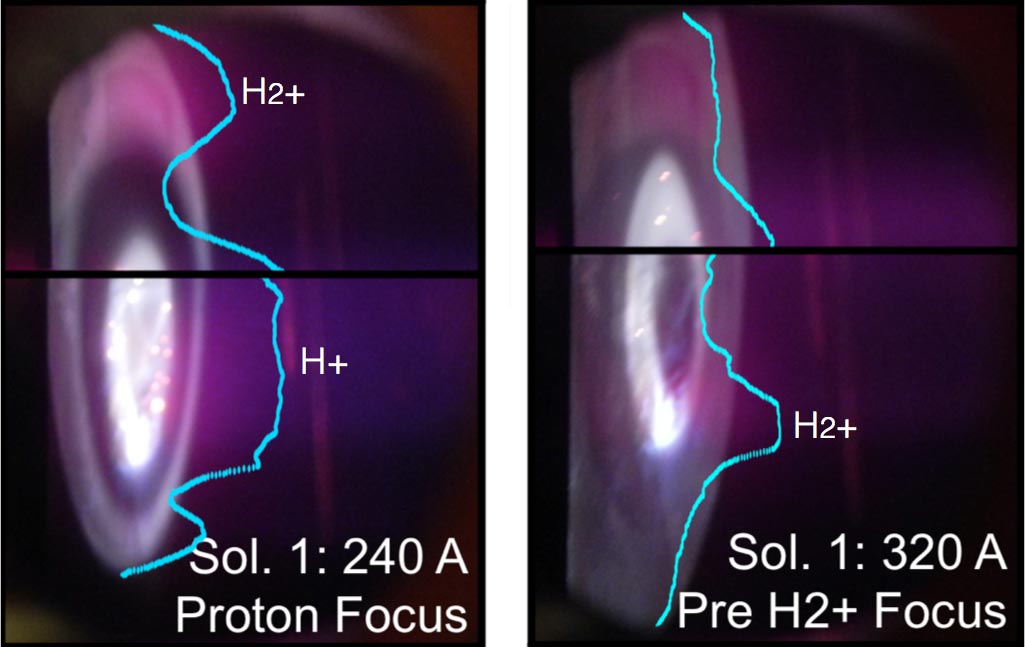}
    \caption{Transverse beam profiles observed on the 4-jaw slits (4Jaw). The 
             beam enters the image from the right and strikes the face of two 
             horizontally separated water-cooled copper plates. Left: The 
             focused proton beam is seen as the central illuminated circle while 
             the H$_2^+$ ions form a concentric outer ring. Right: The bright 
             inner ring is the H$_2^+$ being brought into focus. By focusing the 
             H$_2^+$, the protons are over focused and blast out a hole in the 
             center of the beam, seen as the central intensity deficit. Note the 
             horizontal splice in the image corresponds to the small separation 
             between the water-cooled plates.}
    \label{fig:hollow}
\end{figure} 
\hspace{4ex} Figure~\ref{fig:hollow} shows the beam striking the vertically separated 
water-cooled 4-Jaw collimator (4Jaw) (element [n] of Figure
\ref{beamline_schematic}). An intensity profile through the diameter of the beam 
is superimposed onto the image. The left plot shows the transverse beam profile 
with 240~A supplied to SN1. At this setting, the protons are 
directly focused onto the 4Jaw, 2.30 m away from extraction hole of 
the VIS (see Figure \ref{fig:SN1_230A_Env_horz} for the simulated beam envelop near this setting). At the center of the concentric rings is the focused proton beam. The 
high charge density of the protons repel the H$_2^+$, creating a circular halo 
of H$_2^+$. The right plot shows the scenario for which SN1 is 
supplied with 320 A and the H$_2^+$ is near its focus (see simulation in Figure \ref{fig:SN1_320A_Env_horz}). The strong magnetic field 
on SN1 causes the protons to be over-focused. This creates a region upstream of the 4Jaw with an extremely  high positive charge density which repels the H$_2^+$. An \htp depletion zone is then created at the center of the \htp beam. The formation of this ``hollow beam" was also observed in the beam line simulations and can be compared directly with the results described in Section \ref{marker2}.

\subsection{Beam transport \label{sec:transport}}
\subsubsection{Intensity fluctuations} 
\begin{figure}[!t]
    \centering
    \includegraphics[width=0.8\columnwidth]{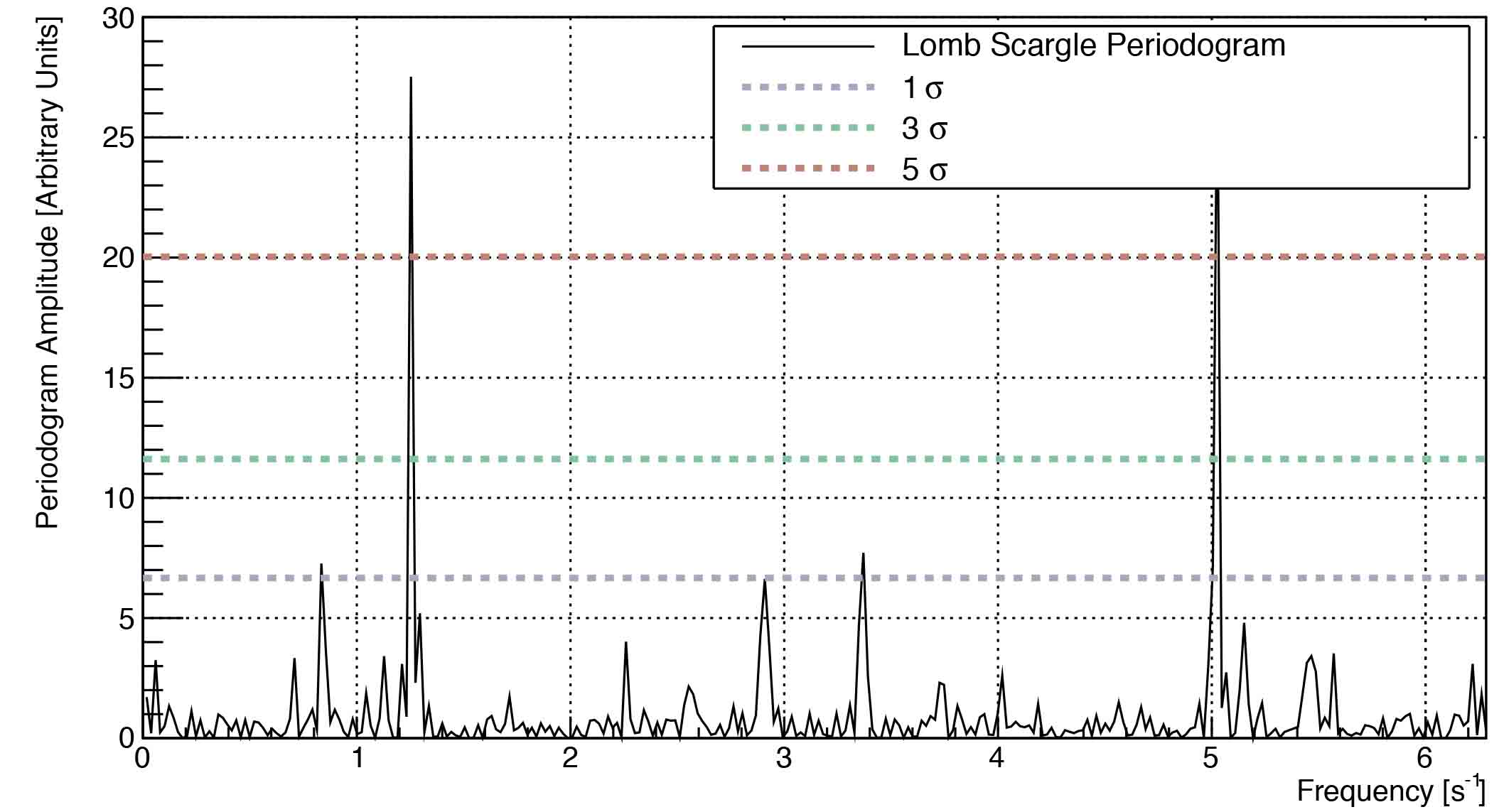}
    \caption{Lomb-Scargle periodogram of the beam line pressure. Two significant 
             periodicities in the data are seen at 1.25 Hz and 5.1 Hz. The 1.25 
             Hz periodicity was identified as the compression cycle of the 
             cryopump. The 5.1 Hz periodicity was not identified, but was 
             assumed to also be attributed to the cryopump.}
    \label{lomb}
\end{figure}

\hspace{4ex} Beam line pressure fluctuations were found to affect the beam current 
measurements. A video of the beam, captured through the quartz observation 
window, indicated that the intensity of the beam fluctuated in sync with the 
1.25 Hz compression cycle of the helium cryopump. A Lomb-Scargle periodogram 
analysis~\cite{scargle:statistics} of beam line pressure data over a period of 5 
minutes indicated that there were two significant oscillation frequencies in 
pressure: one at 1.25 Hz and another at 5.05 Hz, see Figure~\ref{lomb}. The 1.25 
Hz frequency was found to corresponded to the compression cycle mentioned above, 
indicating that the cryopump sorbent was saturated by the pumped out hydrogen gas, 
while the origin of the 5.05 Hz frequency was undetermined. To avoid downtime 
caused by the need to regenerate the cryopump, the beam line cryopump was 
replaced with a turbo pump capable of achieving the same beam line 
pressure. This modification lead to more stable beam transport across the LEBT.

\subsubsection{Emittance measurements}\label{sec:emmitance}
\hspace{4ex} Two sets of emittance measurements are presented. The first will describe the vertical emittance measurements made at the end of Configuration I for a transported \htp beam. The second set include both horizontal and vertical measurements taken at the end of Configuration II, and are of particular interest because of their significant aberrations.

For all the subsequent emittance measurements, the proton contamination has been subtracted via software in order to obtain a more accurate representation of the \htp emittance.

\begin{figure}[!hp]
\centering
\begin{minipage}{.45\linewidth}
  \includegraphics[width=\linewidth]{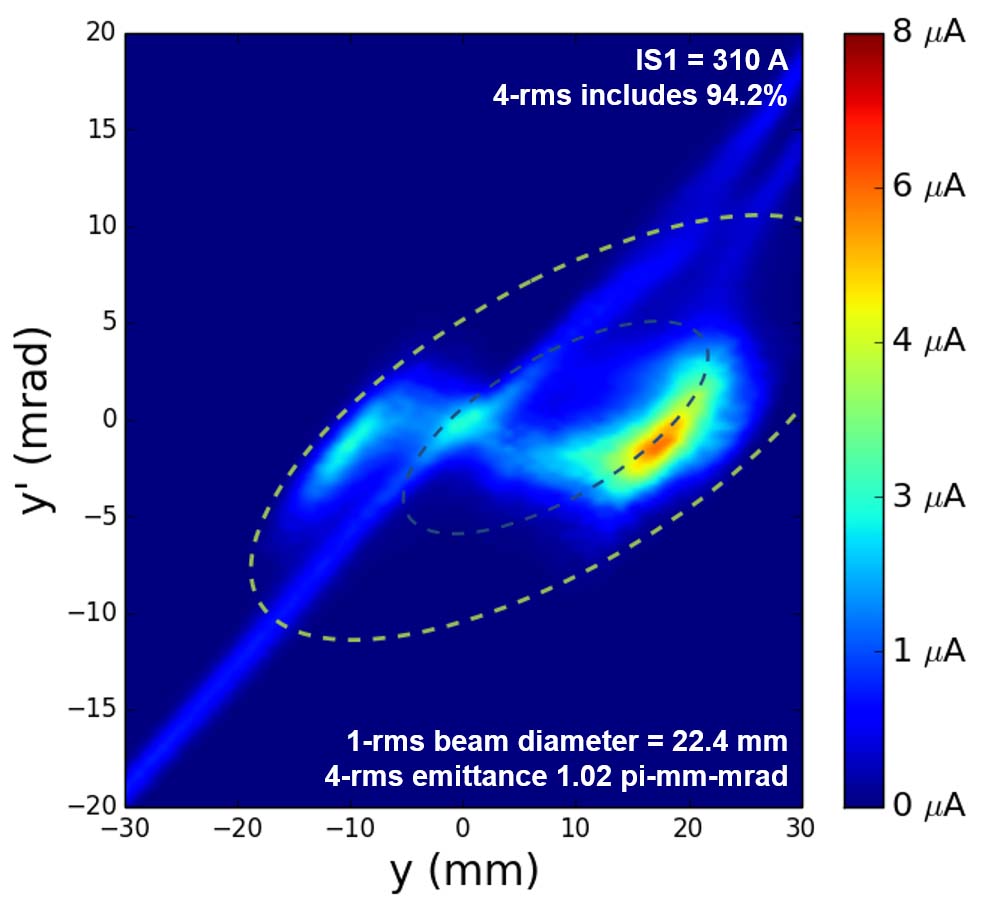}
\end{minipage}
\hspace{.05\linewidth}
\begin{minipage}{.45\linewidth}
  \includegraphics[width=\linewidth]{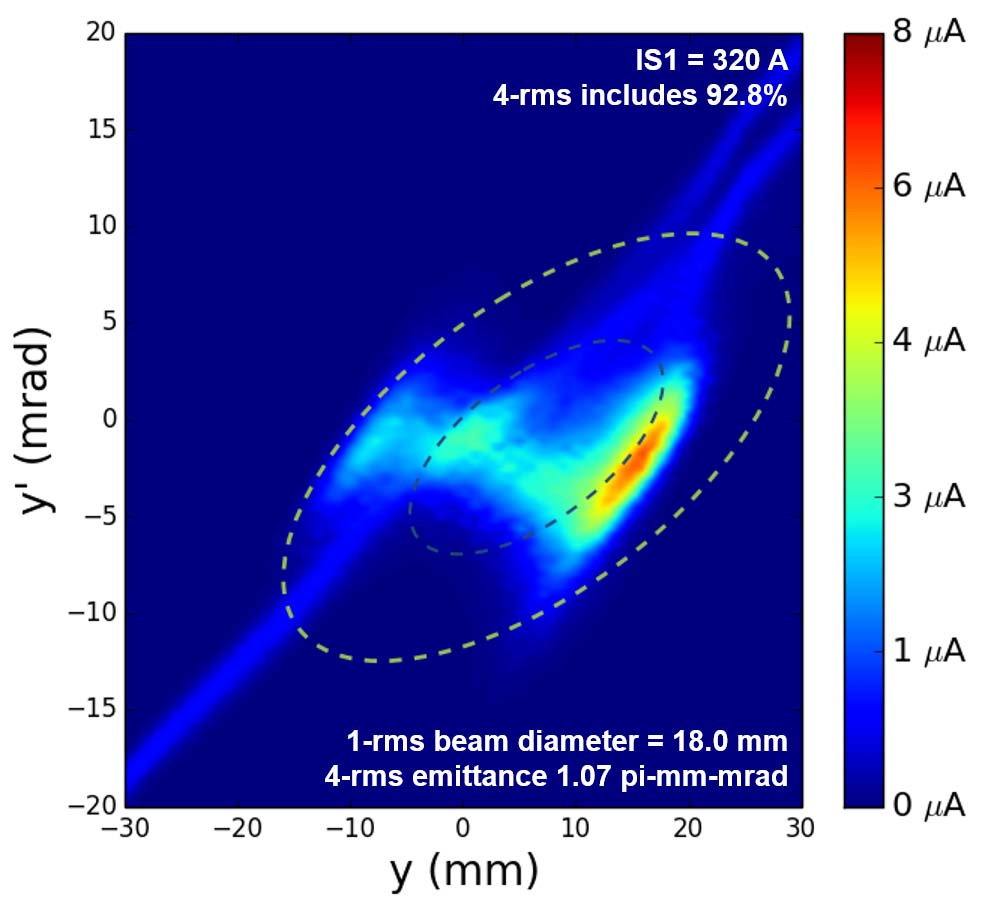}
\end{minipage}
\begin{minipage}{.45\linewidth}
  \includegraphics[width=\linewidth]{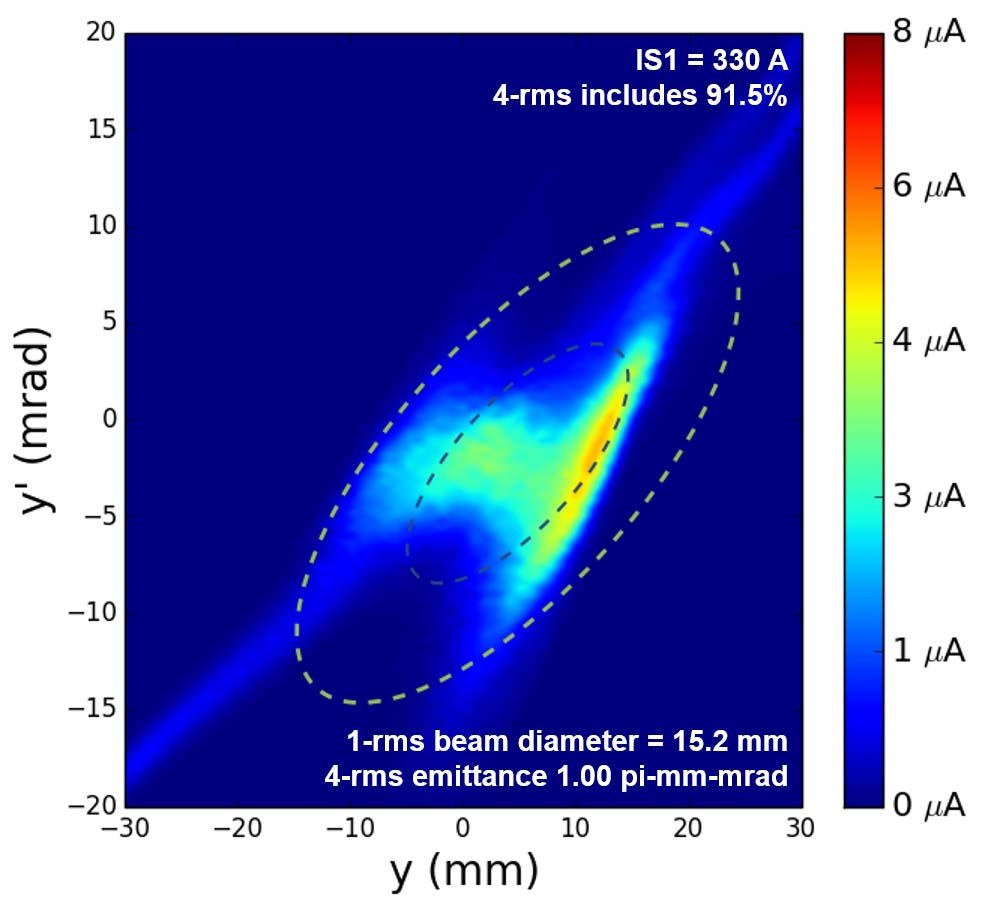}
\end{minipage}
\hspace{.05\linewidth}
\begin{minipage}{.45\linewidth}
  \includegraphics[width=\linewidth]{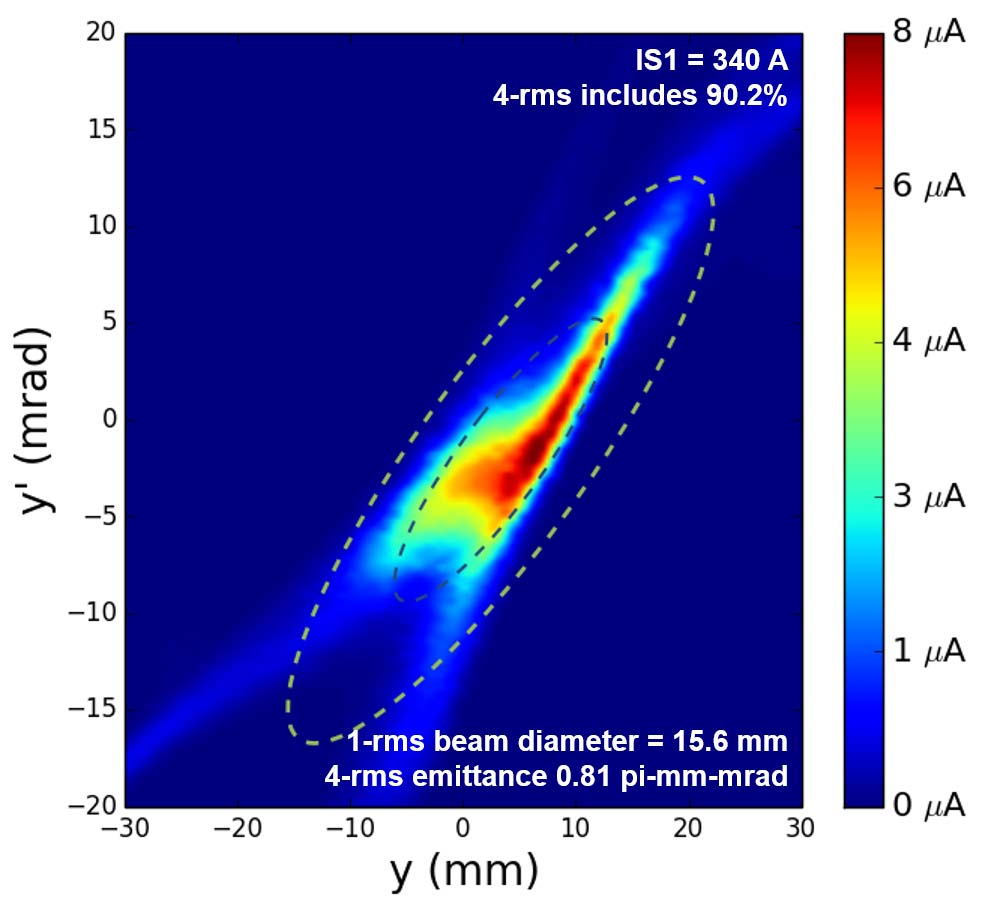}
\end{minipage}
\begin{minipage}{.45\linewidth}
  \includegraphics[width=\linewidth]{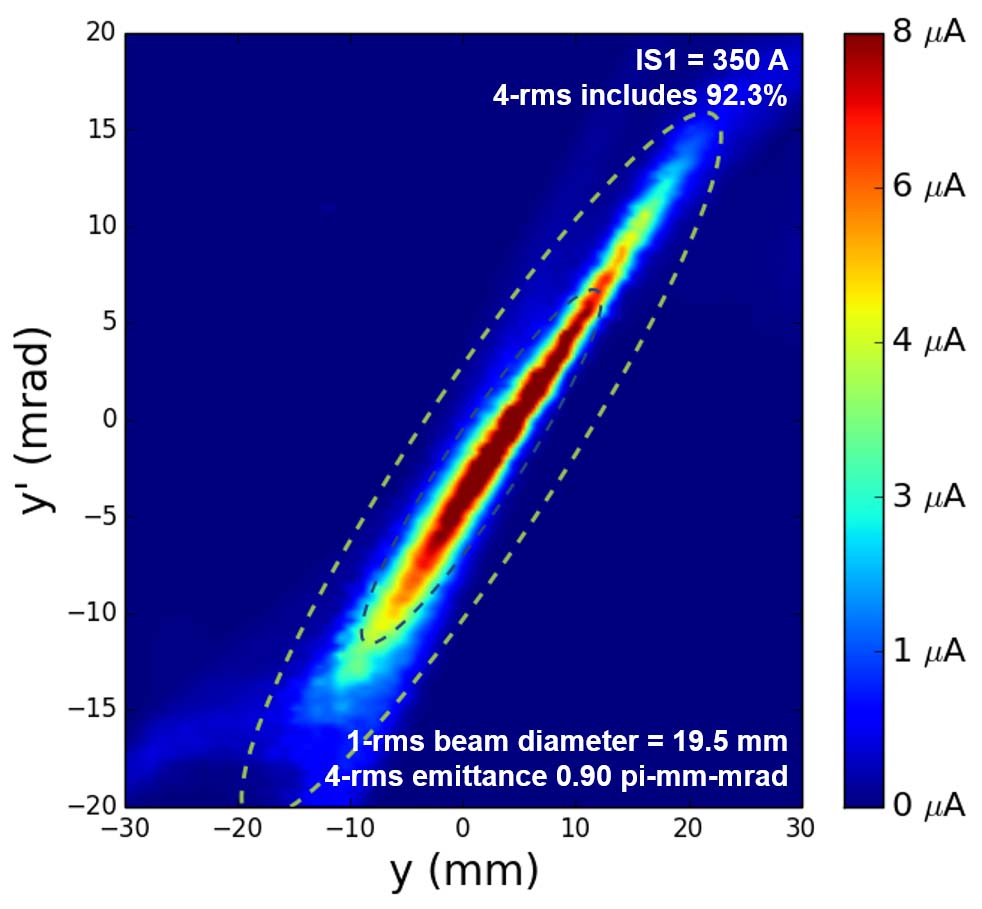}
\end{minipage}
\caption{Configuration I: vertical phase space measurements at several different SN1 currents. The source potential and electron suppression were held at 55.0$\pm0.1$ kV and $-3.0\pm0.1$ kV, respectively. The RF power was maintained at 400 W. The z-axis represents the measured current
per pixel and was not normalized to the total beam current.}
    \label{fig:phase_spaces_con_I}
\end{figure}

\paragraph{Measurements of Configuration I}\mbox{}\\
\hspace{4ex} The Allison-type Emittance Scanner (ES) was placed in the beam line at the end point of Configuration I in order to characterize the beam shape before final focusing and injection into the cyclotron. Figure \ref{fig:phase_spaces_con_I} shows the phase spaces of the beam for SN1 = 310 A (top left) to SN1 = 350 A (bottom). 
The ion source high voltage potential was reduced to 55 kV (55 keV ion beam kinetic energy)
in order to measure the phase spaces of the beam past its focal point. At 55 keV, the \htp 
beam is focused onto the ES entrance aperture between SN1 = 330 and 340 A.
The observed minimum size of the beam was at 330 A, where the 1-rms beam diameter was measured to be 15.2$\pm0.5$ mm. The 4-rms emittance measurements vary from 0.81 to 1.07 $\pi$-mm-mrad ($0.94-1.07$ including the systematic error discussed in Section \ref{sec:errors}). 
The rather strong aberrations in phase space and the resulting large emittances and large 
variation in emittance is believed to be the interplay of the following effects: 
\begin{itemize}
\item By changing the focusing strength of SN1, less and less beam is scraped on 
      upstream collimators (this is directly seen in Figure \ref{fig:phase_spaces_con_I}
      as an increase in intensity).
\item Protons and \htp are focused differently due to their respective magnetic rigidity.
\item Filling a solenoid magnet too much can introduce spherical aberrations into 
      the beam phase space.
\item The hollow beam effect. As the protons are focused further and further upstream,
      they form a tighter and tighter spot and the space charge of the proton beam
      increases significantly. This changes the dynamics of the ions of the \htp beam.
      This can be observed qualitatively in Figure \ref{fig:phase_spaces_con_I} and is in
      agreement with the observation of a hollow beam discussed in Section
      \ref{sec:hollow}. In the figure, this is most evident in the top left image where
      there appears to be an \htp deficit at y$\approx-5$ mm and two \htp lobes on either 
      side (one at y$\approx-15$ mm and the larger lobe at y$\approx 10$ mm). 
\end{itemize}

In Section \ref{sec:configuration_I} the phase space measurements at this location will be compared to simulation furthering our understanding of the effects. 

One particular result of these effects is that the proton beam (not shown) does not 
exhibit similarly strong aberrations. It can therefore be concluded that the emittance of the \htp beam could, in theory, be kept much lower, if one used a dipole magnet
for separation of the ion species thereby avoiding the over-focusing of the protons.

\paragraph{Measurements of Configuration II}\mbox{}\\
\begin{figure}[!t]
    \centering
    \includegraphics[width=0.81\columnwidth]
                    {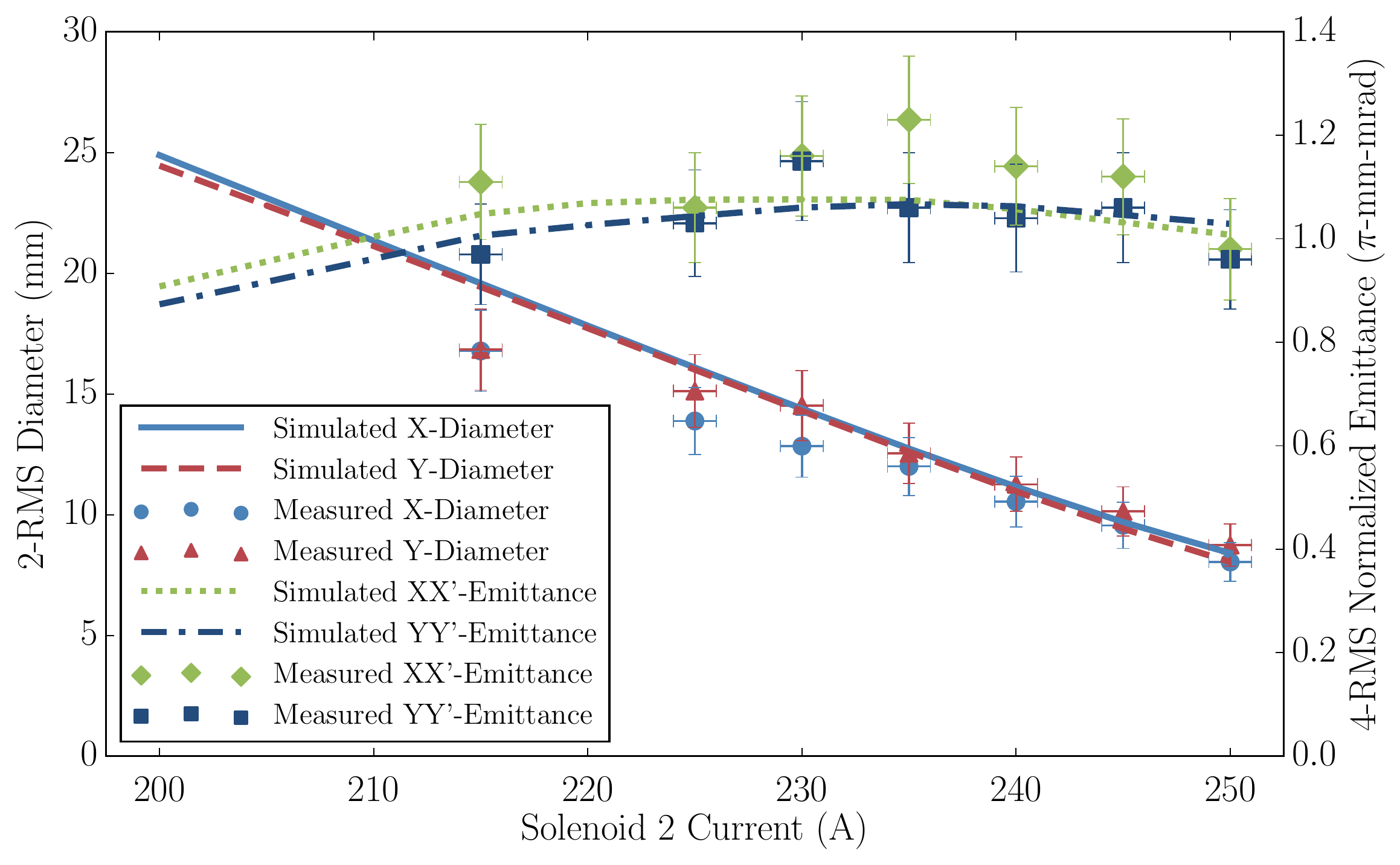}
    \caption{Configuration II: Emittance and beam diameter as a function 
    		 of the current on SN2. 
             It should be noted that,
             as discussed in Section \protect\ref{sec:errors}, 
             there is an additional systematic error due to the
             fact that many of the measured phase spaces used to 
             calculate the emittances did not cover the full beam
             area thus leading to artificially reduced emittance
             values (most notably at the beginning and end of the
             scan). The graph shows the values without the
             systematic errors and the simulated phase spaces,
             truncated with the same limits, yield good agreement
             in emittances and diameters. The systematically corrected
             values are reported in the text.}
    \label{fig:emittance_diameter}
\end{figure} 
An ISO-160 4-way cross containing the emittance scanner (ES) and a Faraday cup
was introduced at the end point of Configuration II (cf. Figure 
\ref{beamline_schematic}) in order to take 
a series of transverse emittance measurements to ensure adequate beam quality to 
inject into the cyclotron. For these measurements, the RF power supplied to the plasma chamber was held constant at 300 W and the extraction potential and electrode bias were held at 62.7 kV and -3.0 kV respectively. SN1 was maintained at 353~A in order to transport the maximum amount of \htp to the ES. The downstream ion gauge measured a beam line pressure of $7.7\pm0.1\times$10$^{-6}$ mbar.  The culmination of these settings yielded a total H$_2^+$ beam current of 5.5$\pm0.1$ mA. The current on SN2 was incrementally increased from 215 A to 250 A. The resulting emittance measurements are summarized and Figure \ref{fig:emittance_diameter}. The horizontal and vertical emittance plots from SN2 = 215 to 230 A are shown in Figure \ref{fig:phase_spaces1} and SN2 = 240 to 250 A in Figure \ref{fig:phase_spaces2}.

The left and right columns of Figures \ref{fig:phase_spaces1} and
\ref{fig:phase_spaces2} show the horizontal and vertical phase space plots respectively,
ranging from SN2 = 215 A to 250 A. Similar to the emittance plots at the end point of Configuration II, the thin streak seen in each graph represents the remaining proton
contamination. For the emittance measurements recorded 
here, the proton contribution was subtracted from the each image to give a more 
accurate representation of the H$_2^+$ beam emittance. As the current is 
increased on SN2 the beam diameter decreases. At 250 A, the measured 
2-rms beam diameters along the x and y axis are $\approx 8$ mm and $\approx 9$ 
mm, respectively. The measured normalized 4-rms emittance remained between 0.96 
and 1.23 $\pi$-mm-mrad, with the 4-rms ellipses containing between $92.6\% - 94.1\%$ 
of the beam (cf. Figure~\ref{fig:emittance_diameter}). Including the systematic error
from cutting away part of the beam, these values increase to $1.15 - 1.23$ $\pi$-mm-mrad.
With the systematic error, the variation goes down and emittance is largely conserved
for the different solenoid currents.
The beam quality measurements shown here met the requirements for clean transport 
through the spiral inflector and provided insight into the transport of the beam 
through the LEBT. A discussion of the origin of the aberrations is presented in 
Section \ref{sec:simulation_of_configuration_II}.

\begin{figure}[!b]
    \centering
    \includegraphics[width=0.99\columnwidth]{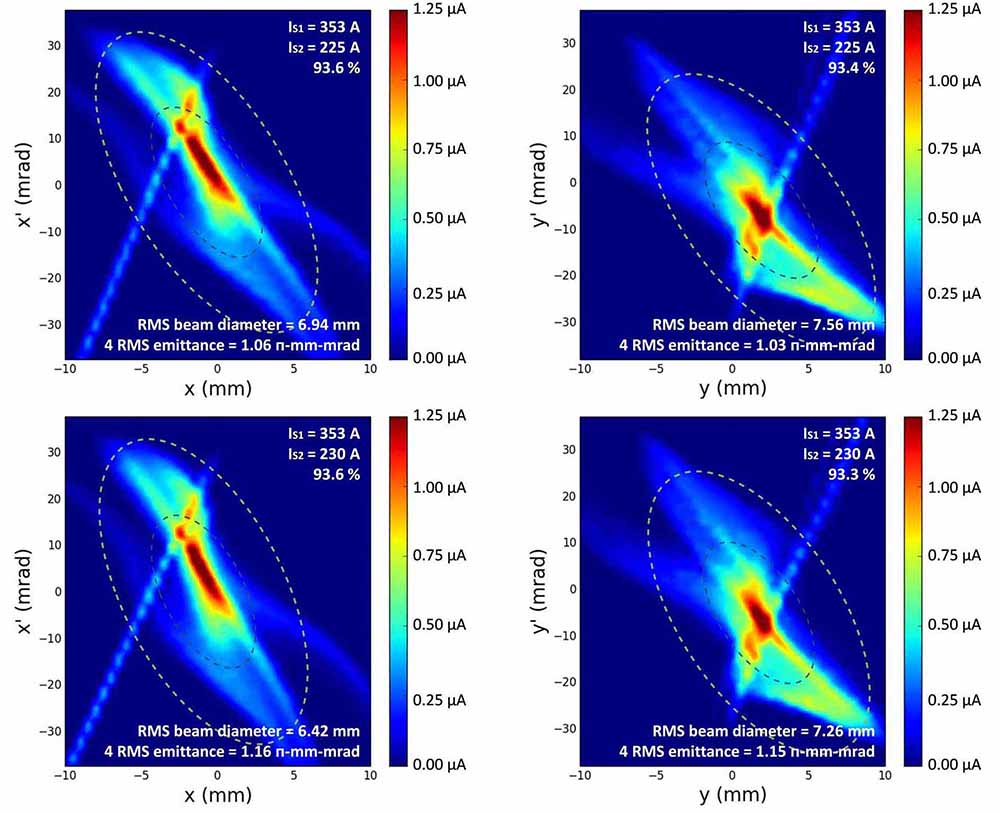}
    \caption{Configuration II: a series of emittance measurements at various 
             currents supplied to SN2. The horizontal emittance plots 
             are shown on the left; the vertical emittance plots are shown on 
             the right. The SN1 current I$_{S1}$, SN2 current 
             I$_{S2}$, and percentage of beam contained in the 4-rms contour 
             (green dashed contour) are shown in the top right of each image. 
             The 1-rms beam diameter and 4-rms emittance measurements are shown
             at the bottom right of each plot. SN2 currents 225 A and 235 A.
             It should be noted that the emittances are without the systematic
             error discussed in Section \protect\ref{sec:errors}. Confer to the
             text for the corrected emittances.}
    \label{fig:phase_spaces1}
\end{figure} 

\begin{figure}[p]
    \centering
    \includegraphics[width=0.99\columnwidth]{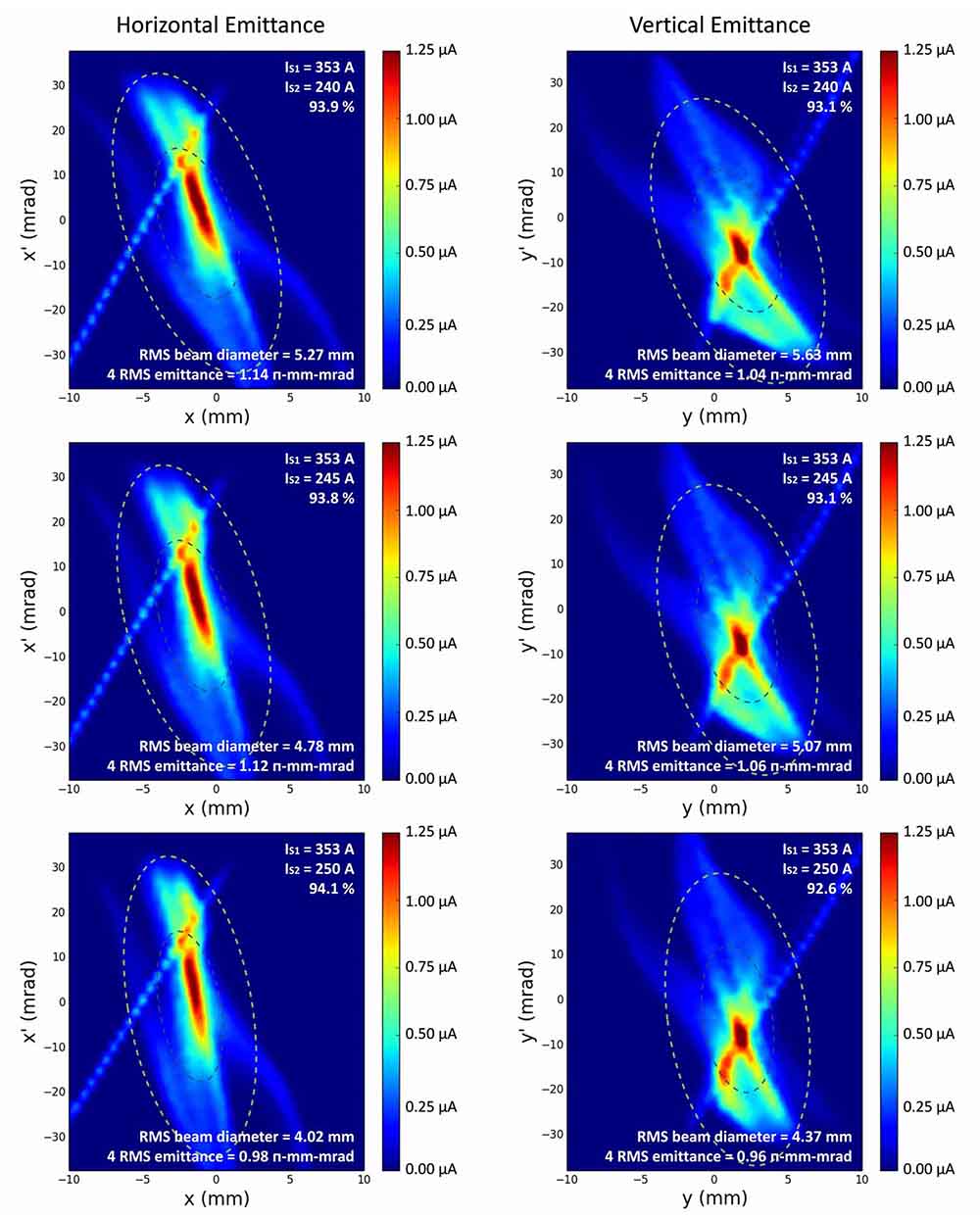}
    \caption{Continuation of Figure \protect\ref{fig:phase_spaces_con_I}.
             SN2 currents from 240 A to 250 A.}
    \label{fig:phase_spaces2}
\end{figure} 

\clearpage
\subsection{Cyclotron injection studies \label{sec:cyclotron_injection}}
\hspace{4ex} The injection into the cyclotron was tested in three steps:
\begin{enumerate}
\item Injection into an axial Faraday cup at the center of the cyclotron
\item Injection through the spiral inflector, with measurement of the beam
      current on a `paddle' radial probe. For this measurement, the central region
      was mounted inside the cyclotron rotated by 90\degree. 
\item Injection through the spiral inflector and acceleration. The currents
      are measured using three different radial probes.
\end{enumerate}
The three steps will now be discussed in the following subsections.

\subsubsection{Injection into an axial Faraday cup inside the cyclotron}
\hspace{4ex} The first injection test was simply to replace the diagnostic box at the end 
of the beam line with the cyclotron as seen in Figure \ref{beamline_schematic}.
Instead of the central region containing the spiral inflector, a Faraday cup 
was mounted axially inside the cyclotron. The reason for this setup was to 
exactly determine the amount of beam current entering the cyclotron through
the spiral inflector aperture. To this end, the Faraday cup was outfitted with 
a similar aperture at the same z position. Electron suppression on the Faraday 
cup was provided by a suppression electrode at -180 V.
The cyclotron magnet was engaged to 100 A ($\approx45\%$ of the nominal current 
for \htp acceleration. This improved beam transport into the Faraday cup significantly.
The result of this first step is summarized in Table \ref{tab:injection_fc}. The 
transport efficiency from the beam stop (end of Configuration 1 in Figure \ref{beamline_schematic})
to the Faraday aperture is close to 100\% (with the current on the aperture being
the main source of error due to missing electron suppression) and the current measured 
inside the cyclotron was $7.25\pm0.36$ mA leading to a final transport efficiency from beam stop into
the future spiral inflector of 89\%. The \htp purity in the Faraday cup was determined
by simulation to be $>98\%$.

\begin{table}[!b]
	\caption{First injection test with axial Faraday cup inside the cyclotron.
	         Parameters and results.}
	\label{tab:injection_fc}
	\centering
    \vspace{5pt}
    \renewcommand{\arraystretch}{1.25}
		\begin{tabular}{lll}
            \hline
            \textbf{Parameter}                  & \textbf{Value} \\
            \hline \hline
            Species                             & \htp     \\
            Beam Energy                         & 62.7 keV \\
            Solenoid 1                          & 356 A    \\
            Solenoid 2                          & 250 A    \\
            Cyclotron Magnet                    & 100 A    \\
            Beam Stop Current                   & $8.2\substack{+0.4\\-1.2}$ mA\\
            Faraday Cup Collimator Current      & $0.5\pm0.2$ mA  \\
            Axial Faraday Cup Current           & $7.25\pm0.36$ mA  \\
            \hline
		\end{tabular}
\end{table}

\subsubsection{Injection through the spiral inflector onto the paddle probe}
\begin{figure}[!t]
    \centering
    \includegraphics[width=0.45\textwidth]{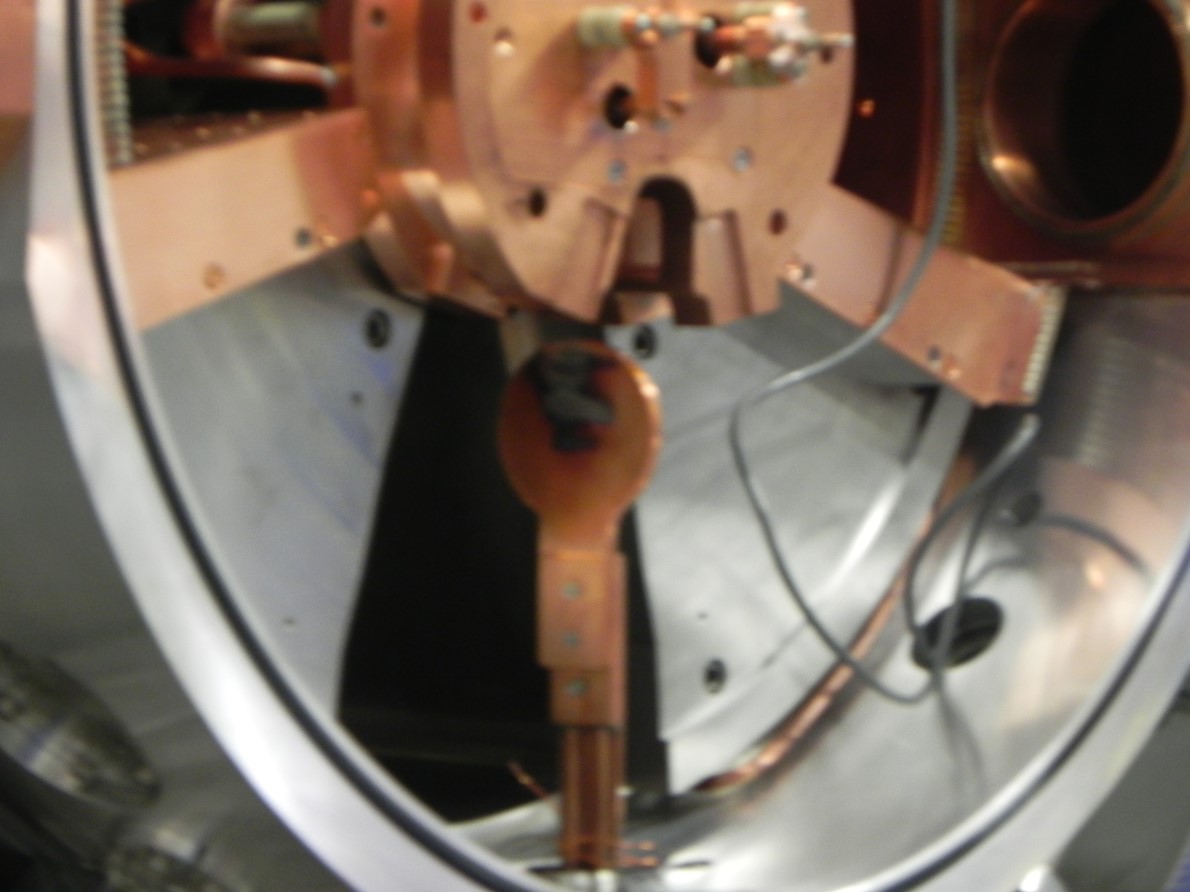}
    \caption{A photograph of the inside of the test cyclotron. The paddle probe
             reaches in from the bottom (currently retracted to see the burn marks)
             to measure beam current transported through
             the spiral inflector (inside copper housing) but not yet accelerated.}
    \label{fig:cyclo_paddle}
\end{figure} 
\hspace{4ex} The axial Faraday cup inside the cyclotron was replaced by the central region piece with
the spiral inflector, but rotated by 90\degree to accommodate a paddle probe reaching
into the dee recess from below (cf. Figure \ref{fig:cyclo_paddle}. 
This way, beam exiting the spiral inflector only traveled about 45\degree before 
hitting the paddle probe, providing a good measure of spiral inflector transmission.
Secondary electrons from the paddle probe were suppressed by the strong magnetic 
field of the cyclotron. The currents on the spiral inflector entrance aperture
were unsuppressed and as per Section \ref{sec:errors}, we are reporting the readings 
as $\mathrm{I}_\mathrm{actual} = 0.5 \cdot \mathrm{I}_\mathrm{meas.} \pm 0.2 \cdot \mathrm{I}_\mathrm{meas.}$. 
The result of this second step is summarized in Table \ref{tab:injection_paddle}.
For this test, the beam was pulsed and thus the additional error from the preglow effect
combined with the low pass filter of the DAQ system is applied as discussed in Section
\ref{sec:errors}. Combining the measured values in Table \ref{tab:injection_paddle} with
the discussed errors, we obtain a transmission of $93.5\substack{+6.5\\-15.6}\%$ through the spiral inflector without subsequent acceleration. This compares well to measurements in the next section and simulations in Section \ref{sec:opera}.
\begin{table}[!b]
	\caption{Injection test with the spiral inflector in place but rotated by 90\degree
	         to accommodate a paddle probe. Parameters and results.}
	\label{tab:injection_paddle}
	\centering
    \vspace{5pt}
    \renewcommand{\arraystretch}{1.25}
		\begin{tabular}{lll}
            \hline
            \textbf{Parameter}                     & \textbf{Value} \\
            \hline \hline
            Species                                & \htp          \\
            Beam Energy                            & 60.0 keV      \\
            Solenoids 1/2                          & 340 A/240 A   \\
            Cyclotron Magnet                       & 223 A         \\
            Spiral Inflector Upper/Lower Electrode & -10 kV/+10 kV \\
            Beam Stop Current                      & $7.5\pm0.8$ mA \\
            Spiral Inflector Aperture Current      & $1.3\pm0.5$ mA \\
            Paddle Probe Current                   & $5.8\pm0.4$ mA \\
            \hline
		\end{tabular}
\end{table}

\clearpage
\subsubsection{Injection through the spiral inflector with acceleration}
\begin{figure}[!t]
    \centering
    \includegraphics[width=1.0\textwidth]{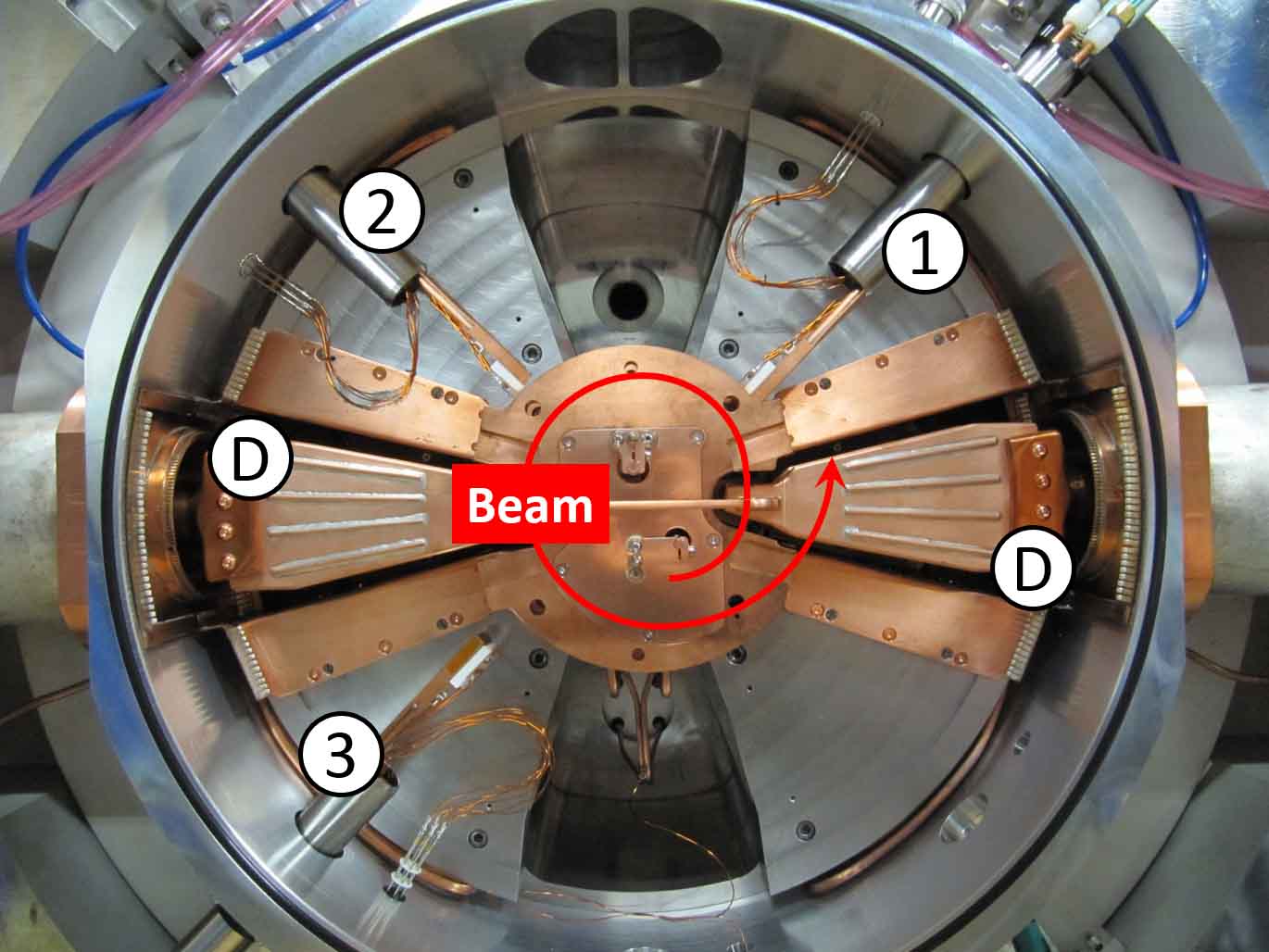}
    \caption{A photograph of the inside of the test cyclotron. the dees are labeled "D"
             and the radial probes are enumerated from (1) to (3). The beam is entering
             the cyclotron axially (towards the reader) is bent into the mid-plane and then 
             follows a trajectory similar to the red spiral.}
    \label{fig:cyclo_inside}
\end{figure}
\begin{figure}[!t]
    \centering
    \includegraphics[width=1.0\textwidth]{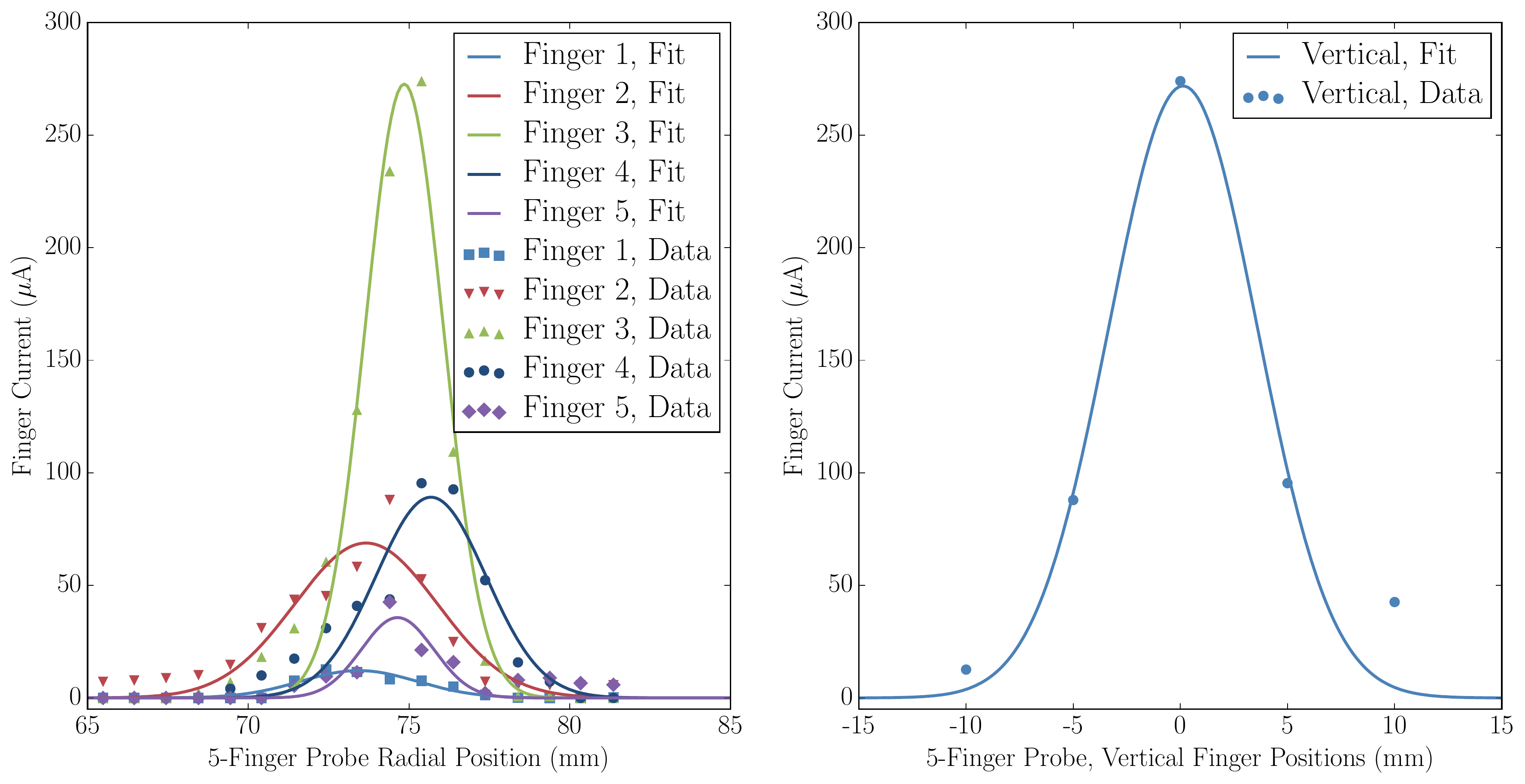}
    \caption{Gaussian fits to the measurement of currents on the 5-finger probe.
             Radial FWHM of the center finger (Finger 3) is $\approx2.85$ mm, 
             vertical FWHM of the beam is $\approx8.2$ mm. The tails on the
             left side of each data set are due to ions not fully accelerated 
             thus having a smaller radius.}
    \label{fig:finger_probe}
\end{figure}

 
\hspace{4ex} The final setup for injection and acceleration tests is shown in Figure \ref{fig:cyclo_inside}. Beam enters the cyclotron axially, is bent into the mid-plane and
follows a counter-clockwise spiral trajectory through the cyclotron. Two dees with
two accelerating gaps can be seen in the image (denoted by the letter 'D'). During
the tests, a function generator was used to pulse the ion source magnetron and the
cyclotron RF amplifier with a 10 ms ON / 10 ms OFF pattern seen in Figure \ref{cap}.
Three radial probes where available (labeled $1-3$ in the figure). These probes 
could be moved via stepper motors. The position accuracy could be determined to 
be better than $\pm1$ mm. Readout from the probes was done using the current readout
channels and methods discussed in Section \ref{sec:daq} with low pass filters attached 
to filter out the preglow peak. Because the RF amplifier could not provide the necessary
power to obtain the full dee voltage of 70 kV, the experiments were conducted at lower 
dee voltages of $\approx50$kV which consequently led to lower injection efficiencies as
the design was optimized for 70 kV. A comparison with OPERA simulations of this 
reduced dee voltage is presented in Section \ref{sec:opera}.
In addition, the RF was unstable and discharges in the resonators occurred frequently.
Large uncertainties are therefore attached to the presented numbers.
In the following, two sets of measurements are presented using the radial probes: Set 1 and Set 2. In these sets, the beam current on each probe was measured, then the probe was retracted until no beam was seen on the probe anymore, indicating it was now just outside of the beam radius, thus allowing the beam to circulate to the next probe. This was done with all three probes sequentially until the 
maximum range of each probe was reached.
\begin{figure}[!t]
    \centering
    \includegraphics[width=0.7\textwidth]{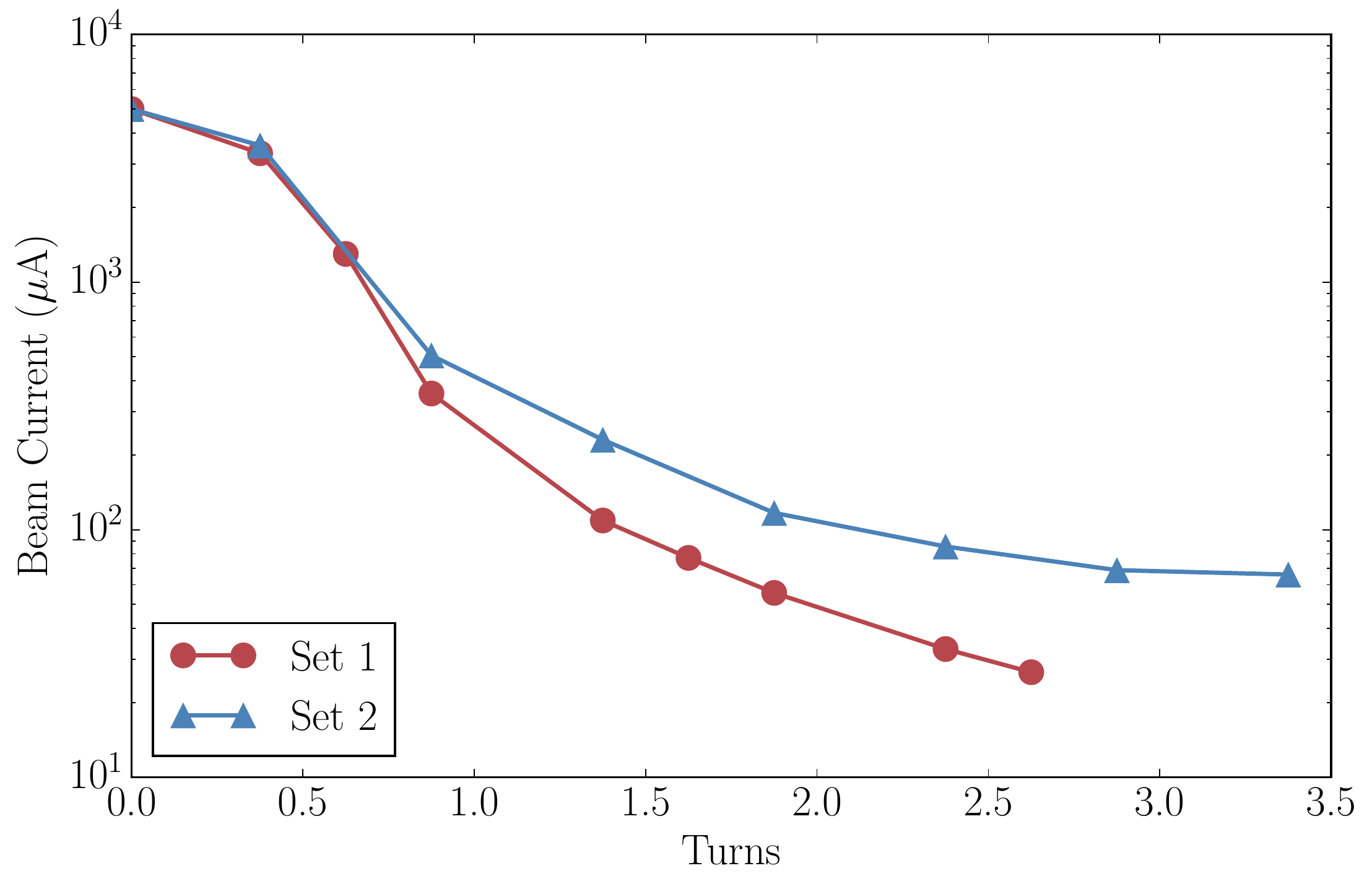}
    \caption{Probe current as a function of turn number for two dee voltages. Turn 0
             starts at the exit of the spiral inflector. The parameters of the
             measurements are listed in Table \protect\ref{tab:injection_final}.}
    \label{fig:accel_probes}
\end{figure}
\begin{table}[!b]
	\caption{Parameters for inflection and acceleration studies.}
	\label{tab:injection_final}
	\centering
    \vspace{5pt}
    \renewcommand{\arraystretch}{1.25}
		\begin{tabular}{lll}
            \hline
            \textbf{Parameter}                       & \textbf{Set 1} & \textbf{Set 2}\\
            \hline \hline
            Species                                  & \htp  & \htp        \\
            Initial Beam Energy                      & 62.7 keV & 62.7 keV      \\
            Solenoids 1 / 2                          & 350 A / 233 A & 350 A / 233 A  \\
            Cyclotron Magnet                         & 218.6 A & 218.0 A        \\
            Spiral Inflector Upper / Lower Electrode & -10.0 kV / +10.15 kV & -10.0 kV / +10.15 kV \\
            Beam Stop Current                        & $6.8\pm0.8$ mA & $6.9\pm0.8$ mA\\
            Spiral Inflector Aperture Current        & $1.0\pm0.4$ mA & $1.1\pm0.4$ mA\\
            Approximate dee voltage                  & $\approx47$ kV & $\approx50$ kV \\
            Captured Beam                            & $\approx 0.4\%$ & $\approx 1\%$ \\
            \hline
		\end{tabular}
\end{table}
The important parameters of Sets 1 and 2 are listed in Table \ref{tab:injection_final}.
In both sets, the dee voltage was estimated using a capacitive pick-up probe. The values fluctuated between 50 and 60 kV. Because of the low reliability of the probe, it is believed that the actual average dee voltage was on the lower end of this interval, 
which fits well with the simulations presented in Section \ref{sec:opera}.
The measured currents as a function of
fractional turn number can be seen in Figure \ref{fig:accel_probes}. the magnetic field of 
the cyclotron and the spiral inflector voltages were optimized individually, but came out to be the same for both sets. It can clearly be seen that the acceptance of the cyclotron
drops with dee voltage deviating more from the design voltage. The beam radii and 
estimated beam energies are listed in Table \ref{tab:injection_results}.
With a five-finger probe in place of Radial Probe 1, it was possible to show that the 
beam was centered vertically in the cyclotron mid-plane. 
With the five-finger probe, the beam size at the position of Radial Probe 1 during the 
first turn could be estimated to be $2.85\pm0.5$ mm FWHM radially and $8.2\pm0.7$ mm FWHM vertically (see Figure \ref{fig:finger_probe}).
The final result of the injection and acceleration studies was that $66~\mu\mathrm{A}$
of \htp could be successfully injected and accelerated to 3 3/8 turns which corresponds
to $\approx410$ keV/amu and an injection efficiency of 1\% which was expected for the 
dee voltage of $\approx50$ kV (cf. Section \ref{sec:opera}).

\begin{table}[!b]
	\caption{Probe 1 radii, currents, and approximate beam energy for the two sets
	         of inflection tests. The approximate energy gain per gap (2 per dee)
	         was estimated to be 47 keV for Set 1 and 50 keV for Set 2.}
	\label{tab:injection_results}
	\centering
    \vspace{5pt}
    \renewcommand{\arraystretch}{1.25}
		\begin{tabular}{lrrrr}
            \hline
            & \textbf{Approx. Radius} & \textbf{Current} & 
              \textbf{Approx. Energy} & \textbf{Average Mag. field}\\
            \hline \hline
            \textbf{Set 1} & 74 mm   & 3314 $\mu$A & 157 keV & 1.1 T \\
                           & 105 mm  & 109  $\mu$A & 345 keV & 1.1 T \\
                           & 135 mm  & 33   $\mu$A & 533 keV & 1.1 T \\
            \hline
			\textbf{Set 2} & 75 mm   & 3564 $\mu$A & 163 keV & 1.1 T \\
                           & 105 mm  & 231  $\mu$A & 363 keV & 1.1 T \\
                           & 135 mm  & 86   $\mu$A & 563 keV & 1.1 T \\
                           & 165 mm  & 66   $\mu$A & 763 keV & 1.1 T \\
            \hline
		\end{tabular}
\end{table}

\clearpage
\section{Simulations\label{sec:simulations}}
\hspace{4ex} One very important aspect of the design and commissioning of a particle 
accelerator system is beam transport simulations and their comparison with 
measurements. For the test runs at BCS in Vancouver, the particle-in-cell (PIC) 
code WARP \cite{grote:warp1} was used to simulate the transport of two ion 
species (protons and \htp) simultaneously through a model of the 
beam line used during the tests (cf. Figure \ref{beamline_schematic}). 
The different beam line elements and how they were included in the simulations
are described in Section \ref{sec:components}.

In high intensity, low energy beams, it is important to include space charge
effects as well as make some assumptions about space charge compensation. 
The way this is done in the WARP simulations is described in more detail in 
Section \ref{sec:warp:scc}. One important parameter in the compensation 
estimation is the pressure along the beam line, because it determines the 
density of neutral gas molecules available for secondary ion and electron
production through residual gas ionization and charge-exchange processes. 
Simulations using the code MOLFLOW were performed in order to obtain pressure
distributions along the beam line for several important cases and are presented
in Section \ref{sec:molflow}.

\subsection{Introduction to the WARP LEBT simulations}

\subsubsection{Basic simulation parameters}
\hspace{4ex} WARP includes a multitude of different field-solvers which are discussed in more 
detail elsewhere \cite{grote:warp1, grote:warp2}. For the presented problem, the 
XY-slice solver was chosen. Here, the beam is transported through a series of 
transversal slices along the z-axis (typically 0.5 mm step size) and the 
transversal self fields are calculated on a 2D grid (typically 512x512) at each 
step. This is a suitable approach for a DC ion beam in which the 
longitudinal self fields are largely negligible. The XY slice solver is robust 
and fast, which makes it an ideal candidate to map a large parameter space.
The number of macro-particles used in these simulations was 60,000 for each 
species (individual species currents are set by charge/macro-particle).

\subsubsection{Beam line components in the simulations \label{sec:components}}
\hspace{4ex} The different configurations (I-III) used during the measurements were discussed in 
Section \ref{sec:bcs_layout} and are shown in Figure~\ref{beamline_schematic}.
The important beam shaping elements were the two solenoids and the various 
apertures and collimators.

The fields of solenoids SN1 and SN2 were calculated in POISSON Superfish 
\cite{menzel:poisson} and saved as axially symmetric 2D field maps for
WARP import. Comparison of the calculated field maps with measurements using
a Gauss-meter yielded excellent qualitative agreement. A scaling factor was
introduced to match the peaks fields quantitatively. As can be seen in Figure 
\ref{fig:fc0_vs_sol1} this lead to good agreement of the proton and \htp focal points.

The numerous apertures and collimators were introduced in WARP as beam scrapers
and care was taken that variable apertures (i.e. 4-Jaw slits) were set correctly.

For simulation purposes, the LEBT in Configuration III is essentially the same as 
in Configuration II with particle distributions saved at slightly different z locations.
The simulation of the cyclotron injection was done with Vector Fields OPERA using the particle distributions obtained from the Configuration II WARP simulations and are treated in Section \ref{sec:opera}.

\subsubsection{Beam line pressure profile simulations \label{sec:molflow}}

\begin{figure}[!t]
    \centering
    \includegraphics[width=1.0\columnwidth]{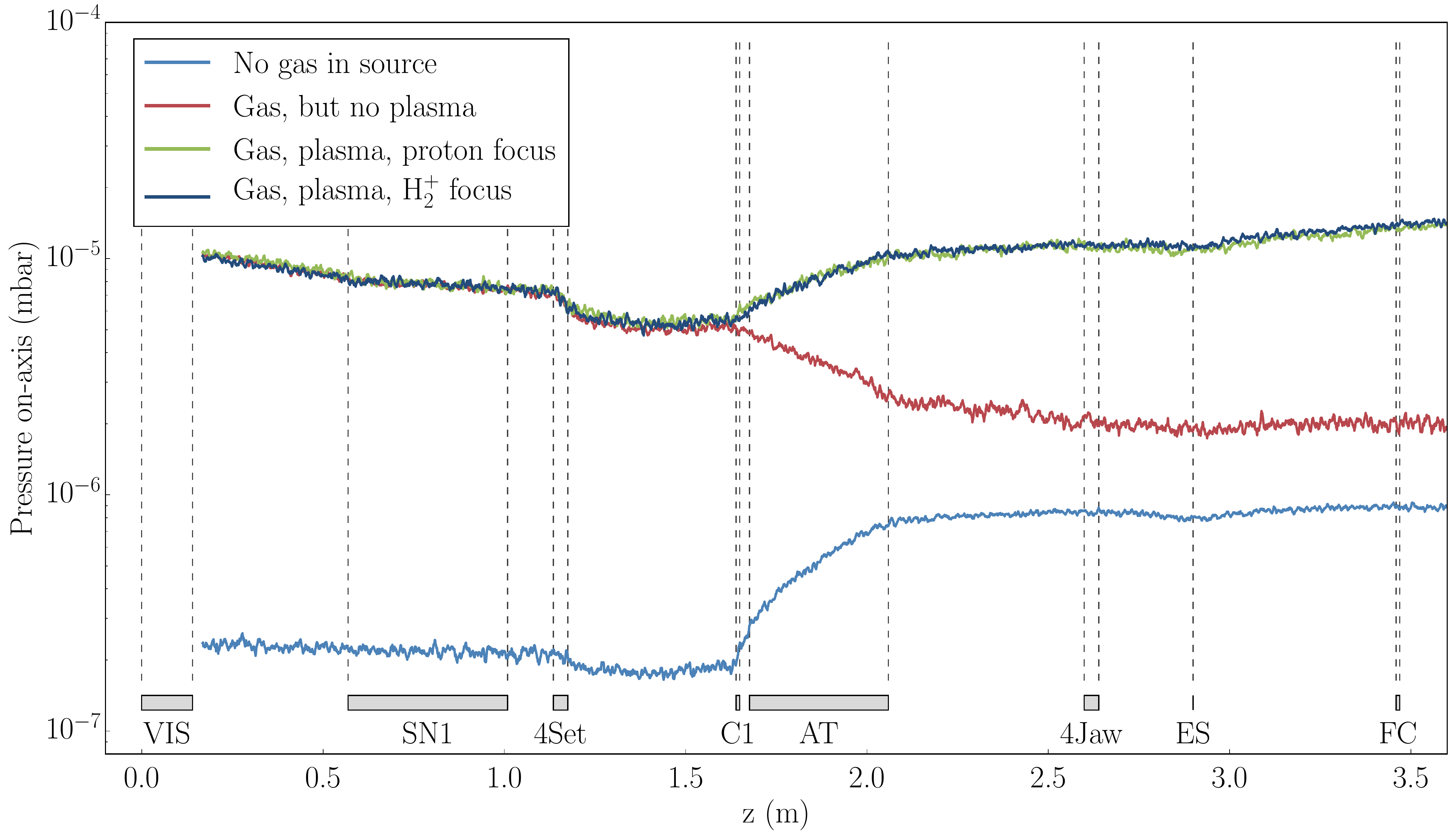}
    \caption{Pressure distributions along the beam line in Configuration I
             calculated with MOLFLOW. It 
             can be seen that the major contribution comes from the ion source
             gas supply with an additional pressure increase downstream from 
             beam striking elements of the beam line and thereby heating and
             sputtering them. Although the pressure distributions for proton
             and \htp foci are slightly different, they result in identical
             beam dynamics during the simulations. Also noteworthy is the effect
             of collimator (C1) and adapter tube (AT), both with small diameter,
             on the pressure distribution. They mainly constrict the flow and
             for the different scenarios, different parts of the beam line 
             contribute more or less to the overall pressure (e.g. beam induced 
             out-gassing on the downstream side).}
    \label{fig:pressure}
\end{figure}

In order to estimate space charge compensation in the WARP LEBT simulations,
axial pressure distributions along the beam line were calculated for several different cases using the MOLFLOW \cite{kersevan1991molflow} simulation 
package. These simulations were matched to recorded upstream and downstream ion gauge readings by varying desorption coefficients
\begin{enumerate}
\item along the beam line (outgassing and leaks),
\item of the ion source aperture (main gas inflow from the source).
\end{enumerate}
Three simplified models of the beam line, corresponding to the three
configurations listed in Section \ref{sec:components} were created in the CAD software AutoDesk Inventor \cite{autodesk2014autocad} and then imported into MOLFLOW.
Configuration I will be used as an example in this section. 

MOLFLOW is a simulation package developed by CERN that uses a Monte Carlo algorithm to calculate characteristics of a system under High and Ultra-High Vacuum 
(HV and UHV) conditions. This pressure regime is called molecular flow regime where collisions between particles are unlikely and where collisions with the walls dominate.
The molecular flow regime is reached when the mean free path of the gas molecules
becomes longer than the vacuum vessel dimensions. At $1\cdot10^{-5}$ Torr, the mean 
free path is on the order of a few meters and so the condition is fulfilled.
This makes Monte Carlo simulations well suited to studying a beam line under HV/UHV conditions. MOLFLOW takes an input geometry and generates particles 
according to the user defined desorption coefficients on the facets. Pumps are
defined by the user by absorption coefficients on the corresponding facets.
Particles are then moved through the system by ray-tracing, having a chance of
reflection or absorption every time they hit a surface. The pressure on a selected facet can be calculated from the number of hits per unit time and the
area of the facet. 

The algorithm of obtaining on-axis pressure distribution for the LEBT is as 
follows (confer also to Figure \ref{fig:pressure} for the different stages 
of this for Configuration I):
Initially, the simulation is run without any gas flow and beam from the ion
source, so that desorption coefficients corresponding to leaks and room-temperature outgassing could be established (the baseline).
Next, the desorption from the ion source aperture facet was increased until
the upstream ion gauge reading corresponded to the measured value (turning on 
the source gas but not the plasma). As expected, the downstream ion gauge value immediately went up to a value very close to the reading during the same stage in the experiment.
To evaluate the system with beam, preliminary LEBT simulations using a constant 
pressure were used to establish facets that were struck by the beam (cf. Figures 
\ref{fig:SN1_230A_Env_horz} and \ref{fig:SN1_320A_Env_horz}), and desorption
coefficients were increased again until the two ion gauges values matched the 
measured ones. It should be noted, that the gas flow from the source was 
reduced slightly, to account for the pumping effect of the plasma (with the 
same gas flow into the source, the gas leaking through the extraction hole will 
be less if plasma is ignited).

For this beam line, it turned out that the variation in on-axis pressure was small enough to be secondary to other effects (interplay between the two ion species, change in beam radius), but for longer beam lines with overall lower pressures and larger variations, this will be an important tool to estimate 
space charge compensation.

\subsubsection{Space charge compensation in the WARP code \label{sec:warp:scc}}

\begin{figure}[!t]
\centering
	\includegraphics[width=1.0\textwidth]{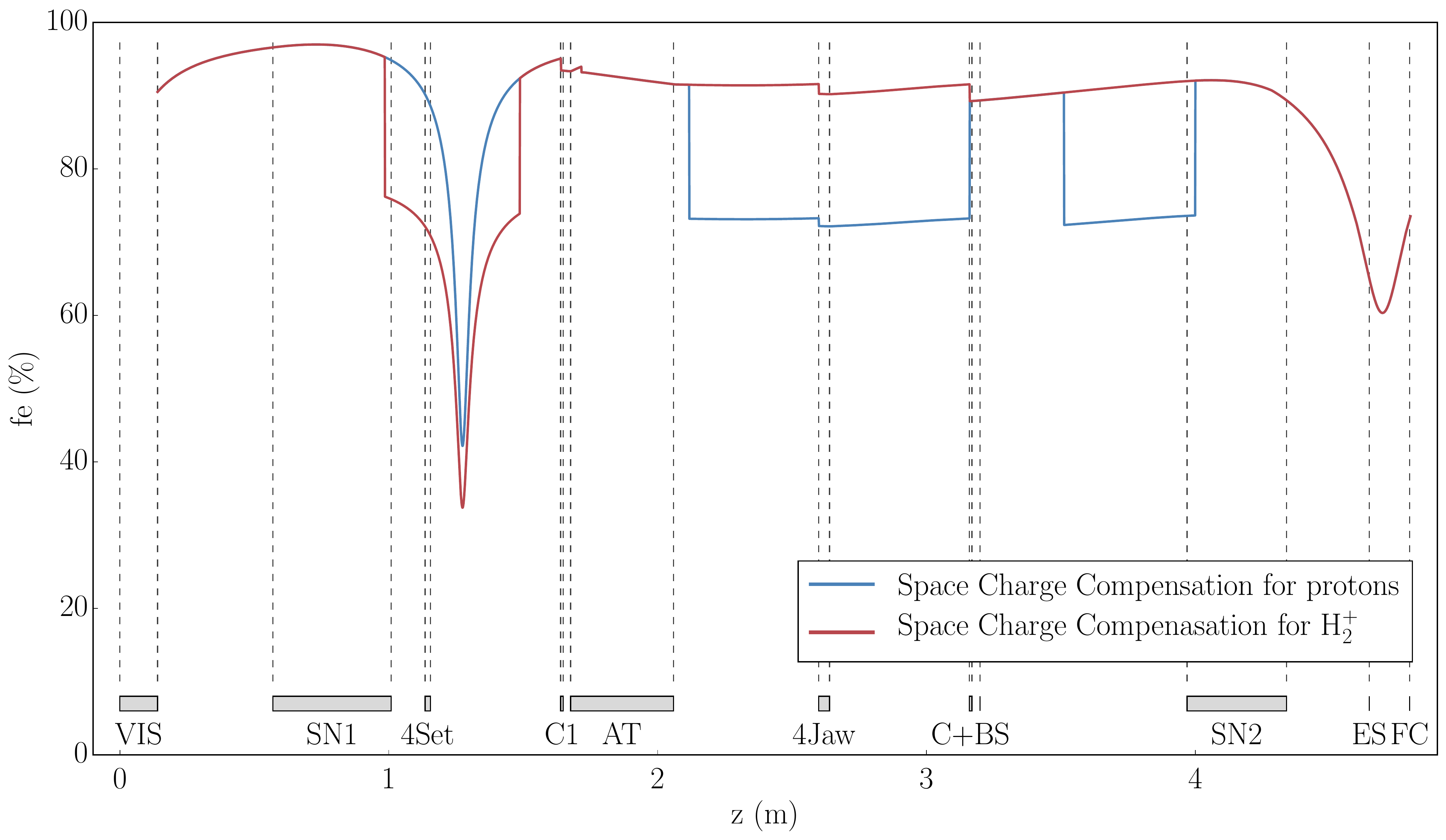}
	\caption{Example of space charge compensation along the beam line for 
	         Configuration II. The smooth dips after the 4-sector collimator 
	         (4-Set) and at the electrostatic emittance scanner (ES) are from 
	         the beam coming to a focus after SN1 and SN2. The sharp changes are from a phenomenological approach
	         to multiple species with different radii and are explained in 
	         the text.
	         \label{fig:fe_example}}
\end{figure}

\hspace{4ex} Space charge compensation arises from the interaction of beam ions with the 
residual gas molecules. There are several processes that can occur, but the
two dominant ones are charge-exchange and ionization. Together, they lead to low energy secondary ions and electrons inside the beam envelope. Because of the 
positive beam potential, electrons are attracted and positive ions are repelled. This 
can lead to a significant lowering of the beam potential. 
In the LEBT simulations, space charge compensation is calculated dynamically 
at each time step. The model is based on the work of Gabovich et al. \cite{gabovich:spacecharge1, gabovich:spacecharge2, soloshenko:spacecharge1,
soloshenko:spacecharge2, soloshenko:spacecharge3} and was recently reviewed 
and updated \cite{winklehner:scc2, winklehner:phd}. In the model, the energy
balance of the secondary electrons is used to calculate a steady-state value
of the space charge compensation factor $\mathrm{f}_\mathrm{e}$. It depends on
the neutral gas density (beam line pressure), the beam currents, the beam
energy, and the beam size. It also relies on knowing the cross sections for
total secondary ion and secondary electron production through the aforementioned
processes. This model 
is a `best case' approximation, where the beam is uniform and round and all
species have the same radius. No collective effects, plasma oscillations, 
and non-linearities are taken into account. In many cases, however, it works 
remarkably well. For the presented application, the difference in beam size had to be taken into account phenomenologically. If we consider for a moment the two species as independent systems, we can immediately see that if one of them is smaller in diameter and has a higher beam potential, it acts like a charged electrode for the second species and will collect electrons that would normally
contribute to the second species' compensation. In the presented simulations,
a threshold $\xi$  was introduced so that if both of  the following two conditions are satisfied simultaneously:
\begin{equation}
\begin{aligned}
r_{1} & \le & \xi \cdot r_{2} \\
I_1 & > & I_2
\end{aligned}
\end{equation}
the space charge compensation factor $\mathrm{f}_\mathrm{e}$ is reduced by a
factor $\eta$:
\begin{equation}
\mathrm{f}_\mathrm{e, new}=\eta\cdot\mathrm{f}_\mathrm{e, old}
\end{equation}
Here, $r_{1,2}$ are the 4-rms beam radii, $I_{1,2}$ the current 
densities, and $\mathrm{f}_\mathrm{e, new}$ the new compensation factor of 
the larger species. In this approach, $\xi$ and $\eta$ are treated as free
parameters. An example of the resulting compensation factors can be seen in
Figure \ref{fig:fe_example}. 
It becomes immediately clear that this is a crude approximation leading to 
(nonphysical) sharp changes in $f_e$. A smoother model or a more involved
analytical solution taking into account different beam radii is desirable 
but not yet developed. For the BCS LEBT simulations, the values of 
$\xi=0.667$ and $\eta=0.8$ were found to lead to good agreement with
experiment. 

In addition, care has to be taken such that
electrostatic elements in the beam line are taken into account. These can 
negate space charge compensation completely as they may collect the 
compensating electrons. 

\subsubsection{Initial particle distribution at the VIS source}
\begin{figure}
\centering
\begin{minipage}{.45\linewidth}
  \includegraphics[width=\linewidth]{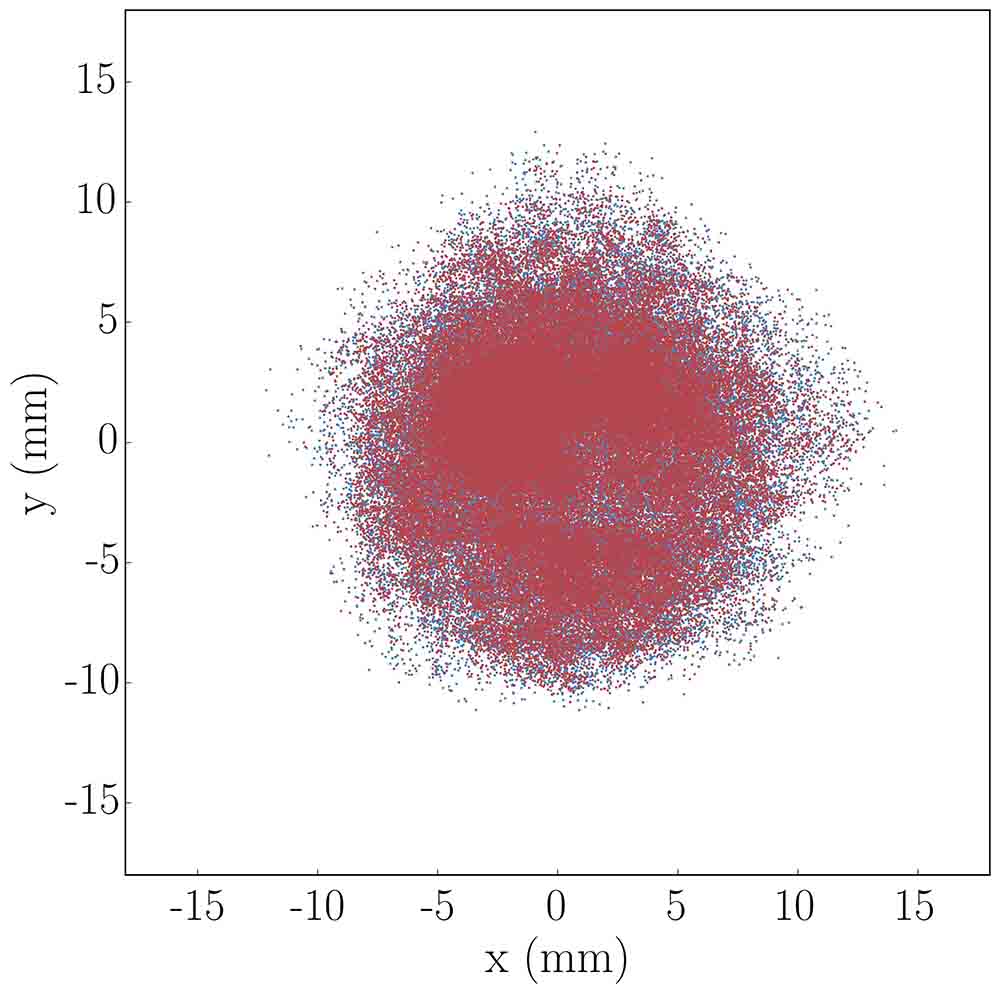}
\end{minipage}
\hspace{.05\linewidth}
\begin{minipage}{.45\linewidth}
  \includegraphics[width=\linewidth]{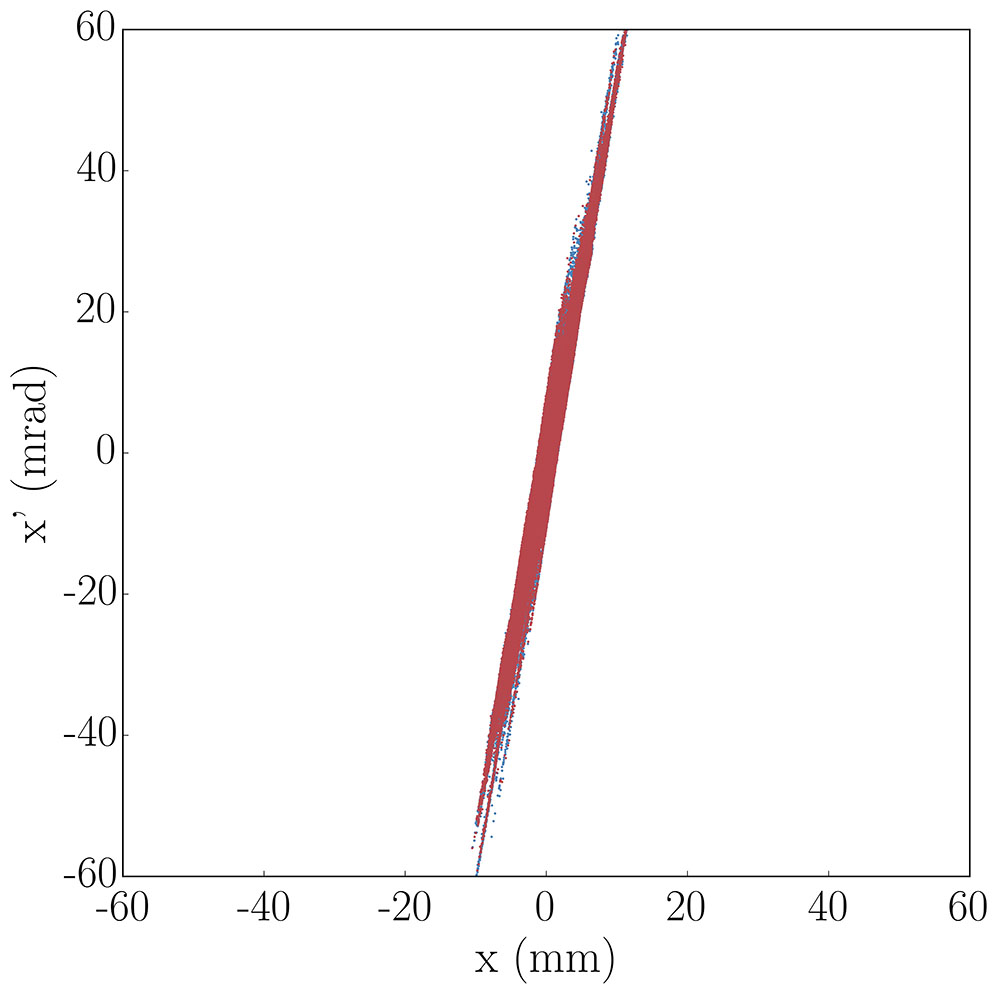}
\end{minipage}
\caption{Raw initial particle distributions used in the WARP simulations,
         obtained with KOBRA-INP. The red markers represent protons, the blue
         markers represent \htp. Left: Initial beam cross-section. Right:
         initial x-x' phase space. There are initial asymmetries, which can
         be attributed to numerical effects in the 3D simulations.}
\label{fig:particles_initial}
\end{figure}

\hspace{4ex} The initial particle distribution was obtained by the Catania group using the 
self-consistent 3D ion source extraction simulation software KOBRA-INP 
\cite{spaedke:kobra}. The beam profile and x-x' phase space 14 cm after the 
extraction aperture of the source are shown in Figure 
\ref{fig:particles_initial}. This is the starting point of the WARP simulations.
An extraction voltage of 60~kV was used to obtain this initial distribution and 
the longitudinal velocity component ($v_z$) of each particle is scaled 
appropriately in the WARP simulation if a different beam energy is desired (i.e.
if the experiment the simulation should be compared to was at a different 
extraction voltage). The initial 2-rms diameters and 4-rms normalized emittances 
are listed in Table \ref{tab:particles_initial}. 

As a benchmark and to make sure the initial asymmetries were not causing the 
effects attributed to the beam line elements, a Gaussian beam with the same 
Twiss parameters was generated and occasionally used instead of the KOBRA-INP 
initial distribution. The results exhibited the same aberrations and effects.

\begin{table}[!hb]
\vspace{10pt}
	\caption{The initial diameters and emittances for protons and 
             \htp obtained by KOBRA-INP extraction simulations and
             used in the WARP beam transport simulations. The 4-rms emittances
             include $\approx85\%$ of the beam.}
	\label{tab:particles_initial}
	\centering
    \vspace{5pt}
    \renewcommand{\arraystretch}{1.25}
		\begin{tabular}{lll}
            \hline
                                               & \textbf{protons}   & $\mathbf{\mathrm{H}_2^+}$ \\
            \hline \hline
            x--diameter (2-rms)                 & 15.2 mm            & 15.0 mm \\
            y--diameter (2-rms)                 & 15.4 mm            & 15.0 mm \\
            x-x'--emittance (4-rms, normalized) & 0.54 $\pi$-mm-mrad & 0.36 $\pi$-mm-mrad\\
            y-y'--emittance (4-rms, normalized) & 0.59 $\pi$-mm-mrad & 0.39 $\pi$-mm-mrad\\
            \hline
		\end{tabular}
\end{table}

\clearpage
\subsection{Comparison of WARP LEBT simulations with BCS results \label{marker2}}

\subsubsection{Simulations of Configuration I}\label{sec:configuration_I}
    
\begin{figure}[!t]
	\centering
	\includegraphics[width=0.95\textwidth]{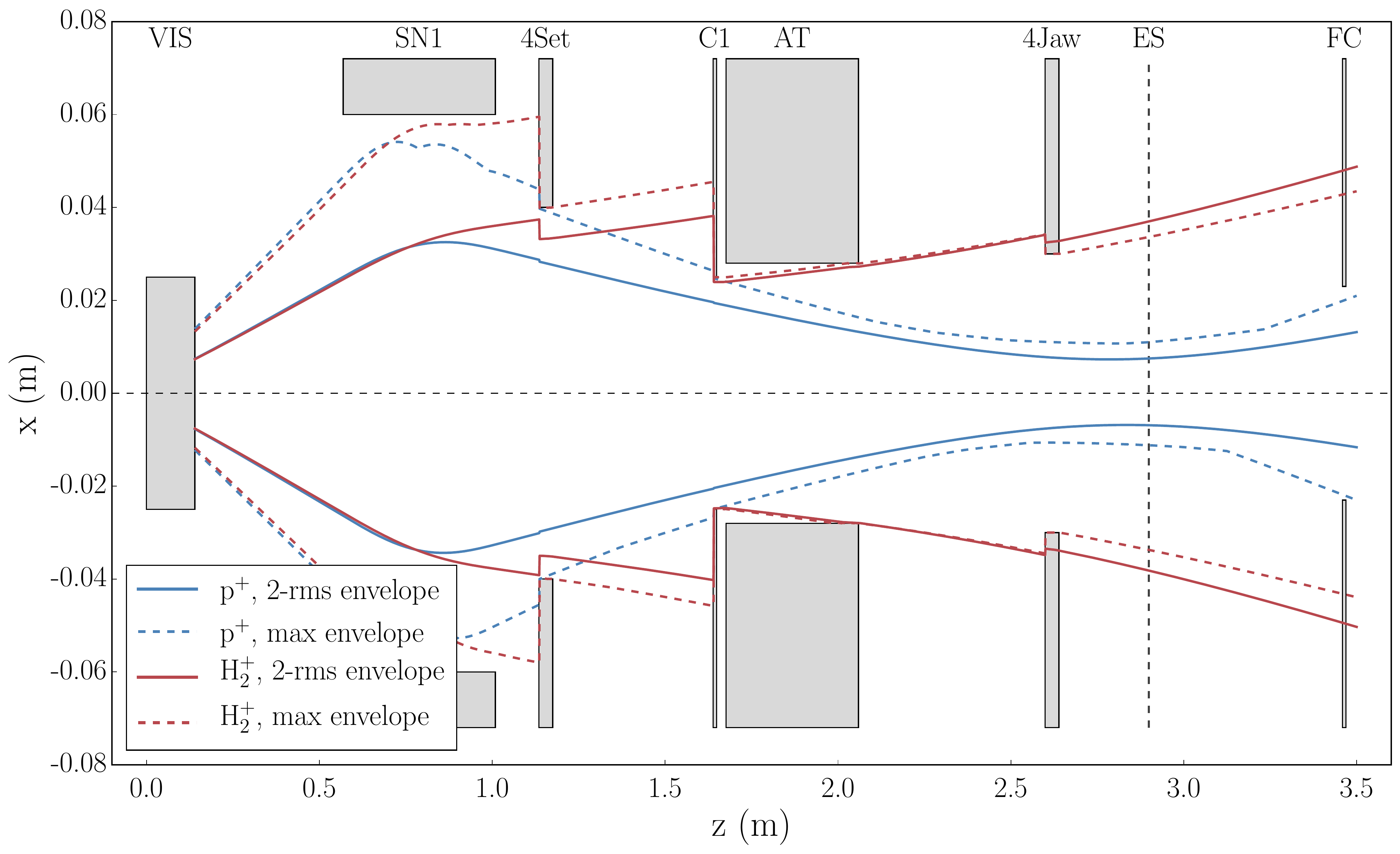}
	\caption{Horizontal beam envelopes of configuration I: 230 A on SN1.
             Protons are focused at the location of the 4-Jaw slits. Beam line 
             elements are: ion source (VIS), solenoid 1 (SN1), segmented
             collimator (4Set), plain collimator (C1), small-ID adapter tube 
             (AT), movable slits (4Jaw), emittance scanner (ES), Faraday cup 
             (FC).}
	\label{fig:SN1_230A_Env_horz}
\end{figure}

\begin{figure}[!t]
	\centering
	\includegraphics[width=0.95\textwidth]{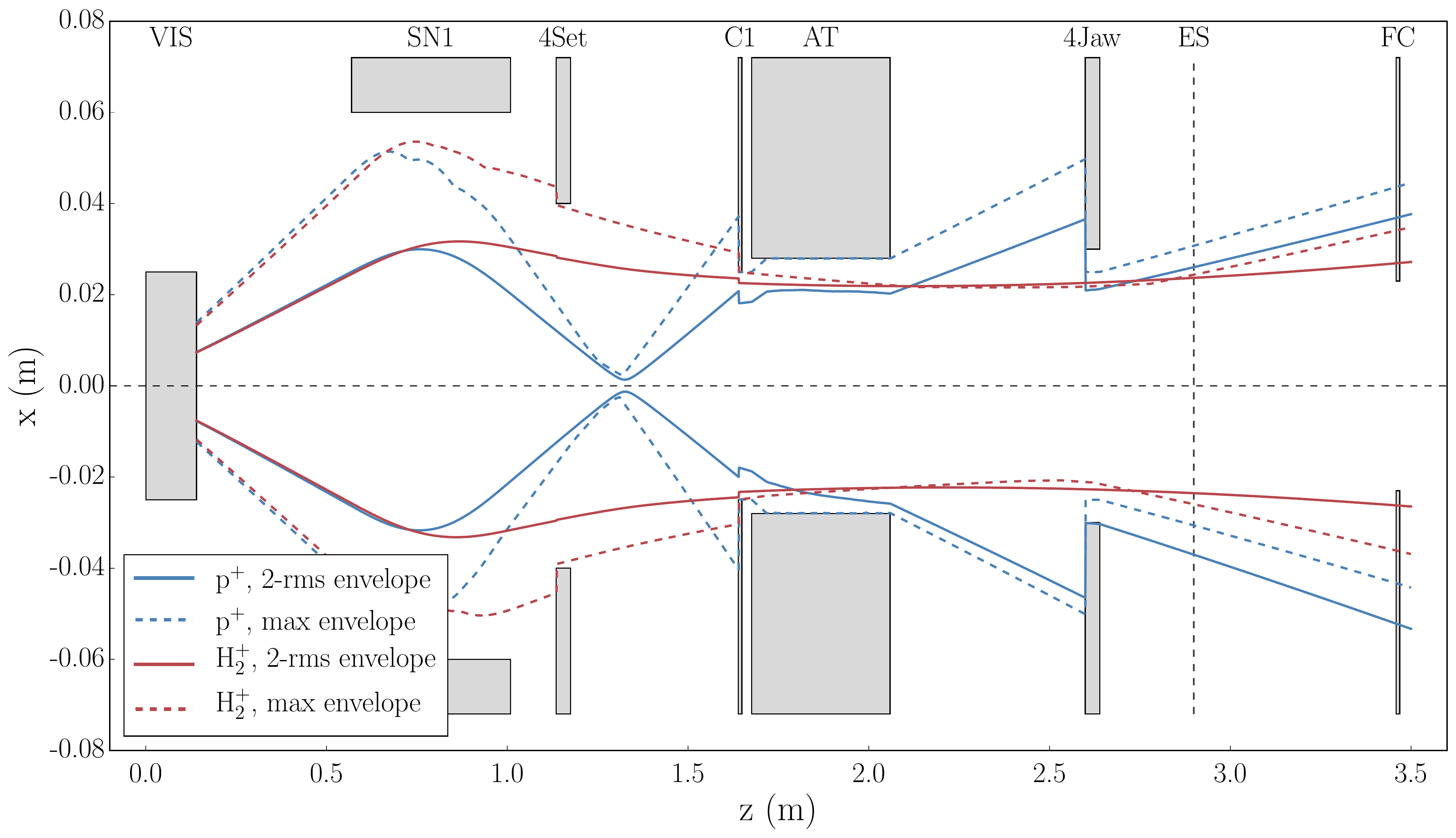}
	\caption{Horizontal beam envelopes of Configuration I: 320 A on SN1.
             \htp is focused at the location of the 4-Jaw slits. A strong
             proton focus at 1.2 m can be seen. This leads to the formation of
             the hollow beam. Beam line elements are listed in Figure 
             \protect\ref{fig:SN1_230A_Env_horz}.}
	\label{fig:SN1_320A_Env_horz}
\end{figure}

\begin{figure}[!t]
    \centering
    \begin{minipage}{.45\linewidth}
        \includegraphics[width=\linewidth]{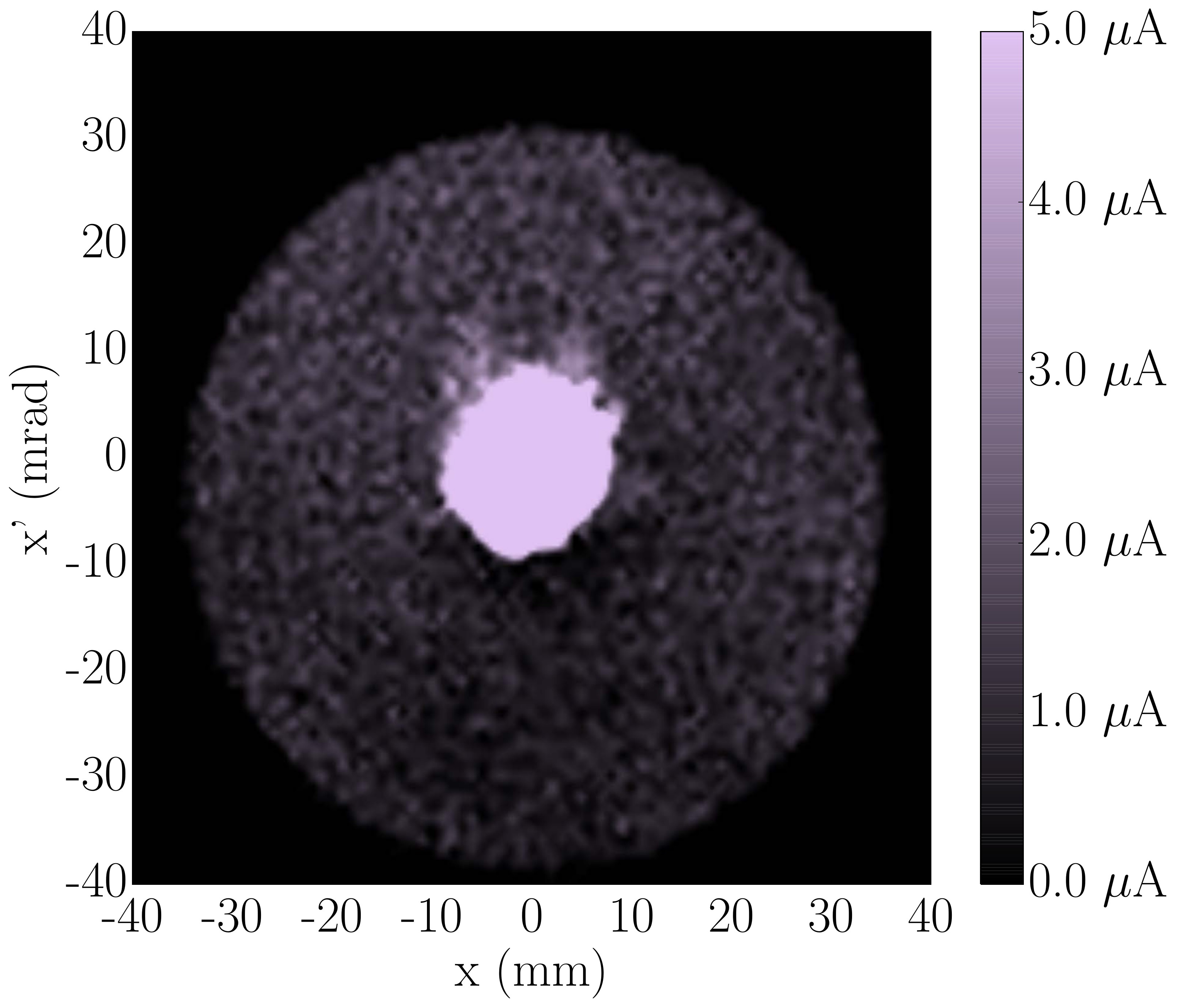}
    \end{minipage}
    \hspace{.05\linewidth}
    \begin{minipage}{.45\linewidth}
        \includegraphics[width=\linewidth]{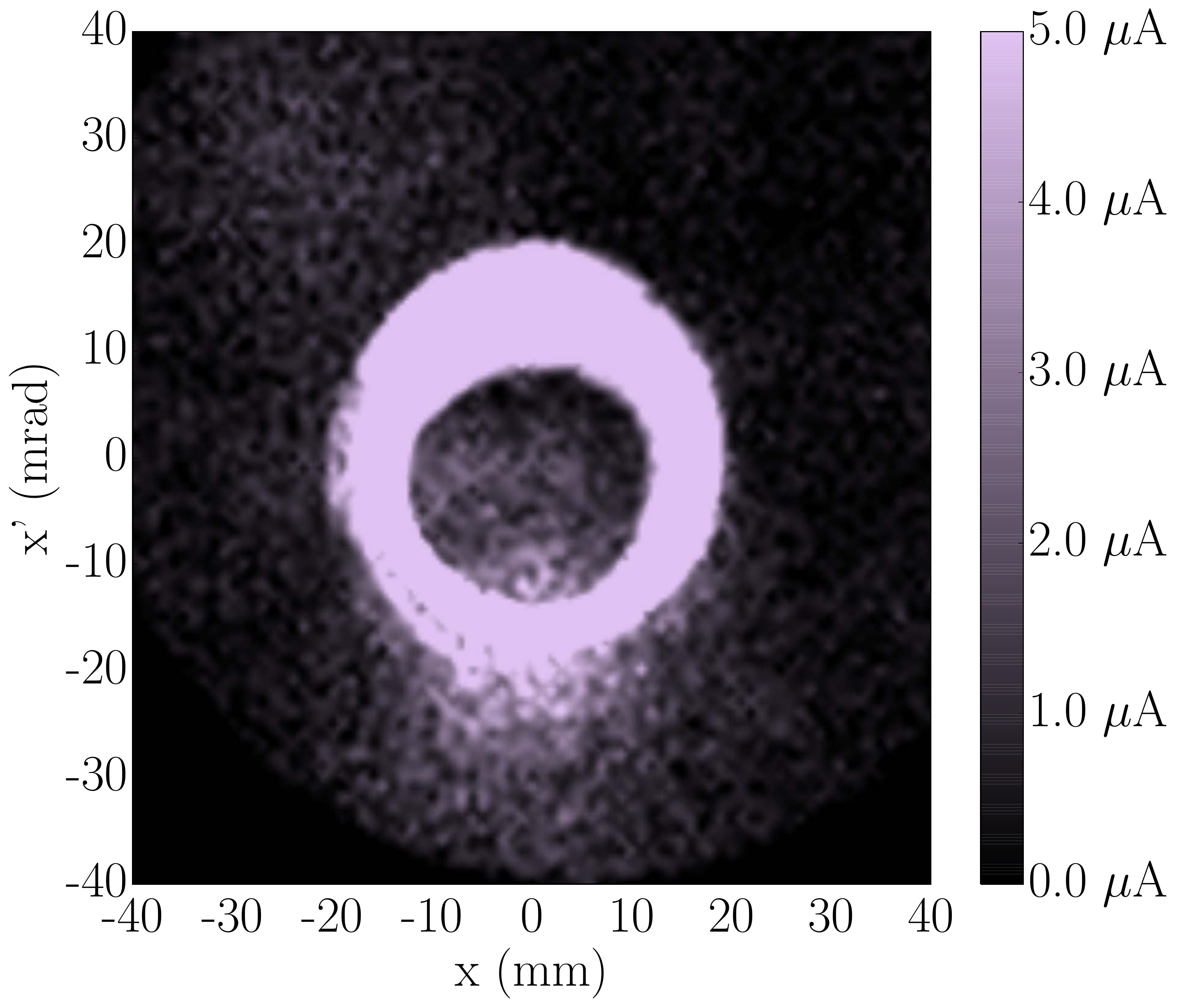}
    \end{minipage}
    \caption{Colored current density plots of the beam cross-sections at the 
             location of the 4-jaw slits (compare to Figure 
             \protect\ref{fig:hollow}). 
             Left: 240 A on SN1, 
             Right: 320 A on SN1. The ring structure of the hollow \htp beam is
             clearly visible on the right.}
    \label{fig:SN1_320A_4Jaw_rastered}
\end{figure}

\begin{figure}[!b]
    \centering
    \begin{minipage}{.45\linewidth}
        \includegraphics[width=\linewidth]{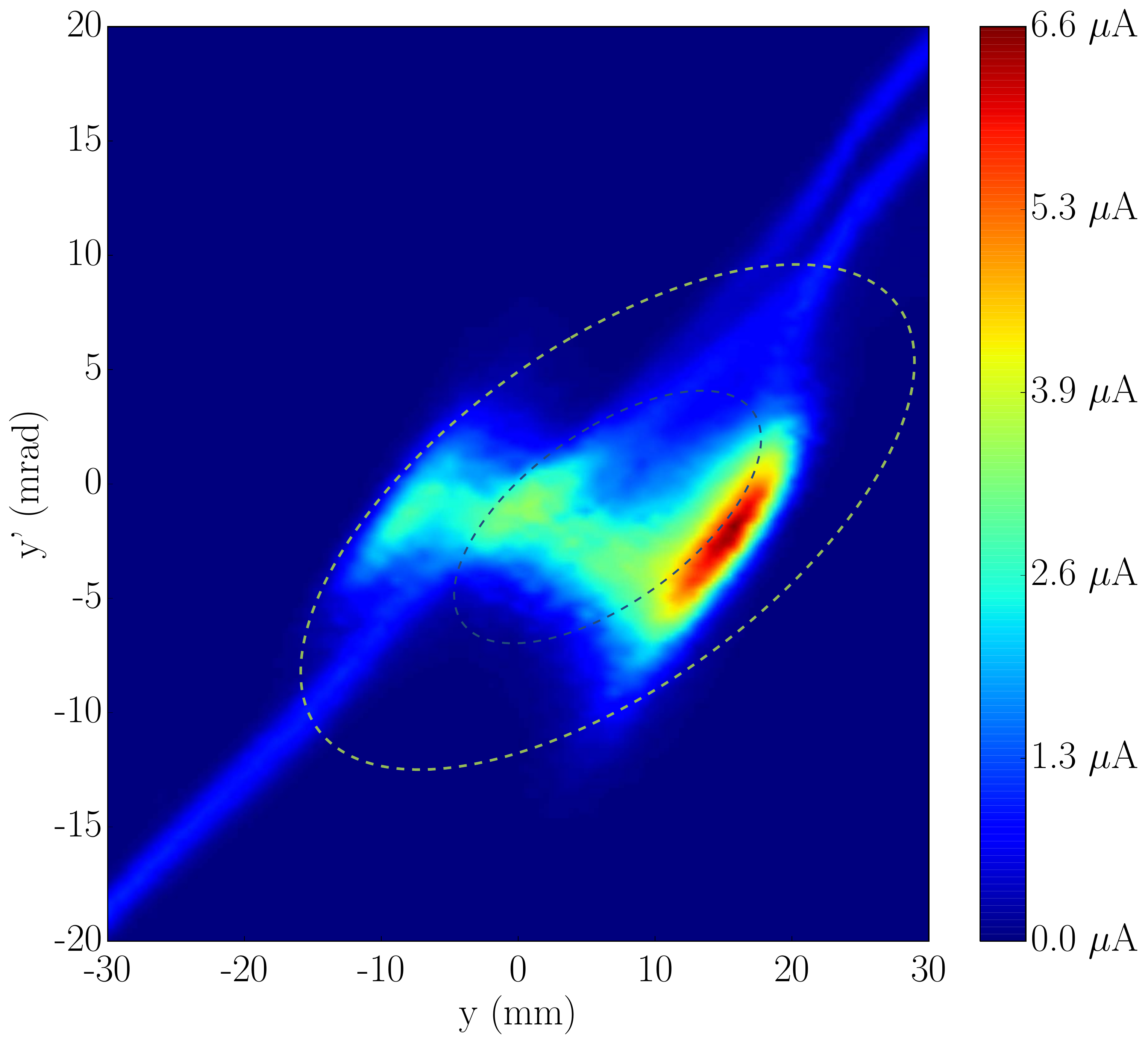}
    \end{minipage}
    \hspace{.05\linewidth}
    \begin{minipage}{.45\linewidth}
        \includegraphics[width=\linewidth]{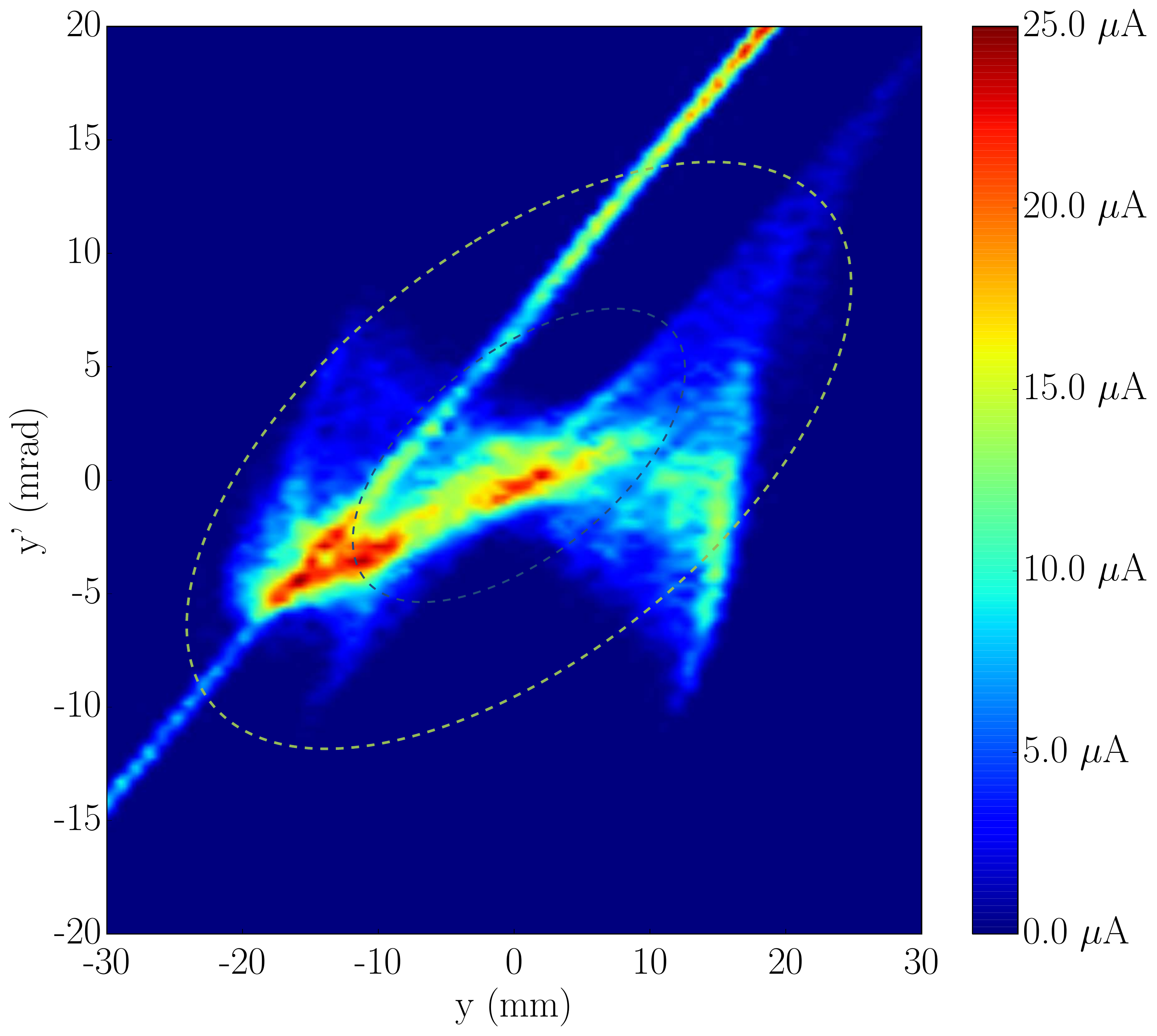}
    \end{minipage}
    \caption{Comparison of measured horizontal phase space (left) and current 
             density plot of simulated horizontal phase space (right) with
             SN1 at 320 A. The 'N'-shape is typical for a hollow beam
             and leads to significantly increased emittance. The beam energy 
             is 55 keV in this case. The different intensity scales arise from 
             the fact that the entrance slit of the emittance scanner was 
             smaller than the step size.}
    \label{fig:SN1_320A_phasespaces}
\end{figure}

\hspace{4ex} The first part of the simulation study was Configuration I, which included the
beam line up to Position I, where a Faraday cup was located. This was
also the first experimental configuration and direct comparisons of beam 
currents, emittance scans, and beam cross-sections are shown in this Section.
The solenoid 1 (SN1) power supply was limited to a current of 350 A and the
beam energy (ion source high voltage) was reduced to 55 keV in most experiments 
to be able to bring the \htp beam to a focus further upstream. Hence the 
simulations were performed at the same reduced energy.
The first experimental study was to increase the current of SN1 to
focus first the protons and then the \htp into the Faraday cup. The result of
this scan, together with the simulated values, was shown in Figure 
\ref{fig:fc0_vs_sol1}. The ratio of protons and \htp in the simulation was
obtained from the measured spectrum. Usually, the drain current on the source
high voltage platform power supply is a good estimate of the total extracted 
current. In the simulation, the measured total extracted current had to be reduced from 29.6 mA to 20 mA in order to get good agreement. The reason 
for this is unclear at this point. Possible explanations are a erroneous 
reading of the drain current (caused by secondary electrons), a larger than 
simulated beam halo, scraped at the 4-sector collimator (4Set), losses through
residual gas interaction, or a combination of underestimated divergence of the
initial beam  and overestimated space charge compensation in the simulation.
Running the simulation with the reduced beam current yielded good agreement 
with the solenoid scan and also with the emittance scans, so it stands to
reason that the results of this section are a good starting point for the 
subsequent simulations presented in the next section. The beam envelopes
corresponding to the proton peak and the \htp peak in Figure
\ref{fig:fc0_vs_sol1} are shown in Figures \ref{fig:SN1_230A_Env_horz} and
\ref{fig:SN1_320A_Env_horz}, respectively.
\begin{table}[!b]
	\caption{Simulated diameters and emittances for protons and 
             \htp 3.5 m downstream of the ion source (location of the
             Faraday cup). Values were obtained at the respective minimum
             diameters (SN1 at 220 A for protons and at 340 A for
             \htp.}
	\label{tab:particles_conf1}
	\centering
    \vspace{5pt}
    \renewcommand{\arraystretch}{1.25}
		\begin{tabular}{lll}
            \hline
                                               & \textbf{protons}   & $\mathbf{\mathrm{H}_2^+}$ \\
            \hline \hline
            Energy                              & 55 keV             & 55 keV \\
            x--diameter (2-rms)                 & 21.2 mm            & 44.1 mm \\
            y--diameter (2-rms)                 & 21.2 mm            & 40.1 mm \\
            x-x'--emittance (4-rms, normalized) & 0.63 $\pi$-mm-mrad & 1.24 $\pi$-mm-mrad\\
            y-y'--emittance (4-rms, normalized) & 0.66 $\pi$-mm-mrad & 1.14 $\pi$-mm-mrad\\
            \hline
		\end{tabular}
\end{table}
Approximately 2.6 m after the source are the 4-jaw slits (4Jaw) which were
completely closed at one point, to obtain photographs of the beam 
cross-section (cf. Figure \ref{fig:hollow}). A comparison can be made with the
simulations by creating x-y density plots from the particle distributions at 
the same location in the two simulation sets for SN1 = 230 A and SN1 = 320A. This
is shown in Figure \ref{fig:SN1_320A_4Jaw_rastered}. Both images agree well with
the photographs in both size and shape of the distributions (note the ring shape
of the \htp beam in the right plot), except for one detail: the \htp halo
around the proton beam in the left photo of Figure \ref{fig:hollow} already 
forms a ring even before the protons come to a tight focus upstream of the 4-Jaw 
slits. This is a hint that space charge compensation could be overestimated in 
the simulation. 
The diameters and emittances of the proton and \htp focus in the Faraday cup (FC)
are listed in
Table \ref{tab:particles_conf1}. While the proton emittances increase by 
17\% and 12\% (x-x' and y-y', respectively), the \htp emittances increase by 
244\% and 192\% (x-x' and y-y', respectively). This is due to the effect that the
protons have on the \htp ions. The severe over-focusing (seen in Figure 
\ref{fig:SN1_320A_Env_horz}) hollows out the beam and leads to a tripling of 
the emittance. This has been seen in the measurements as well.
A comparison of the horizontal phase spaces (measured and from the simulation) is shown in figure \ref{fig:SN1_320A_phasespaces}. Both plots exhibit similar size and structure. The proton intensity in the simulation seems to be higher
and the asymmetry of the \htp beam shifted left rather than right, which we
attribute to steering effects. Overall, the agreement is good.

\clearpage
\subsubsection{Simulations of Configuration II}
\label{sec:simulation_of_configuration_II}

\begin{figure}[!t]
	\centering
	\includegraphics[width=0.95\textwidth]{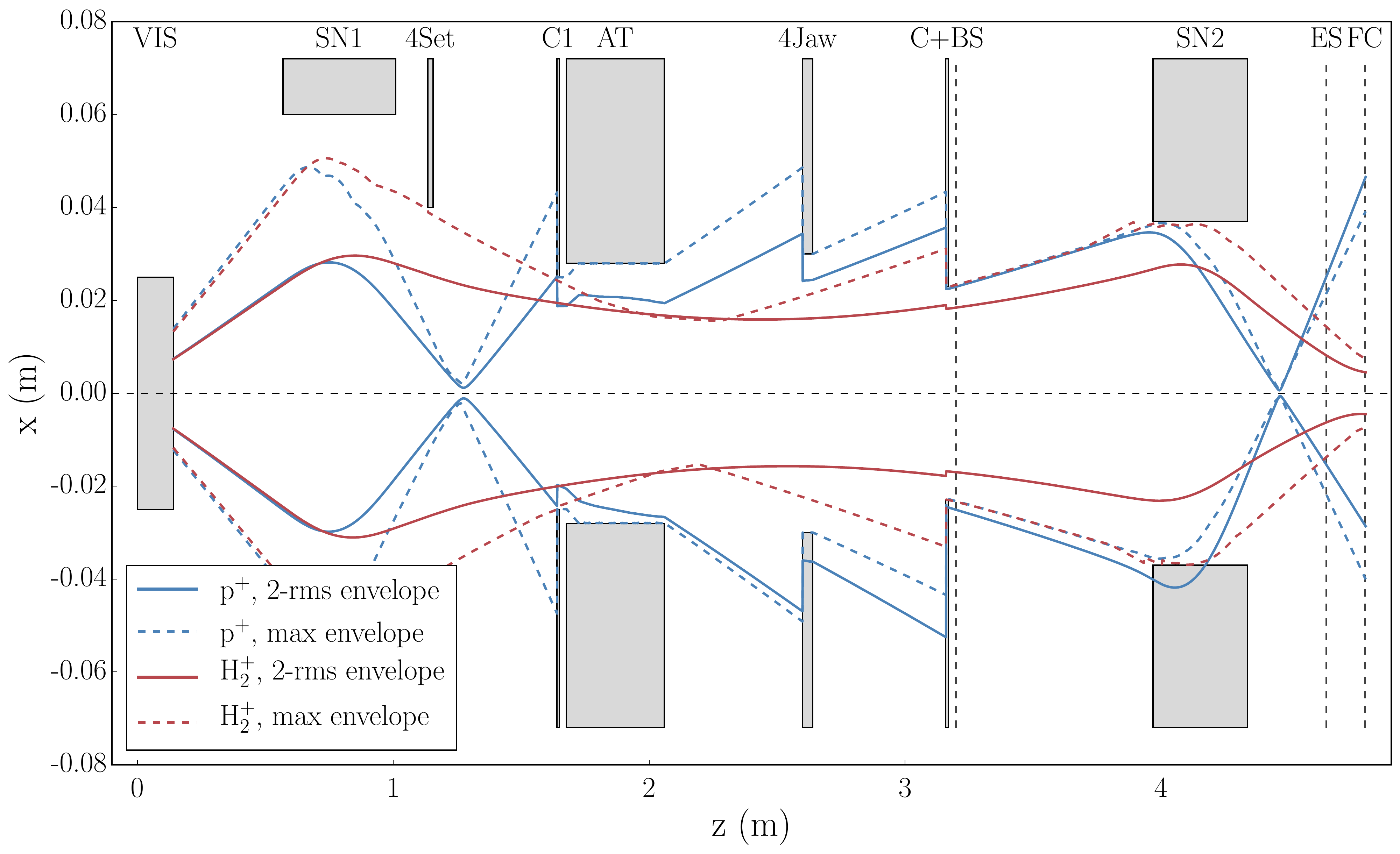}
	\caption{Horizontal beam envelopes of configuration II: 353 A on SN1,
             230 A on SN2.
             \htp is focused at the location of the Faraday cup at the end of
             the beam line which roughly coincides with the entrance aperture 
             of the spiral inflector. Beam line elements are: ion source (VIS),
             SN1, segmented collimator (4Set), plain collimator 
             (C1), small-ID adapter tube (AT), movable slits (4Jaw), 
             pneumatic beam stop + fixed collimator (C+BS), SN2,
             emittance scanner (ES), Faraday cup (FC).}
	\label{fig:SN2_230A_Env_horz}
\end{figure}

\begin{figure}[!hp]
\centering
\begin{minipage}{.45\linewidth}
  \includegraphics[width=\linewidth]{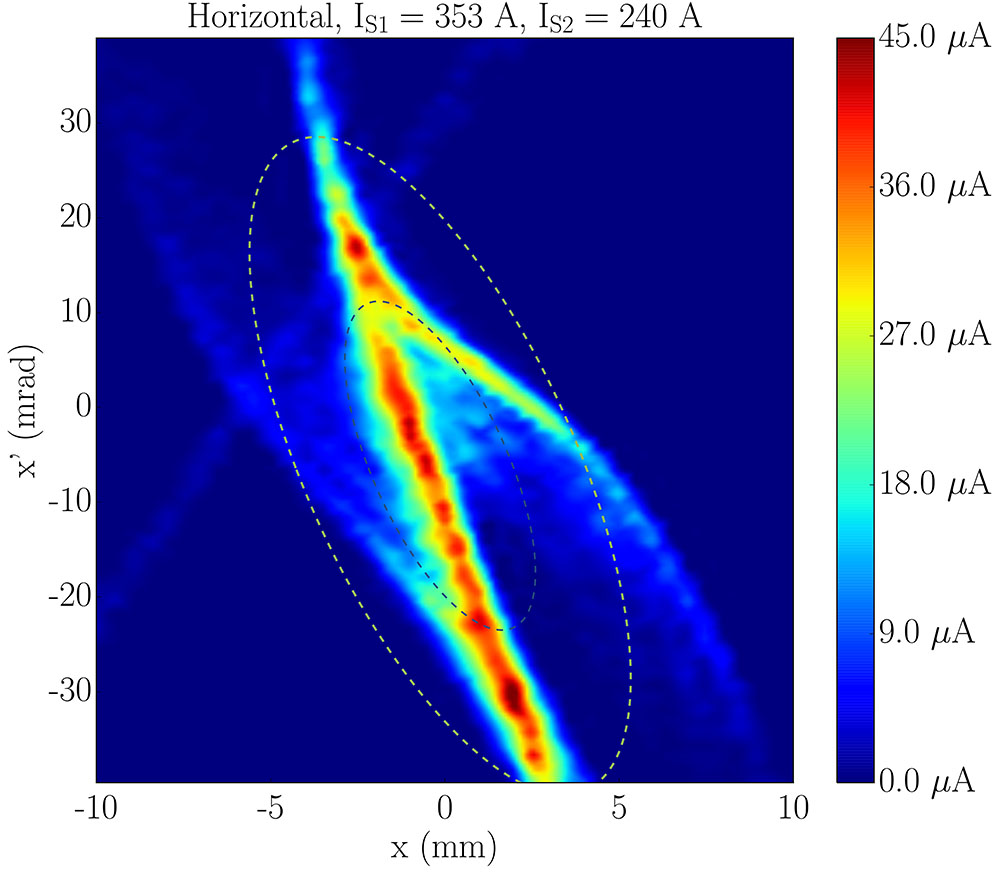}
\end{minipage}
\hspace{.05\linewidth}
\begin{minipage}{.45\linewidth}
  \includegraphics[width=\linewidth]{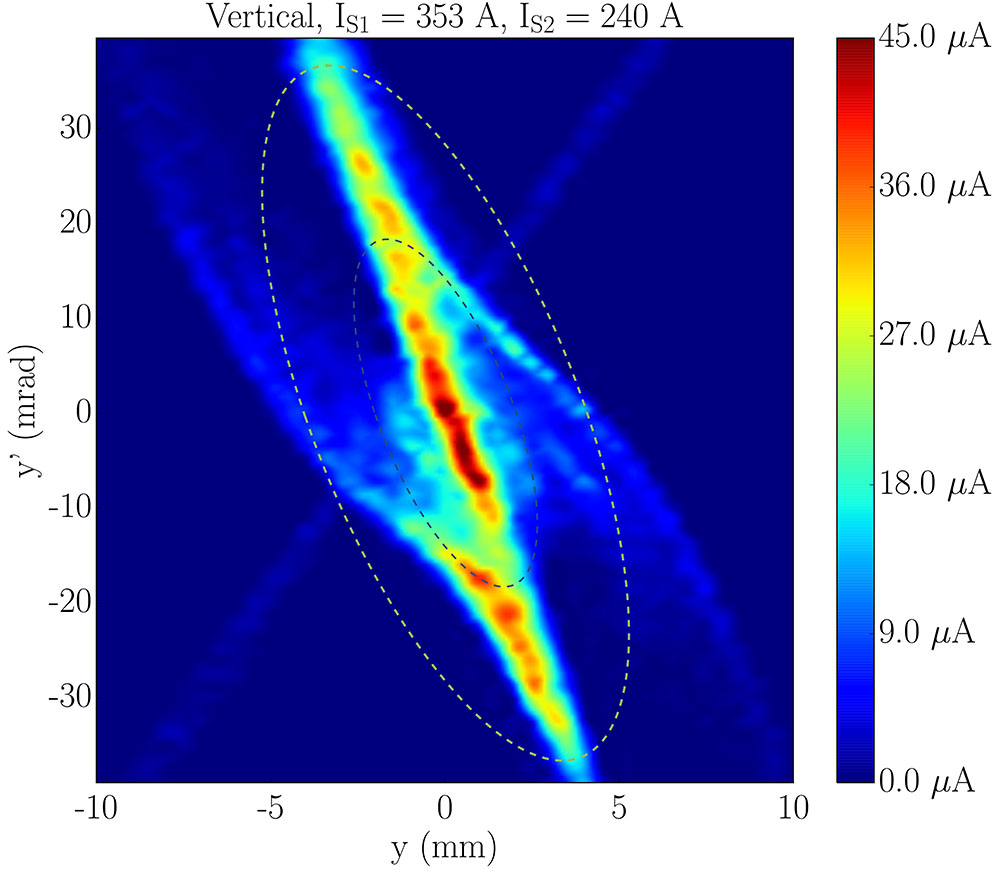}
\end{minipage}
\begin{minipage}{.45\linewidth}
  \includegraphics[width=\linewidth]{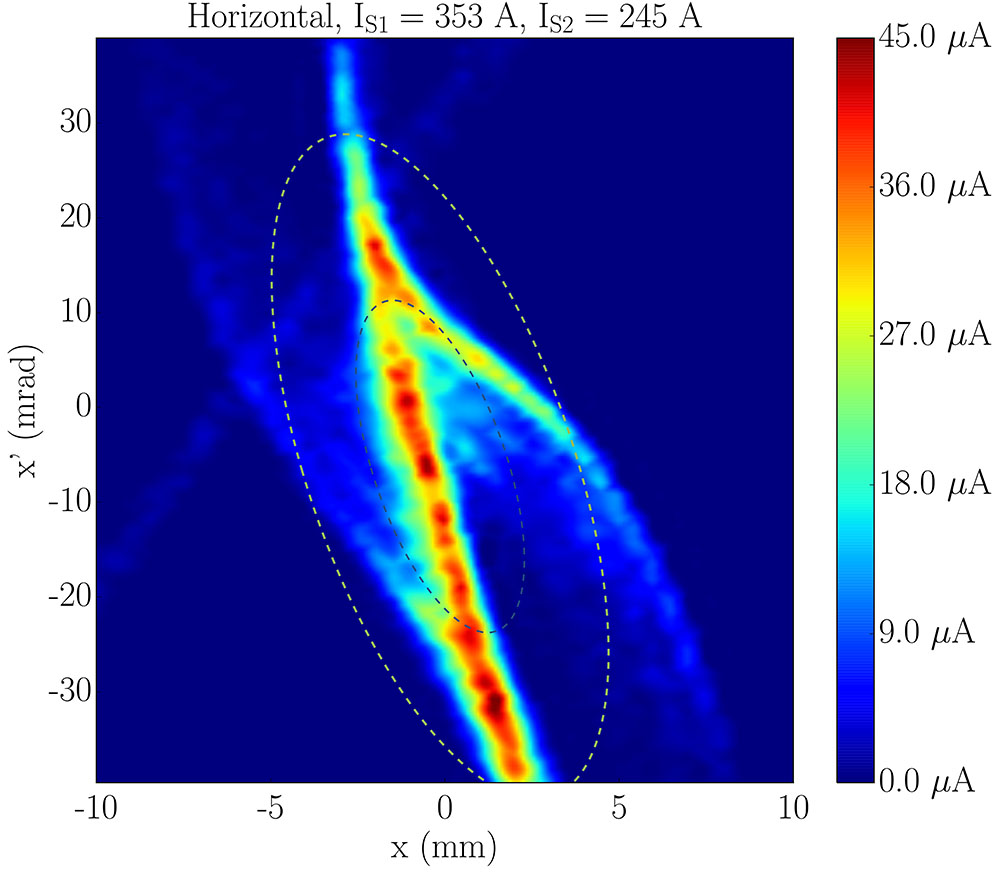}
\end{minipage}
\hspace{.05\linewidth}
\begin{minipage}{.45\linewidth}
  \includegraphics[width=\linewidth]{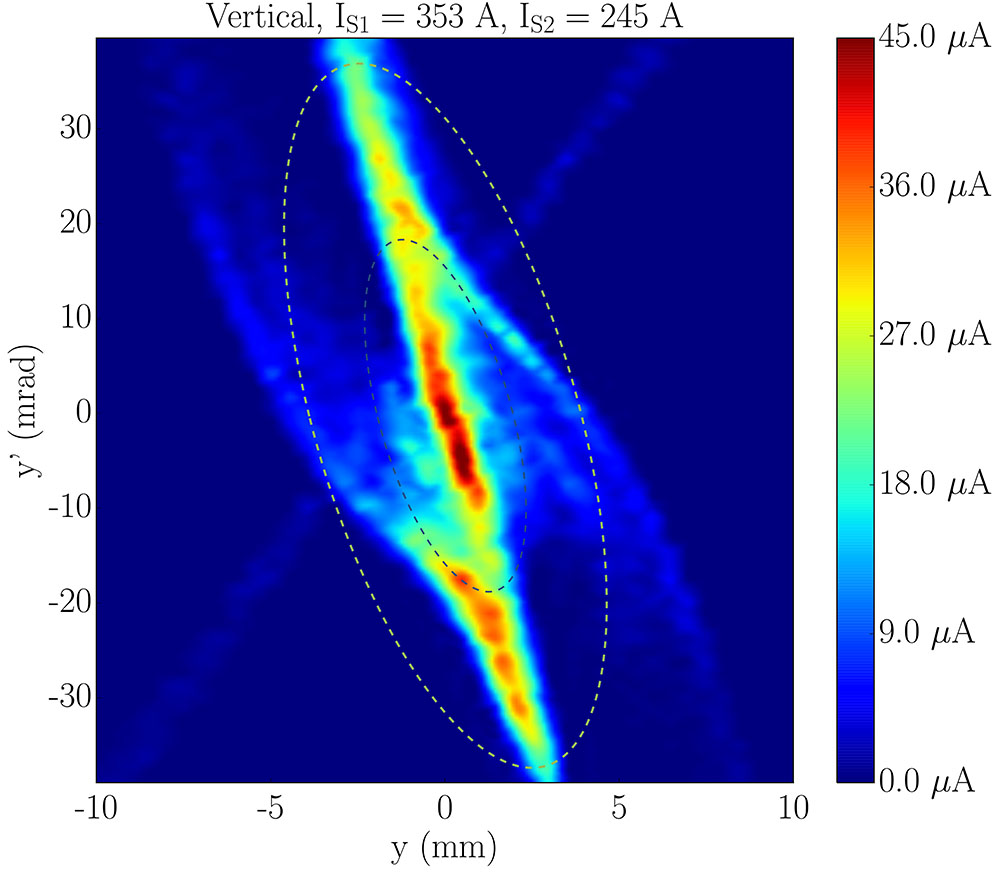}
\end{minipage}
\begin{minipage}{.45\linewidth}
  \includegraphics[width=\linewidth]{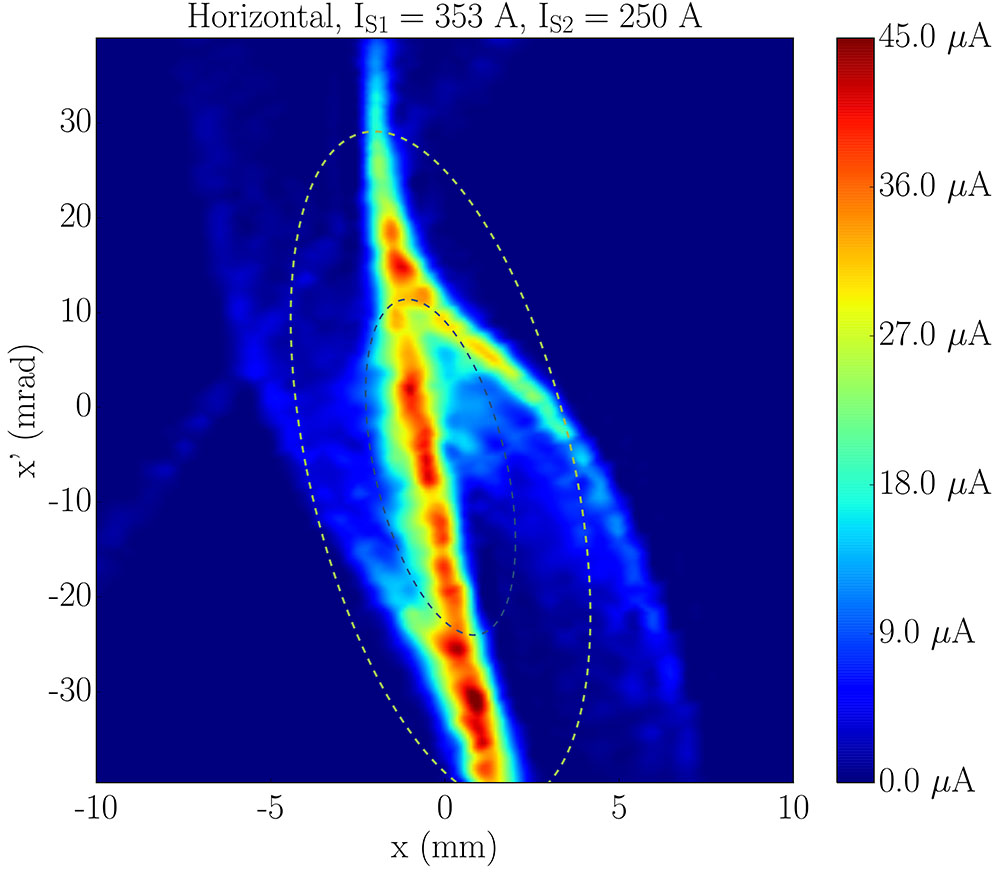}
\end{minipage}
\hspace{.05\linewidth}
\begin{minipage}{.45\linewidth}
  \includegraphics[width=\linewidth]{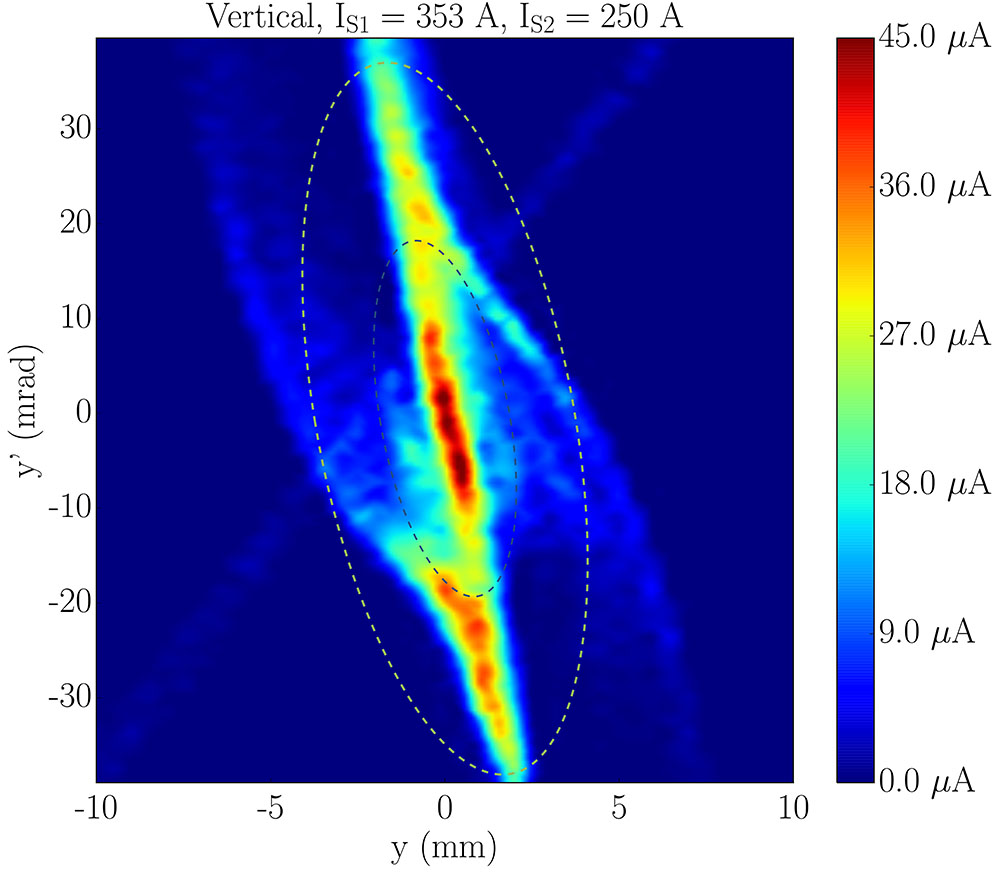}
\end{minipage}
\caption{Current density plots of simulated phase spaces (compare to Figure 
         \protect\ref{fig:phase_spaces2}).}
    \label{fig:phase_spaces_sim}
\end{figure}

\hspace{4ex} Here we present the results of the simulations of the full LEBT guiding the beam
to the cyclotron. The particle distributions at the end of these simulations 
were used as initial conditions for the spiral inflector simulations reported in
Section \ref{sec:opera}. A typical horizontal beam envelope is shown in Figure
\ref{fig:SN2_230A_Env_horz}. As in Configuration I, the tight focal point of 
protons early on can be seen, which leads to strong aberrations in the \htp
beam. Simulated phase spaces at the location of the emittance scanner
("`ES"' in Figure \ref{fig:SN2_230A_Env_horz}) are shown
in Figure \ref{fig:phase_spaces_sim} which should be compared to the measured
phase spaces of Figure \ref{fig:phase_spaces2}. Good qualitative agreement can be seen. Protons are faint streaks in the 
back, while the \htp can be focused to below 11 mm 2-rms diameter. The 
diameters and emittances of the \htp beam for different settings of 
SN2 are compared to the measured values in Figure \ref{fig:emittance_diameter}. It should be noted that, in the measured phase spaces, part of the beam is outside of the axes limits and thus the calculated values are underestimated. In Figures \ref{fig:emittance_diameter} and \ref{fig:phase_spaces_sim} the same limits were applied to the simulation results leading to good agreement. When using the full untruncated particle
distributions to calculate diameters and emittances from the simulations, the
diameters remain largely unchanged, while the emittances increase. The average
increase of emittances from truncated to untruncated phase space is 
$\approx 17.3\%$ for the horizontal phase spaces and $\approx15.4\%$ for the
vertical ones, with a maximum of 35.6\% and 31.4\% for SN2 at 200 A (largest beam). This is treated as a systematic error and is included in the 
errorbars in Figure \ref{fig:emittance_diameter}.

\clearpage
\begin{figure}[!t]
\centering
\begin{minipage}{.45\linewidth}
  \includegraphics[width=\linewidth]{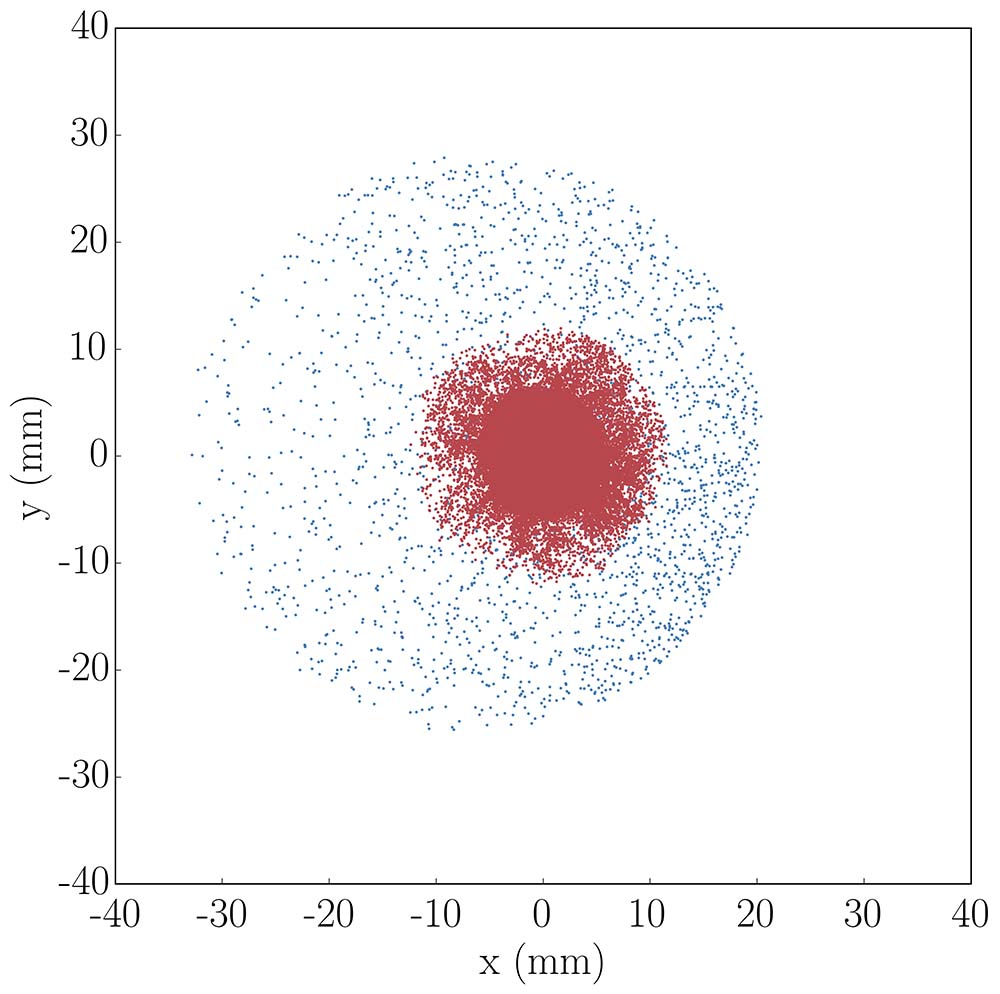}
\end{minipage}
\hspace{.05\linewidth}
\begin{minipage}{.45\linewidth}
  \includegraphics[width=\linewidth]{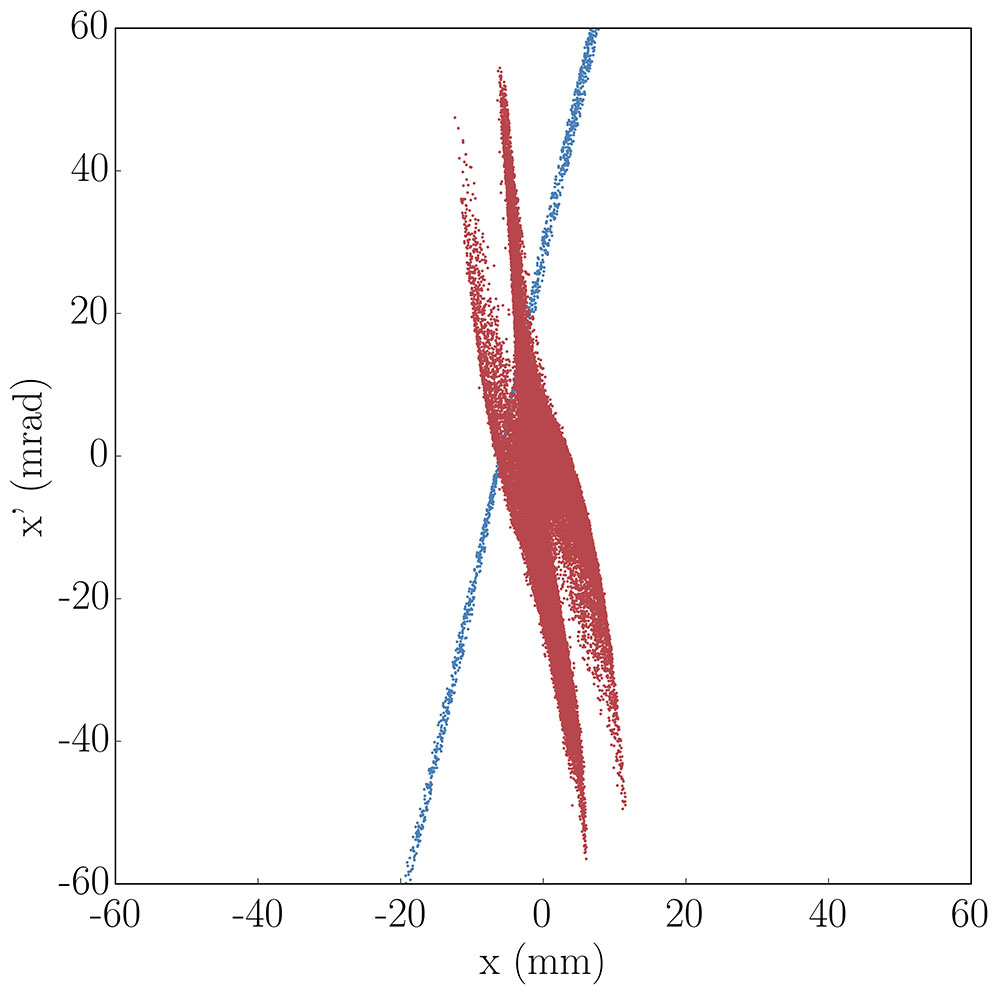}
\end{minipage}
\caption{The final particle distributions obtained by the WARP LEBT simulations,
         Left: Final beam cross-section. Right: Final x-x' phase space. Red: 
         \htp, blue: protons.}
\label{fig:particles_final}
\end{figure}

Using the LEBT settings (solenoid currents and 4-Jaw slit openings) used during
the cyclotron injection tests in the Configuration II simulation, we obtain
realistic particle distributions for further simulation studies of the spiral
inflector transmission and RF capture in the first three turns of the cyclotron.
The cross-section and horizontal phase space are shown in Figure 
\ref{fig:particles_final} and the important beam parameters are listed in 
Table \ref{tab:particles_final}.
Results of OPERA ray-tracing calculations using these distributions are reported
in the following section. 

\begin{table}[!b]
\vspace{10pt}
	\caption{The final beam parameters for \htp at the 
             cyclotron entrance aperture for the beam injected during
             the tests at BCS in Vancouver.}
	\label{tab:particles_final}
	\centering
    \vspace{5pt}
    \renewcommand{\arraystretch}{1.25}
		\begin{tabular}{ll}
            \hline
                                                & $\mathbf{\mathrm{H}_2^+}$ \\
            \hline \hline
            Energy                              & 62.7 keV \\
            x--diameter (2-rms)                  & 10.6 mm \\
            y--diameter (2-rms)                  & 10.2 mm \\
            x-x'--emittance (4-rms, normalized)  & 1.19 $\pi$-mm-mrad\\
            y-y'--emittance (4-rms, normalized)  & 1.16 $\pi$-mm-mrad\\
            4-rms includes                      & 85\%\\
            \hline
		\end{tabular}
\end{table}

\clearpage
\subsection{Comparison of injection measurements with OPERA simulations \label{sec:opera}}

\begin{figure}[!t]
	\centering
		\includegraphics[width=0.45\textwidth]{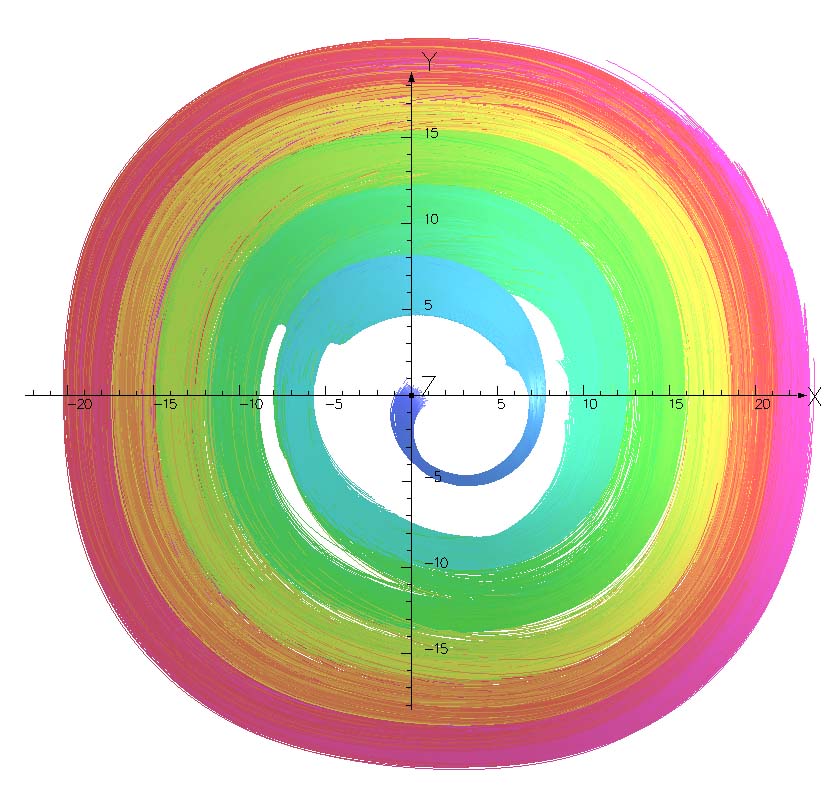}
		\hspace{0.05\textwidth}
		\includegraphics[width=0.45\textwidth]{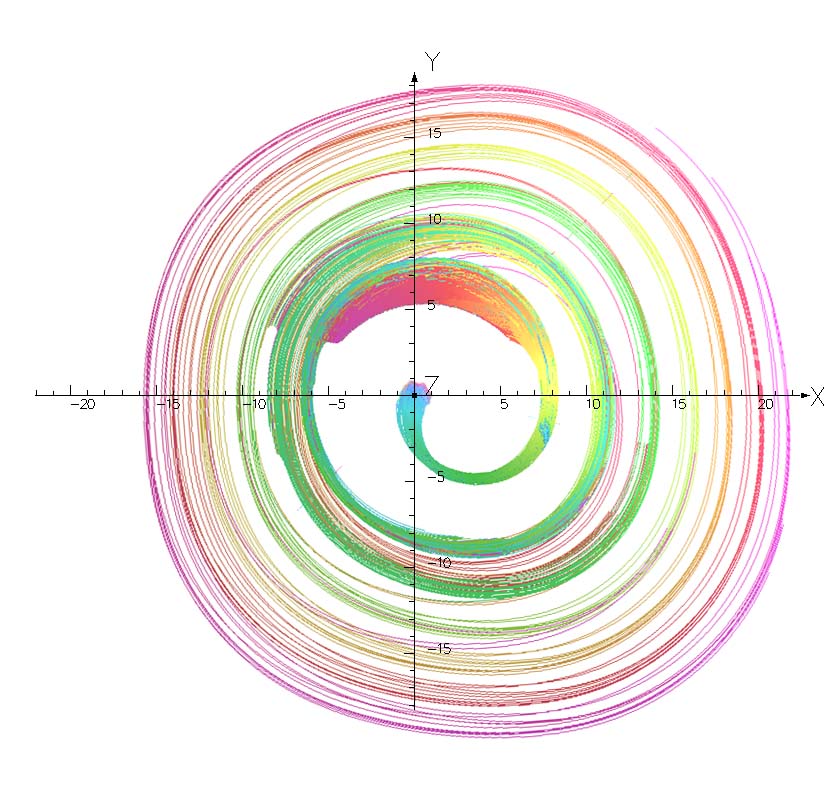}
		\caption{VectorFields OPERA ray-tracing results with different 
		         dee voltages.
		         Left: $\mathrm{V}_{\mathrm{dee}}=70$ kV (nominal value).
		         Right: $\mathrm{V}_{\mathrm{dee}}=45$ kV (minimum simulated).
		         It becomes immediately clear that even perfectly in phase, the
		         low dee voltage is not sufficient to capture the particles and
		         most of them are lost in the first turn. Coloring of the trajectories
		         is loosely based on time-of-flight, but should be considered as a means to 
		         guide the eye only.
		         \label{fig:opera_trajectories}}
\end{figure}

\begin{table}[!b]
	\caption{Parameters for inflection and acceleration simulations using OPERA.}
	\label{tab:opera_parameters}
	\centering
    \vspace{5pt}
    \renewcommand{\arraystretch}{1.25}
		\begin{tabular}{lll}
            \hline
            \textbf{Parameter}                       & \textbf{Value}\\
            \hline \hline
            Species                                  & \htp \\
            Initial Beam Energy                      & 62.7 keV \\
            Solenoids 1 / 2                          & 350 A / 233 A \\
            Cyclotron Magnet                         & 218.6 A \\
            Spiral Inflector Upper / Lower Electrode & -10.0 kV / +10.15 kV \\
            Dee voltage                              & $45-70$ kV \\
            \hline
		\end{tabular}
\end{table}

\hspace{4ex} As mentioned in Section \ref{sec:spiral_inflector}, VectorFields OPERA was used for
the design of the spiral inflector and the central region of the BCS test cyclotron.
OPERA uses a finite elements method to calculate electrostatic and magnetostatic
fields and has a ray-tracing module to track ions through the simulation space.
In this section, we present a set of ray-tracing calculations of a 62.7 keV \htp
beam entering the cyclotron through the spiral inflector and being accelerated for
3.5 turns. The simulation parameters are listed in Table \ref{tab:opera_parameters} 
and are the same as used during the measurements reported in Section
\ref{sec:cyclotron_injection}. 
The initial particle distribution for the OPERA simulation was obtained from the 
WARP LEBT simulation discussed in the previous section. A unbunched beam was assumed 
and a phase resolution of 5 degrees with 500 macro-particles per phase was used 
during the OPERA calculation.
A variety of 6 different dee  voltages $\mathrm{V}_{\mathrm{dee}}$ was used in order 
to investigate the effect of insufficient dee voltage on the transmission (as 
encountered during the experiment).
Figure \ref{fig:opera_trajectories} shows one sample of trajectories each for the 
cases of $\mathrm{V}_{\mathrm{dee}}=45$ kV and $\mathrm{V}_{\mathrm{dee}}=70$ kV,
perfectly in phase with the cyclotron RF.

\begin{figure}[!t]
	\centering
	\includegraphics[width=0.8\textwidth]{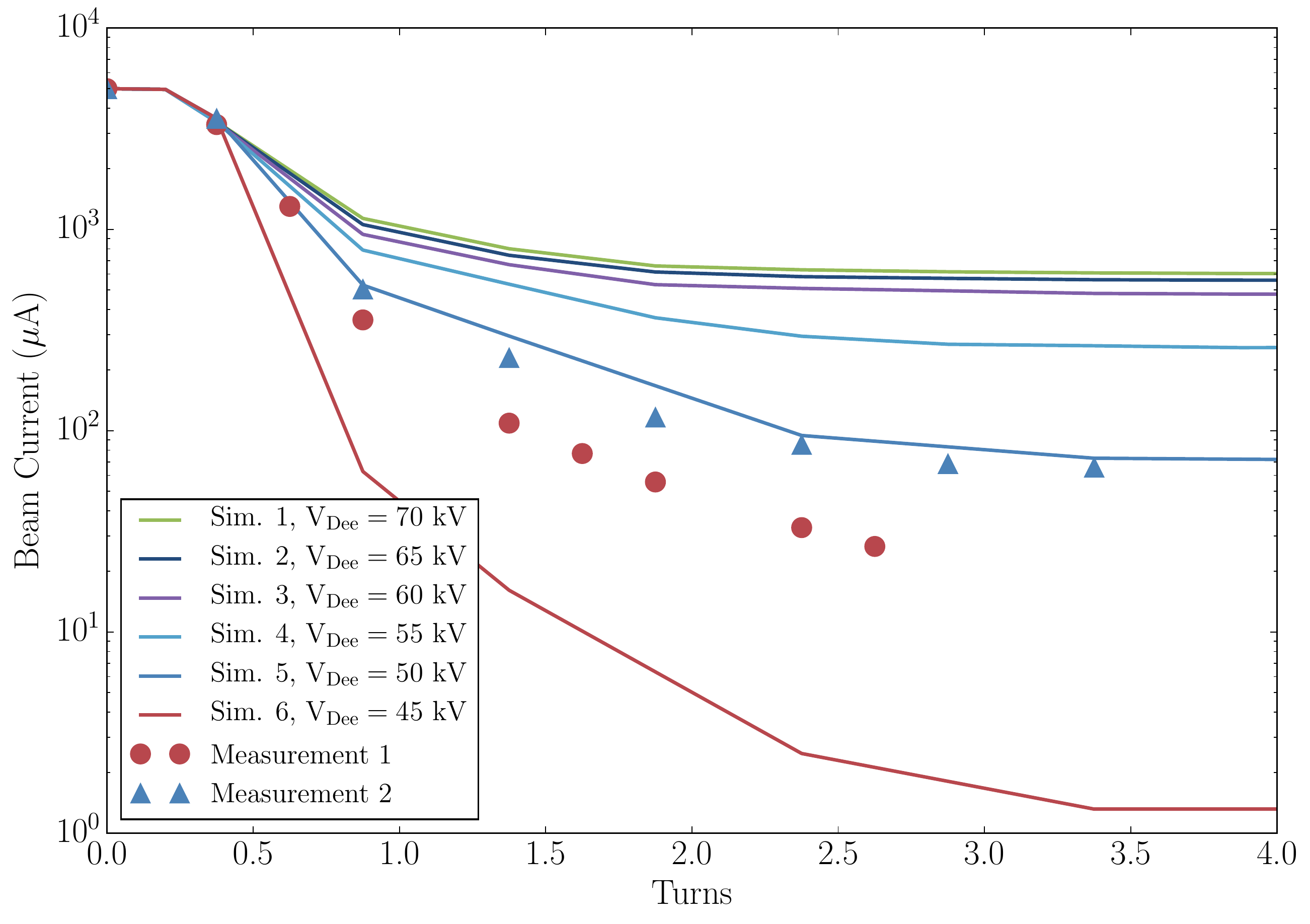}
	\caption{VectorFields OPERA calculation of transmission through the central
	         region of the cyclotron. The counting of completed turns starts at 
	         the exit of the spiral inflector. As comparison, the measured
	         transmission for reduced dee voltages of are also plotted.
	         A clear reduction in transmission can be seen for lower dee
	         voltages.}
	\label{fig:OPERA_result}
\end{figure}

The total transmission through the first turns of the cyclotron for different voltages 
applied to the dees can be seen in Figure \ref{fig:OPERA_result} as well as the 
two measured curves. 
The initial current of the simulations was scaled to 5 mA which is the value assumed
at the exit of the spiral inflector during the measurement.
It is not surprising, that the transmission decreases for dee voltages below the design value of 70 kV. $\mathrm{V}_{\mathrm{dee}}=50$ kV agrees reasonably well with the 
better of the two measured curves. The average dee voltages during the measurements determined by the pick-up probe and from beam radii (radial probe positions) were between 50 and 60 kV. Based on the large variation of the 
pick-up probe signal during the measurements, it is reasonable to assume that the average dee voltage of $48-50$ kV giving good agreement with simulations is what could be achieved with the given experimental setup.
Any discrepancy would be due to the high uncertainties
in the pick-up probe values and the unstable dee voltage. The calculation
of beam energy from the mean beam radii (radial probe positions) and magnetic field data (cf. Table \ref{tab:injection_results}) suggests similar values. 

Space charge effects, which could lead to additional reduction in transmission
during the first turn, were not included in the OPERA calculations.
This will be investigated in the future by using the particle-in-cell (PIC) code OPAL 
\cite{adelmann:opal}, which has a dedicated cyclotron module that has recently been
upgraded to include spiral inflectors.

\clearpage
\section{Conclusion}
\hspace{4ex} This technical report outlined the experimental results obtained at the Best Cyclotron System's test bench, exploring design concepts and hardware relevant to the IsoDAR injection system. In addition we have benchmarked the results against a number of detailed simulation studies. These experimental results represent the early stages of exploration into the development of the IsoDAR injection system and provided excellent information regarding working with high-current \htp beams. Some specific highlights of our measurements, described in this paper are:
\begin{enumerate}
\item The VIS source was able to produce good currents of \htp ions ($9.0\pm0.5$ mA); with the initial (large) plasma chamber the best ratio of \htp/protons was $>40\%$, obtained at microwave power just above the threshold for stable discharge operation (about 300 watts).  This current improved to 12.2 $\pm$ 0.6 mA using a smaller (2.5 cm dis x 10 cm long) plasma chamber, and at the same time the \htp/proton ratio improved to $>50\%$.  
Modeling and systematic measurements had suggested that smaller chambers would 
improve \htp production, due to wall-recombination effects, and this was indeed 
borne out by experiment. We expect further optimization of the chamber shape to yield 
yet improved performance.

\item Separation and elimination of the proton fraction from the \htp beam was attempted by using focusing solenoids instead of magnetic analysis with a bending dipole, because of constraints in the existing test bench at BCS.  This separation was successful in that very few protons were seen at the location of the test cyclotron, however as the separation process involved over-focusing the proton beam, which was then lost on the beam pipe walls, substantial heating of these walls was observed. In addition, the strong electric fields at the tight proton focus had significant disruptive effects on the \htp beam.  In particular, our studies showed that downstream of this proton focus the \htp beam was hollow: its central region had been blown out by the strong fields at the proton focus. We conclude that the use of a dipole magnet would mitigate such effects and improve overall transmission efficiency and \htp beam quality.

\item The importance of space-charge compensation in the transport line was clearly observed, with substantial increases in beam size visually seen when electric fields swept away the compensating electrons. 
Measurements with a retarding field analyzer were performed, and will be reported in a subsequent paper.

\item A well-characterized $6.2\pm0.4$ mA, with measured normalized 4-rms emittance of $1.15\pm0.09$ $\pi$-mm-mrad at the entrance to the spiral inflector was transmitted through the inflector with better than 90$\%$ efficiency. This validates both the design and fabrication of the spiral inflector.

\item Beam was captured and accelerated through 4 turns in the cyclotron, however with very high losses because of inadequate RF accelerating voltage. The problem of sparking across a critical insulator which limited the highest voltage obtainable, and unfortunately this insulator could not be redesigned before the end of our experimental run.
However VectorFields OPERA simulations did provide qualitative fits to the data observed.

\item A very successful set of simulation studies was able to reproduce quite well the 
observed measurements, including the pronounced emittance aberrations and hollow beams;  indicating that the beam dynamics in our transport line can be understood and modeled with these well-established simulation codes. These then are now well validated for 
the design of the injection line for our future \htp experiments.
\end{enumerate}

Building on the base of these measurements, we see several possible clear paths forward for designing the IsoDAR injection line:
\begin{itemize}
\item Ion source: The current generation of non-resonant microwave ECR sources, typified by the VIS we used, is probably capable of up to about 15 to 20 mA of \htp, when fully optimized.  This may be adequate for IsoDAR provided an efficient RF bunching system can be developed.  While it was our initial aim to test such a bunching system, the problems with the cyclotron RF system prevented the installation and running of the buncher that had been built.  If a bunching efficiency of 20 to 25$\%$ can be achieved, the 5 mA injected current requirement could be met. We will endeavor to find a way of conducting these tests in the future.

\item  Inflection:  A key result of our experiments has been the validation of the spiral inflector design for these more rigid, high current beams. It should be noted 
here that the final IsoDAR design will be a further improvement over the BCS test stand
model as it will operate in combination with 4 RF cavities in 4\textsuperscript{th} harmonic mode. This will allow for better vertical focusing and will improve the 
injection efficiency of the cyclotron.

\item  Alternate design:  A new LEBT concept incorporating a low-frequency RFQ (Radio Frequency Quadrupole) structure as a buncher might be an elegant, compact and efficient means of providing beam to the IsoDAR cyclotron.  Such a device \cite{hamm:rfq1}, operating at the cyclotron frequency and placed in the axial bore hole just above the center of the cyclotron could provide in principle bunching of up to 80 or 90$\%$, greatly relaxing the requirements on source performance. In addition, the injection line could be much reduced in size and length, which might also potentially lead to overall cost savings. Initial studies are underway \cite{winklehner:rfq1}.
\end{itemize}

\clearpage
\acknowledgments
\hspace{4ex} This work was supported by grant NSF-PHY-1148134 and MIT seed funding 
(PI: Prof. Janet Conrad). The authors would like to thank Best Cyclotrons 
Systems, Inc. for their support during the measurements.
Furthermore the authors would like to convey their gratitude for being able to
use the Reuse Virtual Cluster at LNS/MIT for the WARP simulation work and 
would like to thank Jan Balewski for his help with the computer system.
Daniel Winklehner is thankful for the continuing support by funds from the MIT 
Bose Fellowship. Spencer Axani would like to express his gratitude to the MIT 
MISTI program for providing travel funds.

\bibliography{Master}{}

\begin{thebibliography}{10}

\bibitem{nakamura:particle_physics}
K.~Nakamura and P.~D. Group, ``Review of particle physics,'' {\em Journal of
  Physics G: Nuclear and Particle Physics}, vol.~37, no.~7A, p.~075021, 2010.

\bibitem{aguilar:lsnd}
A.~Aguilar, L.~B. Auerbach, R.~L. Burman, D.~O. Caldwell, E.~D. Church, A.~K.
  Cochran, J.~B. Donahue, A.~Fazely, G.~T. Garvey, R.~M. Gunasingha, R.~Imlay,
  W.~C. Louis, R.~Majkic, A.~Malik, W.~Metcalf, G.~B. Mills, V.~Sandberg,
  D.~Smith, I.~Stancu, M.~Sung, R.~Tayloe, G.~J. VanDalen, W.~Vernon, N.~Wadia,
  D.~H. White, and S.~Yellin, ``{Evidence for neutrino oscillations from the
  observation of $\overline{\nu}_{\mathrm{e}}$ appearance in a
  $\overline{\nu}_{\mu}$ beam},'' {\em Phys. Rev. D}, vol.~64, p.~112007, 11
  2001.

\bibitem{aguilar:miniboone}
A.~A. Aguilar-Arevalo, B.~C. Brown, L.~Bugel, G.~Cheng, E.~D. Church, J.~M.
  Conrad, R.~Dharmapalan, Z.~Djurcic, D.~A. Finley, R.~Ford, F.~G. Garcia,
  G.~T. Garvey, J.~Grange, W.~Huelsnitz, C.~Ignarra, R.~Imlay, R.~A. Johnson,
  G.~Karagiorgi, T.~Katori, T.~Kobilarcik, W.~C. Louis, C.~Mariani, W.~Marsh,
  G.~B. Mills, J.~Mirabal, C.~D. Moore, J.~Mousseau, P.~Nienaber, B.~Osmanov,
  Z.~Pavlovic, D.~Perevalov, C.~C. Polly, H.~Ray, B.~P. Roe, A.~D. Russell,
  M.~H. Shaevitz, J.~Spitz, I.~Stancu, R.~Tayloe, R.~G. Van~de Water, D.~H.
  White, D.~A. Wickremasinghe, G.~P. Zeller, and E.~D. Zimmerman, ``{Improved
  search for $\overline{\nu}_{\mu} \rightarrow \overline{\nu}_{\mathrm{e}}$
  Oscillations in the MiniBooNE Experiment},'' {\em Phys. Rev. Lett.},
  vol.~110, p.~161801, 4 2013.

\bibitem{mention:reactor_anomaly}
G.~Mention, M.~Fechner, T.~Lasserre, T.~A. Mueller, D.~Lhuillier, M.~Cribier,
  and A.~Letourneau, ``Reactor antineutrino anomaly,'' {\em Phys. Rev. D},
  vol.~83, p.~073006, 4 2011.

\bibitem{bungau:daedalus}
A.~Bungau, A.~Adelmann, J.~R. Alonso, W.~Barletta, R.~Barlow, L.~Bartoszek,
  L.~Calabretta, A.~Calanna, D.~Campo, J.~M. Conrad, Z.~Djurcic, Y.~Kamyshkov,
  M.~H. Shaevitz, I.~Shimizu, T.~Smidt, J.~Spitz, M.~Wascko, L.~A. Winslow, and
  J.~J. Yang, ``{Proposal for an Electron Antineutrino Disappearance Search
  Using High-Rate $^{8}\mathrm{Li}$ Production and Decay},'' {\em Phys. Rev.
  Lett.}, vol.~109, p.~141802, 10 2012.

\bibitem{abe:kamland}
S.~Abe, T.~Ebihara, S.~Enomoto, K.~Furuno, Y.~Gando, K.~Ichimura, H.~Ikeda,
  K.~Inoue, Y.~Kibe, Y.~Kishimoto, M.~Koga, A.~Kozlov, Y.~Minekawa, T.~Mitsui,
  K.~Nakajima, K.~Nakajima, K.~Nakamura, M.~Nakamura, K.~Owada, I.~Shimizu,
  Y.~Shimizu, J.~Shirai, F.~Suekane, A.~Suzuki, Y.~Takemoto, K.~Tamae,
  A.~Terashima, H.~Watanabe, E.~Yonezawa, S.~Yoshida, J.~Busenitz, T.~Classen,
  C.~Grant, G.~Keefer, D.~S. Leonard, D.~McKee, A.~Piepke, M.~P. Decowski,
  J.~A. Detwiler, S.~J. Freedman, B.~K. Fujikawa, F.~Gray, E.~Guardincerri,
  L.~Hsu, R.~Kadel, C.~Lendvai, K.-B. Luk, H.~Murayama, T.~O'Donnell, H.~M.
  Steiner, L.~A. Winslow, D.~A. Dwyer, C.~Jillings, C.~Mauger, R.~D. McKeown,
  P.~Vogel, C.~Zhang, B.~E. Berger, C.~E. Lane, J.~Maricic, T.~Miletic,
  M.~Batygov, J.~G. Learned, S.~Matsuno, S.~Pakvasa, J.~Foster, G.~A.
  Horton-Smith, A.~Tang, S.~Dazeley, K.~E. Downum, G.~Gratta, K.~Tolich,
  W.~Bugg, Y.~Efremenko, Y.~Kamyshkov, O.~Perevozchikov, H.~J. Karwowski, D.~M.
  Markoff, W.~Tornow, K.~M. Heeger, F.~Piquemal, and J.-S. Ricol, ``Precision
  measurement of neutrino oscillation parameters with kamland,'' {\em Phys.
  Rev. Lett.}, vol.~100, p.~221803, 6 2008.

\bibitem{conrad2014precision}
J.~Conrad, M.~Shaevitz, I.~Shimizu, J.~Spitz, M.~Toups, and L.~Winslow,
  ``{Precision $\nu_{\mathrm{e}}$ - electron scattering measurements with
  IsoDAR to search for new physics},'' {\em Physical Review D}, vol.~89, no.~7,
  p.~072010, 2014.

\bibitem{aberle2013whitepaper}
C.~Aberle, A.~Adelmann, J.~Alonso, W.~Barletta, R.~Barlow, L.~Bartoszek,
  A.~Bungau, A.~Calanna, D.~Campo, L.~Calabretta, {\em et~al.}, ``Whitepaper on
  the daedalus program,'' {\em arXiv preprint arXiv:1307.2949}, 2013.

\bibitem{adelmann:isodar}
A.~Adelmann, J.~R. Alonso, W.~Barletta, R.~Barlow, L.~Bartoszek, A.~Bungau,
  L.~Calabretta, A.~Calanna, D.~Campo, J.~M. Conrad, Z.~Djurcic, Y.~Kamyshkov,
  H.~Owen, M.~H. Shaevitz, I.~Shimizu, T.~Smidt, J.~Spitz, M.~Toups, M.~Wascko,
  L.~A. Winslow, and J.~J. Yang, ``Cost-effective design options for
  {I}so{DAR},''

\bibitem{reiser:beams}
M.~Reiser, {\em Theory and design of charged particle beams}.
\newblock Weinheim: Wiley-VCH, 2~ed., 2008.

\bibitem{celona:trips}
L.~Celona, G.~Ciavola, S.~Gammino, F.~Chines, M.~Presti, L.~And\`o, X.~H. Guo,
  R.~Gobin, and R.~Ferdinand, ``Status of the trasco intense proton source and
  emittance measurements,'' {\em Review of Scientific Instruments}, vol.~75,
  no.~5, pp.~1423--1426, 2004.

\bibitem{castro:vis2}
G.~Castro, G.~Torrisi, L.~Celona, D.~Mascali, L.~Neri, G.~Sorbello,
  O.~Leonardi, G.~Patti, G.~Castorina, and S.~Gammino, ``{A new
  $\mathrm{H}_2^+$ source: Conceptual study and experimental test of an
  upgraded version of the VIS - Versatile Ion Source},'' {\em PRSTAB}, 2015.

\bibitem{xu:current}
Y.~Xu, S.~Peng, H.~Ren, J.~Zhao, J.~Chen, A.~Zhang, T.~Zhang, Z.~Guo, and
  J.~Chen, ``High current h2+ and h3+ beam generation by pulsed 2.45 ghz
  electron cyclotron resonance ion sourcea),'' {\em Review of Scientific
  Instruments}, vol.~85, no.~2, pp.~--, 2014.

\bibitem{zhang:ion}
H.~Zhang, {\em Ion Sources}, vol.~1.
\newblock Springer-Verlag Berlin Heidelberg, 1999.

\bibitem{aseev:track}
V.~Aseev, P.~Ostroumov, E.~Lessner, B.~Mustapha, {\em et~al.}, ``Track: The new
  beam dynamics code,'' in {\em PAC05}, 2005.

\bibitem{grote:warp1}
D.~P. Grote, A.~Friedman, J.-L. Vay, and I.~Haber, ``The {WARP} code:
  {M}odeling high intensity ion beams,'' in {\em 16th {I}ntern. {W}orkshop on
  {ECR} {I}on {S}ources} (M.~Leitner, ed.), vol.~749, AIP, 2004.

\bibitem{jongen:cyclotron1}
Y.~Jongen, M.~Abs, A.~Blondin, W.~Kleeven, S.~Zaremba, D.~Vandeplassche, A.~V.,
  S.~Gursky, O.~Karamyshev, G.~Karamysheva, N.~Kazarinov, S.~Kostromin,
  N.~Morozov, E.~Samsonov, G.~Shirkov, V.~Shevtsov, E.~Syresin, and A.~Tuzikov,
  ``{IBA-JINR} 400 {M}e{V}/u superconducting cyclotron for hadron therapy,'' in
  {\em Proceedings of CYCLOTRONS 2010}, 2010.

\bibitem{bellomo:axial_injection}
G.~Bellomo, D.~Johnson, F.~Marti, and F.~Resmini, ``On the feasibility of axial
  injection in superconducting cyclotrons,'' {\em Nuclear Instruments and
  Methods in Physics Research}, vol.~206, no.~1--2, pp.~19--46, 1983.

\bibitem{toprek:spiral_inflector}
D.~Toprek, ``Theory of the central ion trajectory in the spiral inflector,''
  {\em Nuclear Instruments and Methods in Physics Research Section A:
  Accelerators, Spectrometers, Detectors and Associated Equipment}, vol.~440,
  no.~2, pp.~285 -- 295, 2000.

\bibitem{opera:online}
OPERA3D, ``Cobham plc: Aerospace and security, antenna systems, kidlington.''
  \url{http://www.cobham.com/}, 2 2013.

\bibitem{alonso:vis1}
J.~R. Alonso, L.~Calabretta, D.~Campo, L.~Celona, J.~Conrad, R.~G. Martinez,
  R.~Johnson, F.~Labrecque, M.~H. Toups, D.~Winklehner, {\em et~al.},
  ``{C}haracterization of the {C}atania {VIS} for {H}$_2^+$,'' {\em Review of
  Scientific Instruments}, vol.~85, no.~2, p.~02A742, 2014.

\bibitem{scargle:statistics}
J.~D. Scargle, ``Studies in astronomical time series analysis.
  {II}-{S}tatistical aspects of spectral analysis of unevenly spaced data,''
  {\em The Astrophysical Journal}, vol.~263, pp.~835--853, 1982.

\bibitem{grote:warp2}
D.~P. Grote, {\em {WARP} {M}anual}, 2000.

\bibitem{menzel:poisson}
M.~Menzel and H.~Stokes, ``Users guide for the {POISSON}/{SUPERFISH} group of
  codes,'' tech. rep., Los Alamos National Lab., NM (United States), 1987.

\bibitem{kersevan1991molflow}
R.~Kersevan, ``Molflow user's guide,'' {\em available from one of the authors
  (RK)}, 1991.

\bibitem{autodesk2014autocad}
I.~Autodesk, ``Autocad,'' {\em Sausalito, CA}, 2009.

\bibitem{gabovich:spacecharge1}
M.~D. Gabovich, L.~P. Katsubo, and I.~A. Soloshenko, ``Selfdecompensation of a
  stable quasineutral ion beam due to coulomb collisions,'' {\em Fiz. Plazmy},
  vol.~1, pp.~304--309, 1975.

\bibitem{gabovich:spacecharge2}
M.~D. Gabovich, ``Ion-beam plasma and the propagation of intense compensated
  ion beams,'' {\em Soviet Physics Uspekhi}, vol.~20, no.~2, p.~134, 1977.

\bibitem{soloshenko:spacecharge1}
I.~A. Soloshenko, ``Physics of ion beam plasma and problems of intensive ion
  beam transportation (invited),'' {\em Review of scientific instruments},
  vol.~67, no.~4, pp.~1646--1652, 1996.

\bibitem{soloshenko:spacecharge2}
I.~A. Soloshenko, ``Space charge compensation of technological ion beams,'' in
  {\em Discharges and Electrical Insulation in Vacuum, 1998. {P}roceedings
  {ISDEIV}. {XVIII}th International Symposium on Discharges and Electrical
  Insulation in Vacuum}, vol.~2, pp.~675--678, IEEE, 1998.

\bibitem{soloshenko:spacecharge3}
I.~A. Soloshenko, ``Transportation of intensive ion beams,'' {\em Review of
  scientific instruments}, vol.~69, no.~3, pp.~1359--1366, 1998.

\bibitem{winklehner:scc2}
D.~Winklehner, D.~Leitner, D.~Cole, G.~Machicoane, and L.~Tobos, ``Space-charge
  compensation measurements in electron cyclotron resonance ion source low
  energy beam transport lines with a retarding field analyzer,'' {\em Review of
  Scientific Instruments}, vol.~85, no.~2, 2014.

\bibitem{winklehner:phd}
D.~Winklehner, {\em Ion Beam Extraction from Electron Cyclotron Resonance Ion
  Sources and the Subsequent Low Energy Beam Transport}.
\newblock PhD thesis, Michigan State University, 2013.

\bibitem{spaedke:kobra}
P.~Sp\"adtke, ``Kobra3-inp user manual, version 3.39,'' 2000.

\bibitem{adelmann:opal}
A.~Adelmann, A.~Gsell, C.~K. (PSI), Y.~I. (IBM), S.~R. (LANL), Y.~Bi, C.~Wang,
  J.~Y. (CIAE), H.~Z.~T. University), S.~Sheehy, C.~R. (RAL), and C.~M.
  (Cornell), ``{The OPAL (Object Oriented Parallel Accelerator Library)
  Framework},'' Tech. Rep. PSI-PR-08-02, Paul Scherrer Institut, (2008-2013).

\bibitem{hamm:rfq1}
R.~Hamm, D.~Swenson, and T.~Wangler, ``Use of the radio-frequency quadrupole
  structure as a cyclotron axial buncher system,'' in {\em 9. International
  conference on cyclotrons and their applications}, 1982.

\bibitem{winklehner:rfq1}
D.~Winklehner, R.~Hamm, J.~Alonso, and J.~Conrad, ``{AN RFQ DIRECT INJECTION
  SCHEME FOR THE ISODAR HIGH INTENSITY $\mathrm{H}_2^+$ CYCLOTRON},'' in {\em
  6th International Particle Accelerator Conference (IPAC2015)}, 2015.

\end{thebibliography}
\bibliographystyle{ieeetr}

\end{document}